\documentclass[draft]{agujournal2019}
\usepackage{url} 
\usepackage{lineno}
\usepackage[inline]{trackchanges} 
\usepackage{soul}

\usepackage{bm} 

\addeditor{AG}

\usepackage{enumerate}
\usepackage{amsmath} 
\usepackage{epstopdf}
\graphicspath{{fig/}}
\draftfalse

\journalname{Water Resources Research}

\begin{document}


\title{Diagnosis of model-structural errors with a sliding time-window Bayesian analysis}

%
%

\authors{Han-Fang Hsueh\affil{1,2}, Anneli Guthke\affil{1}, Thomas W{\"o}hling\affil{3,4}, Wolfgang Nowak\affil{1}  } 

\affiliation{1}{Department of Stochastic Simulation and Safety Research for Hydrosystems (IWS/LS$^3$), University of Stuttgart, Stuttgart, Germany}
\affiliation{2}{Center for Applied Geoscience, University of T\"ubingen, Tübingen, Germany} 
\affiliation{3}{Department of Hydrology, Technical University of Dresden, Dresden, Germany}
\affiliation{4}{Lincoln Environmental Research, Lincoln Agritech, Hamilton, New Zealand}

\correspondingauthor{Han-Fang Hsueh}{han-fang.hsueh@iws.uni-stuttgart.de}




\begin{keypoints}
\item We propose a data-driven method for model-structural error detection. 
\item Our method rests on a statistically rigorous Bayesian framework without prior assumptions about error sources or patterns.
\item We confirm successful error detection on various temporal scales in synthetic test cases and present insights from a real-world case study.
\end{keypoints}

%
%

%
%



\begin{abstract} 

Deterministic hydrological models with uncertain, but inferred-to-be-time-invariant parameters typically show time-dependent model structural errors. Such errors can occur if a hydrological process is active in certain time periods in nature, but is not resolved by the model. Such missing processes could become visible during calibration as time-dependent best-fit values of model parameters. We propose a formal \emph{time-windowed Bayesian analysis} to diagnose this type of model error, formalizing the question ``In which period of the calibration time-series does the model statistically disqualify itself as quasi-true?'' Using Bayesian model evidence (BME) as model performance metric, we determine how much the data in time windows of the calibration time-series support or refute the model. Then, we track BME over sliding time windows to obtain a dynamic, time-windowed BME (tBME) and search for sudden decreases that indicate an onset of model error. tBME also allows us to perform a formal, sliding likelihood-ratio test of the model against the data. Our proposed approach is designed to detect error occurrence on various temporal scales, which is especially useful in hydrological modelling. We illustrate this by applying our proposed method to soil moisture modeling. We test tBME as model error indicator on several synthetic and real-world test cases that we designed to vary in error sources and error time scales. Results prove the usefulness of the framework for detecting structural errors in dynamic models. Moreover, the time sequence of posterior parameter distributions helps to investigate the reasons for model error and provide guidance for model improvement.
\end{abstract}

\section{Introduction} 

Water resources research and management is increasingly relying on mathematical models to gain a deeper system understanding, to predict a system's current state (now-casting), and to predict a system's state under future conditions (forecasting). Unfortunately, from a philosophical point of view, all models are wrong because they are simplifications of the real system \cite{1976_Box}. Model-structural errors generally arise when the dominant processes are not comprehensively and correctly represented in a model. It is important to investigate the occurrence of structural errors for several reasons: (1) Model errors lead to biased predictions \cite{2014_Brynjarsd,2007_Ajami} and hence compromise the usefulness of a model for practically all modelling purposes (process identification, nowcasting, forcasting) and subsequent management decisions \cite{2019_Sargsyan,2015_Steinschneider}. (2) Model errors lead to biased parameter identification through calibration or Bayesian parameter inference \cite{2015_Xu} or cause non-identifiability of parameters \cite{2019_Menberg, 2009_Reichert}, because parameters are optimized in the wrong context; they are adjusted regardless of their physically feasible ranges to obtain a satisfying agreement to observed data despite the model error \cite{1994_Oreskes}. (3) Given model-structural errors, uncertainty analysis within a conventional Bayesian framework yields misleading results, because Bayesian theory assumes an error-free (``true'') model \cite{2018_Fenicia, 2013_Gelman, 1973_Box, 2017_Xu}. 

Examples of misspecified mechanisms or missing mechanisms in hydrological models are given by \citeA{2012_Gupta}. Model uncertainties and errors typically exist in model inputs (forcings) \cite{2006_Kuczera,2015_DelGiudice, 2013_Giudice}, model building \cite{2008_Clark}, and observations of model output \cite{2010_Renard}. Working with a model with structural errors becomes a problem, if we are interested in true parameter values that relate to property or process identification, or if we make predictions by extrapolating a physically-based model beyond calibration conditions. When parameters compensate for model structural errors, they lose their physical meaning \cite{2017_Xu,2012_Reichert}. Such errors can also show more subtle, e.g. through the fact that dynamic parameters or a nonstationary model status improve model performance \cite{2016_Wagner,2016_Pianosi,2015_Getahun,2015_Moran}. Examples for processes that often cause time-dependent model structures in hydrological systems are soil cracking-shrinking and freezing-thawing cycles, complex vegetation dynamics, or transient changes of river morphology. Model-structural errors can also be caused by neglecting major processes (e.g. macro-pore flow) or by the application of ill-suited model concepts.

Model-structural errors are currently addressed by methods that typically follow one of two primary directions: error detection or error mitigation. To detect error occurrence, the temporal change of model performance is often used as an indicator. \citeA{1986_Klemes} presented a generalization of the routine split-sample test. Their study shows varied model performance in the calibration period, which is associated with varied model conditions or model inadequacy. Following this concept, \citeA{2012_Coron} proposed the generalised split-sample test that uses a sliding window to calibrate the model with subsets of the time series data and to study model deficiencies by analysing the parameter transferability. \citeA{2019_Motavita} proposed the comprehensive differential split-sample test (CDSST) to study (non-)stationarity of calibrated parameters. This research even suggests that, for a model with errors, the choice of hydrological conditions in calibration is more important for model reliability than the length of the calibration dataset. \citeA{2014_Westra} proposed to use parameter nonstationarity as a hint to diagnose model-structural errors. \citeA{2016_Pathiraja} allow parameters to vary in time and use data assimilation to detect temporal patterns. While the mentioned approaches rely on common model evaluation metrics, \citeA{2019_Ruddell} propose to diagnose structural error by quantifying the trade-off between functional and predictive performance with information-theoretic measures. 

Another approach to detect structural errors is to compare the performance of competing model candidates in calibration on varied datasets. The underlying assumption is that structural errors show by a significant decrease in prediction quality of any specific data type when calibrated on a different data type - parameter compensation is preventing the model from finding a robust best-fit parameter set for all relevant data types. For example, \citeA{2013_Eddy} use multiobjective calibration as a diagnostic tool to identify structural errors in coupled soil-plant models. In a later study, \citeA{2015_Eddy} use Bayesian model selection \cite{1995_Raftery} to evaluate the worth of different calibration data types. Bayesian model selection (BMS) works under the assumption that one of the models in the set is true; if the identified (pseudo-)true model changes depending on the chosen calibration dataset, this is an indication of varying expression of model-structural errors with varied calibration strategy. A review of BMS in a more general model context is given by \citeA{2019_Marvin}. \citeA{2015_DelGiudice} assess the decrease of model-structural errors with increasing model complexity and draw conclusions about the relative importance of model-structural errors vs. input errors for the quality of model predictions. 

To mitigate the impact of model-structural errors on model performance, the majority of approaches treat the apparent bias in model output, i.e., these approaches weaken the commonly violated assumption of uncorrelated errors in the likelihood function/objective function during calibration. For example, \citeA{2019_Ammann} investigate various error models and report that including non-stationary autocorrelation improves model performance. In contrast, however, \citeA{2011_Eddy} reported unacceptably large prediction errors when  using the $AR-1$ model in the likelihood formation. \citeA{2015_Smith} summarize the performance of eight residual-error models and report significant change in the posterior mode of their key model parameter as a function of the error model. Generalized likelihood uncertainty estimation (GLUE) \cite{1992_Beven} is a variation of the likelihood formulation, which explicitly expresses the autocorrelation in errors and widens the formal likelihood function to subjective limits of acceptability based on expert knowledge \cite{2020_Ragab,2015_Mirzaei}. Instead of correcting the likelihood function, \citeA{2009_Demissie} propose a data-driven error model to reduce the bias in groundwater model predictions. 

Another branch of studies aims to tackle structural deficits \emph{within} the model, not only ``at the end of the pipe''. For example, \citeA{2015_Xu} infer a Gaussian process error model to improve parameter estimation and model prediction quality. \citeA{2009_Reichert} introduce stochastic time-dependent parameters to correct for structural errors within the model. Their goal is to achieve more realistic prediction uncertainty intervals. 

In this work, we are focusing on model error \emph{detection}, because we believe that effective model improvement (making the model do the right things for the right reasons) has to rely on solid error detection as a foundation. This sequential procedure seems scientifically more satisfying than aiming for better model performance without knowing the reasons for the model's misbehavior. We therefore present a method for error detection here, and leave an extension toward error mitigation for future work. 
       
Despite the increasing scientific interest in model-structural errors, existing methods still suffer from major limitations. Current methods for diagnosing model-structural errors require much a-priori, expert knowledge about model errors \cite{2014_Westra, 2014_Hrachowitz}, trade statistical rigor for easy applicability and/or suffer from numerical inefficiency \cite{2009_Reichert}. The role and limitations of hypothesis testing for identifying of model-structural errors has been recently discussed by, e.g., \citeA{2016_Nearing,2017_Neuweiler,2018_Beven,2020_Nearing}. Nevertheless, these studies agree that, until now, the hydrological community is lacking a rigorous framework specifically designed for the diagnosis of model-structural errors. 



We intend to fill this gap by proposing a time-windowed Bayesian analysis framework to detect model-structural errors. We target time-dependent errors in dynamic models, but do not assume any a-priori knowledge about error sources or error patterns. In the spirit of Bayesian hypothesis testing, we identify periods in the calibration time series that feature statistically significant model error.

We start from the assumption that the model can describe the real world system sufficiently well for most of the time, except when some unresolved mechanism occasionally occurs. This means that we can claim the model to be quasi-true for specific periods in ''time windows'', in the sense that it fits the observation data sufficiently well within, e.g., one or two days or a week. Only when we increase that time window, we put the model into potential trouble: if the increased time window now contains data impacted by any unresolved mechanism, the calibration goodness-of-fit will deteriorate, and parameter identification will become visibly ambiguous. At that stage, the model parameters are forced to move away from their previous (allegedly) well-calibrated state, in order to compensate for the now occurring structural error. 

We detect this problematic phase by calibrating the model within a pre-defined shorter time window and moving this time window through the whole data time series (hence the name of our method, ``moving time-window analysis''). While moving the time window, we track model performance and search for sudden decreases that would indicate an onset of model error. 

We built our framework on the mathematically rigorous grounds of Bayesian probability theory and use Bayesian model evidence (BME) as the model performance metric. BME is perfectly suited to judge a model's performance under parameter and data uncertainty, because it evaluates the model's likelihood to have generated the data, averaged over the model's parameter space. We use a brute-force Monte Carlo implementation of BME \cite{2014_Anneli} as a basis for our proposed routine. This numerical implementation is computationally heavy, but often practically feasible. The additional effort for model error detection is moderate compared to the baseline effort of setting up a predictive ensemble for brute-force Monte Carlo Bayesian updating, because it does not require any additional model runs. 

To identify statistically significant drops in performance that are triggered by an onset of model error, we introduce a \emph{tBME reference distribution} as sampling distribution. We start by asking, what should the tBME value be if the model was true (i.e., error-free)? We randomly sample a realization of model output as synthetic dataset to generate a series of theoretical tBME values. We repeat this random sampling procedure many times to produce a random sample of the tBME series that accounts for the full output distribution of the model. This procedure is similar to the bootstrapping analysis on BMS results introduced by \citeA{2015_Anneli}, with the synthetic data here being drawn from the model's output distribution instead of being resampled from noisy observed data. Sampling the distribution of model states to perform ``hypothetical calibration'' is known from Bayesian pre-posterior analysis \cite<see, e.g.,>[]{2012_Philipp,2016_Wolfgang}. With that tBME sampling distribution, we have a reference for each considered time window. If the tBME value given real data in a specific time window falls outside of the sampled range, it is very unlikely that the model is true, and hence it is very likely that we spotted the onset of model error. If, in contrast, the tBME value given a field dataset falls within the reference interval, we conclude that the model fulfills the statistic requirement of a true model at this moment. This model can be considered as a quasi-true model with (statistically) negligible structural error. Technically, this is a fully data-driven likelihood ratio test, i.e., free of parametric assumption about the distributional shape of model errors. 

Our Bayesian approach does not presume any model error pattern, and it can handle any choice of formal likelihood function. The method provides an informative and intuitive visualization of model error occurrence on varied temporal scales, simply by changing the size of the sliding window. We will further show that this method is able to detect superposed errors, if they occur on different temporal scales (e.g. daily and seasonal patterns). In contrast to most existing methods, our approach follows a data-driven bottom-up perspective: there is no need to make parametric assumptions about the type of model error first and then to formulate and perform a parametric hypothesis test as in \citeA{2017_Neuweiler}; rather, we let the observed data speak to identify the onset (and offset) of error in a data-driven hypothesis test. Only in subsequent steps, the modeller may investigate the error period more closely and try to formulate hypotheses about the reasons for model error.  
    
By detecting model errors with our proposed analysis, we set the grounds for improved system understanding and potential model improvement in further model building. Detection of errors on different temporal scales is especially useful in hydrology, since it helps improve our conceptual understanding in general and because it improves hydrological predictions and their usefulness in all practical aspects. Nevertheless, the application of this method can be generalized to any dynamic system with long time series available for calibration. 

In summary, the main contributions of this study are:
\begin{enumerate}
    \item We develop a data-driven method for model structural error detection in a statistically rigorous Bayesian framework, with no prior assumptions about error sources or patterns.
    \item By introducing a reference range (sampling distribution) for the model performance metric (BME), we formalize the question: ``In which periods does the model statistically disqualify itself as quasi-true?'' to a data-driven likelihood ratio test.  
    \item Our proposed approach is designed to detect error occurrence on various temporal scales, with the choice of temporal scales to be investigated left to the modeller. Hence, our framework generalizes well to arbitrary dynamic systems with long calibration time series, but allows for individual adaptation to specific applications and scientific research questions. 
\end{enumerate}

The only restriction to our procedure is that the time scale resolved by the available data time-series must be smaller, and the total length of the available data must be larger than the time scale of errors to be diagnosed. The article is structured as follows: Section \ref{sec:Material}  introduces the proposed method, especially the time-windowed Bayesian analysis and the reference range. Section \ref{sec:Application} presents the model and the field data used in our case study, and defines several synthetic test cases to validate our method. We discuss the implications of the synthetic cases and interpret the time-windowed analysis for the real data in Section \ref{sec:Results}. Finally, we conclude about the strengths and limitations of this method in Section \ref{sec:Conclusion}.

\section{Methods} 
\label{sec:Material} 

To investigate the temporal occurrence of model error, we propose to perform a time-window-based analysis of model performance, with the performance metric being BME. Assuming that the model is generally fitting the calibration data well, it should receive a high BME score most of the time; only in periods with missing/misspecified mechanisms acting, the model's performance will deteriorate significantly. This sudden misfit will cause BME to decline, and hence we can use the time-window based BME (tBME) as model error indicator. 

The challenge is now to distinguish ``noisy fluctuations'' in BME from statistically significant declines that actually represent model error. Since there is no meaningful interpretation of BME values on an absolute scale, we need a reference to compare the obtained BME value to. We therefore determine a reference interval which contains hypothetical BME values under the assumption that the model is in fact true and error-free (the fact that we still see an interval and not just a single value is due to parameter uncertainty in our model). We can then perform a time series of hypothesis testing: if the model is error-free per window, the corresponding tBME value given observed data should fall within the hypothetical interval. This means at this time scale (of the time window size) the model can be considered ``pseudo-true'', i.e. the model is representative of the true system in a statistical sense.

\begin{figure}[ht]
\centering
\includegraphics[scale=0.5]{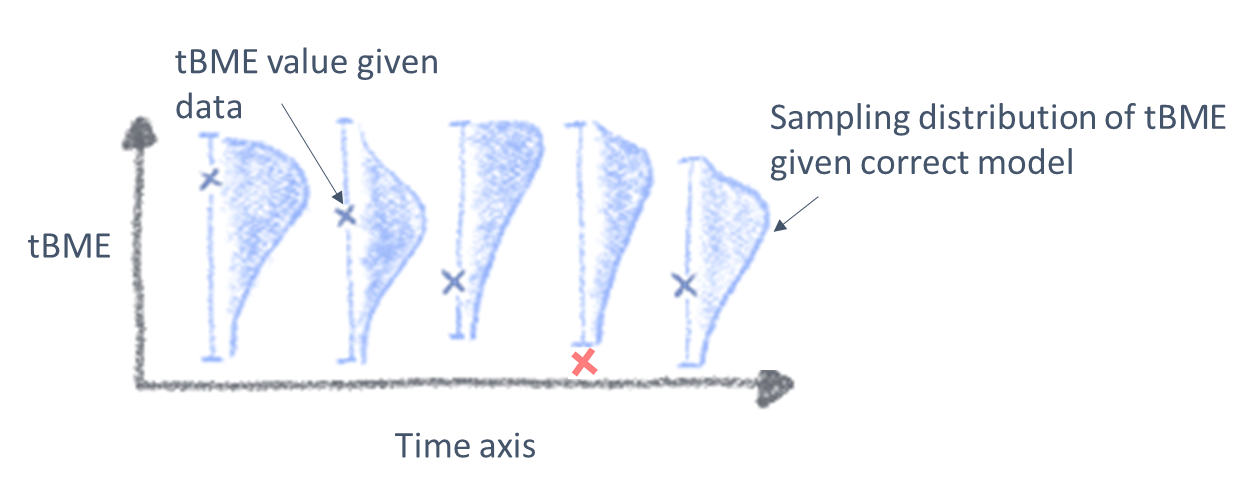}
\caption{Conceptual illustration of the proposed method: a time series of hypothesis tests with time-windowed BME (tBME). If the tBME value given data falls within the sampling distribution of tBME, the model can be considered ``pseudo-true'' during this period (blue crosses); if the value falls outside the sampling distribution, this is an indication of model error (red cross).}
\label{fig:HypoTestTau}
\end{figure}

If the tBME value drops outside the interval, we can conclude that the current model setup does not accurately describe the observed system state in a given time window, i.e. that model error occurs. This procedure is schematically illustrated in Fig. \ref{fig:HypoTestTau}.

In the following, we will explain the theoretical development of our proposed method. Since the method is settled in a Bayesian framework, we start with fundamentals of Bayesian model evaluation in Section \ref{sec:Bayes}. Then, we explain the moving time-window Bayesian analysis in Section \ref{sec:tBME}, mention relevant time scales in Section \ref{sec:Time_Scale}, and discuss the theoretical behavior of the tBME curve for varied time scales in Section \ref{sec:tBME_interp}. Section \ref{sec:BME_reference} describes the construction of the reference interval for tBME interpretation. As a side product of time-windowed calibration, we obtain dynamic posterior parameter distributions introduced in Section \ref{sec:Dynamic_Posterior}.


\subsection{Bayesian Model Evidence (BME)} 
\label{sec:Bayes}

The Bayesian framework allows to quantify an uncertain state of knowledge, which can be updated in the light of observed data. An uncertain quantity of interest $\textbf{y}$ predicted by a model $M$ can be expressed as a random variable with a prior PDF $p(\textbf{y}|M)$, where PDF stands for probability density function. By determining the likelihood of the model prediction to have generated the data, $p(\textbf{D}|\textbf{y},M)$, this prior PDF is updated to a posterior PDF $p(\textbf{y}|\textbf{D},M)$: 

\begin{equation}
    \label{eqn:Bayes1}
    p(\textbf{y}|\textbf{D},M) = \frac{p(\textbf{D}|\textbf{y},M)p(\textbf{y}|M)}{p(\textbf{D}|M)}
\end{equation}

The term in the denominator, $p(\textbf{D}|M)$, is a model-specific normalizing constant that marginalizes over the posterior PDF, and represents the model’s marginal likelihood. This term describes how likely it is that the model $M$ has generated the observed data $\textbf{D}$. It is therefore also called Bayesian model evidence (BME), which can be expressed as 

\begin{eqnarray}
    \label{eqn:BME} 
    \mathrm{BME} & \equiv & p(\textbf{D}|M) \\ \nonumber
    & = & \int_{ -\infty }^{ \infty } p(\textbf{D}|\bm{\omega},M)p(\bm{\omega}|M)d\bm{\omega} \\ \nonumber 
    & \approx & \frac{1}{N_{MC}} \sum_{i=1}^{N_{MC}} p(\textbf{D}|\bm{\omega}_i ,M),
\end{eqnarray}
with the last row showing a brute-force Monte Carlo numerical approximation of BME \cite{2014_Anneli}. A large number $N_{MC}$ of random parameter realizations $\bm{\omega}_i$ is drawn from the parameters' prior PDF $p(\bm{\omega}|M)$ to form a Monte Carlo ensemble, the corresponding model predictions $\bm{y}_i$ are determined, and their likelihoods are averaged to approximate the Bayesian integral. This approximation is assumption- and bias-free, but computationally expensive. For other possible approximations of BME, readers are referred to  \citeA{2014_Anneli,2016_Liu} and \citeA{2019_Sergey}.

BME is perfectly suited as a model performance metric in the face of uncertainty (in parameters, forcings, and output data), because it balances a model's goodness-of-fit with a penalty for overcomplexity, see \citeA{2018_Marvin} for a review on this topic. To give a simplistic explanation, think of a 1D prediction: BME is exactly the PDF value of the model's predictive distribution read off at the observed data value. The numerical approximation in Eq. \ref{eqn:BME} avoids the problem of estimating high-dimensional PDFs (as would be necessary for long time-series of calibration data) by sampling the low-dimensional parameter space instead. Thus, it obtains a sample of the predictive PDF by running the forward model on the parameter sample.



As likelihood function to represent measurement error, we choose a Gaussian distribution centered about the observed data $\textbf{D}$: 

\begin{equation}
    p(\textbf{D}|\bm{\omega}_i,M) =   \frac{1}{ \sqrt{ (2\pi)^{N_o} |\textbf{R}| }  }   \exp \big( -\frac{1}{2}  ( \textbf{D} - \textbf{y}_i  )^T\textbf{R}^{-1} ( \textbf{D} - \textbf{y}_i )   \big),
    \label{eqn:likelihood}
\end{equation}
with $N_o$ representing the number of observations (i.e., the length of $\textbf{D}$). The model misfits (residuals) $\textbf{D}-\textbf{y}_i$ are attributed to uncorrelated measurement errors, i.e., error covariance matrix $\textbf{R}$ of size $N_o$ x $N_o$ has non-zero elements only on its main diagonal. It is clear that model errors will show a complicated pattern of correlation \cite<see, e.g.,>[]{2013_Evin}, and many authors used the likelihood function to also represent model errors \cite{2019_Ammann, 2015_Smith}. We are not doing so, because we do not want to prescribe any specific model error pattern a priori. Instead, we let the data speak and identify the model errors for us. We therefore propose to start with the simplified assumption of uncorrelated errors in the model error detection phase, and recommend to use the result of this analysis as input for more realistic modelling of model errors (or even better: improvement of the underlying conceptual or physics-based model) in a subsequent phase of model error mitigation. A modeler could also treat random measurement error and known systematic model error in a lumped fashion, in order to detect further yet unknown errors. Besides, the choice of likelihood function is not limited to the one we show here, and does not influence the functioning of our framework.

\subsection{BME with a Moving Time-Window (tBME)} 
\label{sec:tBME}
 
Recall that we assume that the model under investigation is generally correct except for some unresolved processes that occasionally occur. Under this assumption, the goodness-of-fit of a model given a short dataset should be high, unless this subset of the calibration time series includes the occurrence of unresolved processes. Based on this concept, we propose to calibrate the model with a moving window, using BME as a model performance metric. 

The analysis starts with the selection of a fixed window size. This window size determines the length of the time-series subset chosen for BME evaluation. With a time-window size of $\tau$, BME is evaluated on a dataset of $\tau$ time steps. Eq. \ref{eqn:Bayes1} then rewrites as:  

\begin{equation}
    \label{eqn:Bayes2}
     p(\textbf{y}_{T_j(\tau)}|\textbf{D}_{T_j(\tau)}) = \frac{p(\textbf{D}_{T_j(\tau)}|\textbf{y}_{T_j(\tau)})p(\textbf{y}_{T_j(\tau)})}{p(\textbf{D}_{T_j(\tau)}|M)}
\end{equation}

which transforms BME (Eq. \ref{eqn:BME}) to be tBME and writes as:

\begin{equation}\label{eqn:BME2}   
    \mathrm{tBME} =  p(\textbf{D}_{T_j(\tau)}|M).
\end{equation}

Note that evaluating $p(\textbf{D}_{T_j(\tau)}|M)$ according to Eq. \ref{eqn:BME} means to apply the likelihood function (Eq. \ref{eqn:likelihood}) to the data subset $\textbf{D}_{T_j(\tau)}$ instead of $\textbf{D}$.

\begin{figure}[ht]
\centering
\includegraphics[scale=0.8]{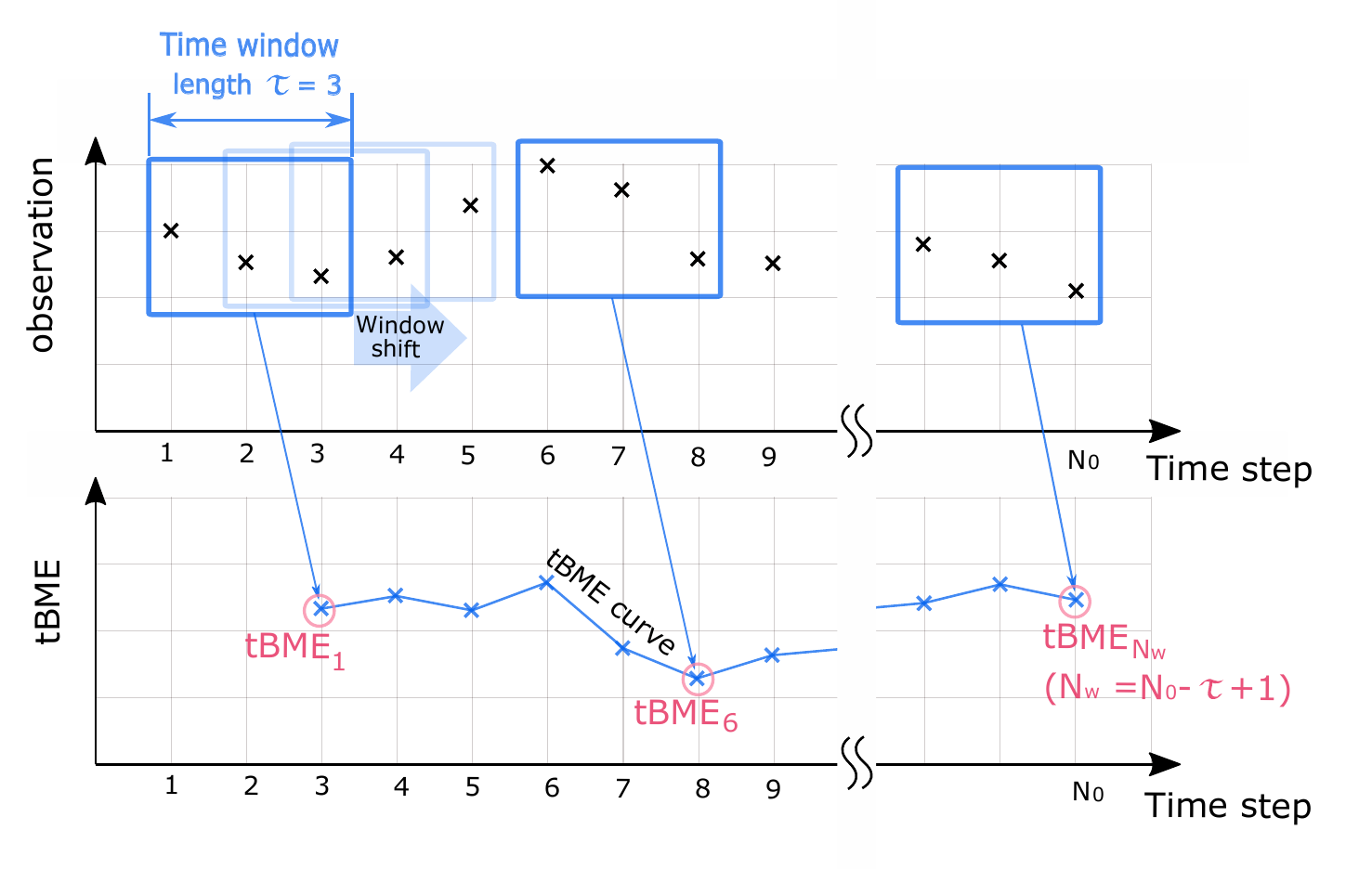}
\caption{Illustration of tBME evaluation with a moving time-window. A series of BME evaluation constructs the tBME curve, which is later used as an indicator for model structural error occurrence.}
\label{fig:movingTau}
\end{figure}

This time window is then sliding through the observed time series to form a tBME curve as a function of time, which we call tBME curve (see Figure\ref{fig:movingTau}). 
We obtain a number of $N_w$ individual tBME values that make up this curve. $N_w$ is equal to $N_o-\tau+1$, i.e. the first time window ``eats up'' $\tau$ data points of the original data time series, and then this window is shifted time step by time step ($j=\tau,\tau+1,\tau+2,\ldots,N_o$) until the final window contains the last data point with index $N_o$  (cf. Figure\ref{fig:movingTau}). The choice of shifting by a certain time step (e.g. typically a day in hydrology) yields the maximum number of possible windows; computational time constraints could justify a shifting of more than one time step at once and thus reducing the number of tBME values to be evaluated. The length of the time step can be arbitrary, depending on the frequency of the measurements. In our application (\ref{sec:Cases}) we choose one day for one time step.

\subsection{Relevant Time Scales for tBME Analysis}
\label{sec:Time_Scale}

Before discussing the different time scales that are relevant for the tBME analysis, we wish to differentiate between the terms \emph{error period} and \emph{residual period}: With \emph{error period}, we will refer to the time span during which model error occurs (e.g., model lacks a snow-melt routine), while we define the \emph{residual period} as the time span during which the \emph{effect} of model error can be observed in form of model residuals (e.g., modeled soil conditions are too dry because input from snow-melt is missing). Therefore, the length of residual periods does not necessarily coincide with the length of the error period itself; instead, residual periods are typically longer in hydrological modeling because errors tend to accumulate over time. Since we do not know the source of model error in real-world settings, we can only detect \emph{residual periods} with the proposed method and hypothesize about the responsible error periods based on our expert knowledge. 

\begin{figure}[ht]
\centering
\includegraphics[scale=0.6]{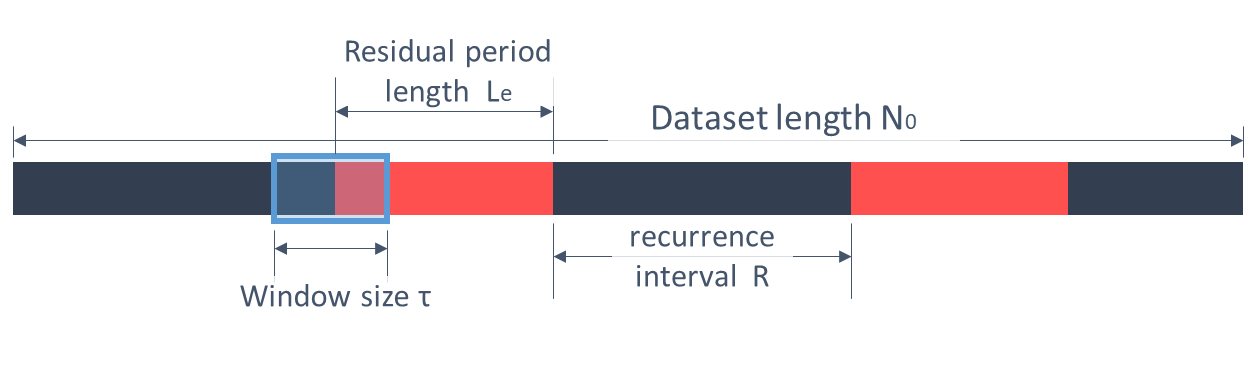}
\caption{Schematic illustration of the four time scales relevant in tBME analysis.}
\label{fig:4time_Scale}
\end{figure}

Besides the residual period with length $L_e$, three additional time scales are relevant in our proposed method: the window size $\tau$, the residual recurrence interval $R$, and the total dataset length $N_o$ (see Figure\ref{fig:4time_Scale}). The term ``residual recurrence interval'' stands for the residual-free time interval between two distinguishable residual periods. Note that the two residual periods in Fig. \ref{fig:4time_Scale} do not have to be caused by identical sources. The only condition for errors to be detectable is simply the residuals' significance.

The observed dataset should be long enough to at least contain a complete residual period $L_e$. Further, the residual recurrence interval $R$ should be larger than the window size $\tau$, so that the residuals caused by separate events/hydrological conditions are not included in the same time window at any point in time. This ensures identifiability of multiple error segments as individual residual periods. Nonetheless, we will demonstrate the usefulness of our proposed method even for superimposed errors (Section \ref{sec:Cases}), since this is a phenomenon often anticipated in real-world applications.

\subsection{Theoretical Behavior of tBME Curves}
\label{sec:tBME_interp}

\begin{figure}[ht]
\centering
\includegraphics[scale=0.9]{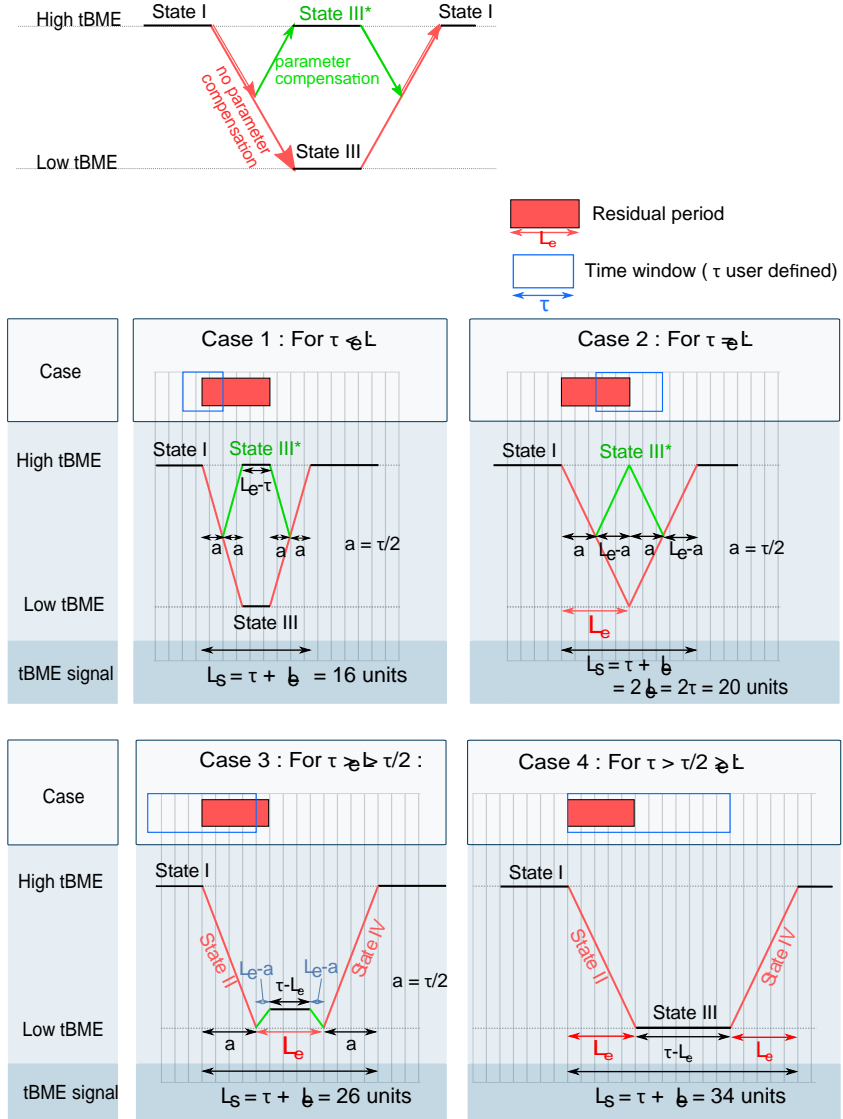}
\caption{Schematic illustration of expected tBME signal as a function of residual period length $L_e$ (here: fixed at 10 exemplary units) and window size $\tau$ (here: varied from 6, 10, 16, to 24 units).}
\label{fig:tBME_schematic}
\end{figure}

Fig. \ref{fig:tBME_schematic} schematically illustrates the theoretically expected behavior of tBME for different window size $\tau$ and residual period length $L_e$. The sketch at the top depicts how tBME curves behave generally with and without parameter compensation. In each of the four panels below, the top row identifies the respective case as a function of the length of the time window $\tau$ (blue box) relative to the residual period length $L_e$ (red-shaded box). The plots illustrate the corresponding theoretical behavior of tBME curves. The bottom row summarizes the relationship between the total tBME signal length, denoted as $L_s$, time-window length $\tau$, and residual-period length $L_e$. We provide exemplary numbers for $L_s$, with one unit corresponding to half a gridline increment. In each case, the blue window slides from the residual-free period (high tBME) through the residual period of length $L_e$ into the next residual-free period (high tBME again). We assume here that residuals are pronounced enough to produce a clear signal, and that conditions are exactly the same before and after the residual period, so that the tBME curve will climb back to its previous level.

In each panel of Fig. \ref{fig:tBME_schematic}, four major states of the theoretical tBME curve can be identified, as the blue time window would slide along the solid red residual period:
\begin{itemize}
    \item State I: The time window is still outside of a residual period and produces a high tBME value.    
    \item State II: The time window starts to contain the residual period and tBME drops to a low value. 
    \item State III: For the time during which the time window completely captures the residual period, the tBME curve will remain at a low plateau. Potentially, if parameters can successfully compensate model error, tBME temporarily returns back to its previous high value (state III$^*$). 
    \item State IV: The time window starts to move into a residual-free period again and the tBME curve rises to the pre-residual value (state I).
\end{itemize}

Error signals are defined in State II, when a tBME curve drop can be recognized. It occurs when the time window partially covers the residual period. At this state, the observed data points inside the window are obtained from two different system conditions. Even if there are realizations that fit segments of the data well, there still exists another part of the dataset that cannot be fitted with the same parameter values. As a consequence, goodness-of-fit is reduced for that window, and tBME drops.

In the following, we discuss the differences between the four cases distinguished in Fig. \ref{fig:tBME_schematic}. If we assume that parameters cannot compensate for model error at all (there is no parameter realization in the model ensemble that can fit the data well during the residual period), we see a single drop to the low tBME value, no matter whether the window is longer or shorter than the residual period (following the red lines to the low tBME level in Fig. \ref{fig:tBME_schematic}). This is clearly visible in Case 1, 2, and 4. The relationship between two specific time scales defines how long the low tBME value will persist: it can be a valley of length $L_e - \tau$ for $\tau < L_e$ (Case 1) , it can be a single inflection point in the case of $\tau = L_e$ (Case 2), or it can be a valley of length $\tau - L_e$ in the case of $\tau > L_e$ (in Case 3 and 4).

To some degree, modified parameter values might be able push the model's predictions back into the right direction (for the wrong reasons): parameters have a tendency to ``absorb'' model error. The degree to which this effect is happening depends on the flexibility of the model structure and the type of error. If the model parameters are able to compensate for model error, tBME may temporarily return to its previous high value (state III$^*$). In this state, the observed data cannot be fitted well by the model with the original parameter set, but the model still can describe the data well with other parameter values. This hypothesis suggests that the tBME curve will recover \emph{within} an residual period, if the data before and during the residual period have been generated under the same ``model setup''. Such type of model error could be treated well with, e.g., time-dependent parameters. Generally, this state is somewhat dangerous, because it might be interpreted as the end of the residual period, although in fact conditions have not changed. It is therefore important to keep an eye on the further development: Our basic assumption is that, during most of the time, there is no significant model error, so typically a stable residual-free period should follow after a tBME climb. If, instead, we observe a second dive right after, we should be warned and suspect state III$^*$ instead of state IV. 

The timing of the temporary tBME rise can be determined as follows: when the residual period covers half of the time window, this is the worst state in terms of model performance - because parameters have to fit two conflicting states (error-free and error-prone model) simultaneously, which is impossible to achieve. As soon as the window slides further into the residual period, error conditions take up a larger fraction in tBME calculation, and now parameter compensation can start to come into effect, so the tBME value has a chance to increase. It may rise until the window covers the full residual period, and it will stay at this high level until the window starts leaving the residual period. This results in the fact that, just like the lower valley, also the upper intermediate peak (state III$^*$) might turn into a plateau of different length, depending on the relationship between window size $\tau$ and residual period length $L_e$ (see Fig. \ref{fig:tBME_schematic}).   

Finally, once the window has moved out of the residual period completely, the tBME curve fully recovers, i.e. it returns to its pre-residual-period high value and remains there in a stable fashion (state I).

We find that the total signal length $L_s$ equals $L_e+\tau$, no matter if the window size $\tau$ is longer or shorter than the residual period length $L_e$. Hence, by observing the total signal length $L_s$ from the analysis with a specific $\tau$, we can directly infer the residual period length to be $L_s - \tau$. This has a large practical value and will help our interpretation of tBME curves over varied window sizes. We will illustrate and discuss this further on our case studies in Section \ref{sec:Application}.

What we did not take into account when drawing this schematic Figure is the fact that tBME calculation will yield larger differences between high and low values for larger datasets, i.e. for larger $\tau$. Therefore, the signals become more evident from larger $\tau$. Finding an optimal window size $\tau$ is therefore a compromise between a long enough window to obtain a clear signal, a short enough window to only capture individual error events and avoid ``smearing'' of different types of model error, and a short enough window to enable fast convergence of the tBME sampling distribution. 

In practice, we find that measurement noise and other overlaying effects may distort a clear error indication for window sizes smaller than the residual period length. Our investigations suggest that increasing the window size gradually enlarges true error signals and is hence advisable for reliable error detection (Section \ref{sec:Results_test_cases}). This is apparent from Fig. \ref{fig:tBME_schematic} especially for $\tau/2 \geq L_e$ (case 4): the signal is enlarged to a wider valley, because effective error compensation by parameters is not possible anymore, since the residual-free period always takes up a larger fraction within the moving window than the residual period. Shifting parameter values would lead to a lower performance in reproducing the error-free data, and hence the overall skill metric (here: tBME) would decrease. Naturally, there is an upper limit to increasing the time-window size: once we arrive at $\tau=L_d$, there will be only a single window left, tBME will be equal to BME, and we will not have any error signal. When moving from $\tau \approx L_e$ toward this upper limit, the signal will start to become weaker at some point because the residuals within the residual period are diluted too much in the residual-free data. Additionally, the computational effort becomes increasingly challenging. Hence, we recommend to test window sizes starting from the anticipated temporal scale of noise and going up to roughly $\tau \approx 2L_e$.

\subsection{Sampling Distribution of tBME as Reference}
\label{sec:BME_reference}

To distinguish a ``low tBME value'' from a ``high tBME value'' as simplistically labelled in Fig. \ref{fig:tBME_schematic}, we wish to construct a reference for the obtained tBME values. If a tBME value falls into the model's ``personal BME range'', the model's behavior in this time window $j$ is interpreted to be quasi-true; if the tBME value falls outside of that range, statistically-significant model error is detected.

To obtain this range, we randomly sample tBME per time window by randomly picking a model realization as synthetic data $\textbf{D}_{T_j(\tau)}$. We then evaluate the corresponding tBME value by brute-force integration with the complete ensemble minus the picked realization, i.e. with an ensemble of size $N_{MC}-1$. Repeating this leave-one-out procedure with the full ensemble (and hence representing the full predictive uncertainty in the synthetic data) will yield a tBME distribution that reflects the predictive variability in the model. This variability is typically due to measurement noise and parameter uncertainty (which is the source of uncertainty that we consider in our case study in Section \ref{sec:Application}), but could also reflect other sources such as uncertain drivers and boundary conditions.    

With the tBME sampling distribution as reference, we perform a series of hypothesis tests: for any specified position of the moving window, does the tBME value given real data fall within the high-density region of the sampling distribution, or is it at the low-BME low-probability tail, or does it even fall outside of the sampled range? It is in the modelers hands to decide for a significance threshold to reject the hypothesis that the model has generated the data (e.g., tBME lower than the $xy^{th}$ quantile of the sampling distribution, or tBME smaller than the smallest sampled BME, etc.). We do not give a general recommendation here, because this significance threshold should be chosen context- and study-specific. We will discuss this further during the interpretation of our test case results, see Section \ref{sec:Results}.

By observing the dynamic trend of the tBME curve compared with the tBME reference distribution, we can monitor how model performance varies and thus diagnose temporal model error occurrence as shown in Fig. \ref{fig:HypoTestTau}. Note that the tBME curve and the reference distribution should be evaluated with the time window of same length to ensure a consistent comparison. 

\subsection{Dynamic Posterior Parameter Distribution}
\label{sec:Dynamic_Posterior}

When a model is calibrated with a sliding time-window, a series of posterior parameter PDF samples is obtained ``on-the-fly''. If varying conditions in different time windows favor significantly different parameter values, we expect to see significant shifts in the posterior parameter samples, e.g. visualized as histograms. Such shifts provide information about potential parameter compensation effects. 

We recommend inspecting the series of posterior parameter distributions as a diagnostic tool to learn about the type of model error that was detected with the tBME analysis. If parameters are able to compensate for model error, the model structure and the prescribed parameter bounds seem to be adequate (enough) in principle, and the reason for this apparent time-dependence of parameter values should be investigated in detail. If, instead, parameters are not able to effectively compensate for model error, the model structure seems to be too rigid and inaccurate, which points the modeller to missing processes, wrong assumptions about mass balance components, or similar misspecifications. In this case, the posteriors either run up against bounds of their priors (as parameters \emph{could} in principle compensate at the cost of taking implausible values), or they do not react at all (as parameters simply cannot address the model error).

\section{Demonstration on a Hydrological Case Study}
\label{sec:Application}

We demonstrate our proposed approach to model error detection on a hydrological case study from the field of soil hydrological modelling. We first introduce the model and field data in Sections \ref{sec:Model} and \ref{sec:FieldData}, then explain the numerical implementation in Section \ref{sec:Implementation}, define synthetic test cases for our analysis in Section \ref{sec:Cases}, and finally present and discuss the results in Section \ref{sec:Results}.

\subsection{Soil Hydraulic Model}
\label{sec:Model}

We use HYDRUS-1D \cite{2005_Hydrus1D} to numerically solve the Richards equation, which describes water movement in a one-dimensional vadose zone. It writes as:
\begin{equation}
\frac{\partial \theta }{\partial t} = \frac{ \partial}{\partial z}  \left[ K \Big( \frac{\partial h}{\partial z}-1  \Big)  \right]  -R,
\end{equation}

where $\theta$ is the volumetric water content [-], $h$ is the soil water pressure head $[cm]$, $t$ is time $[day]$, $z$ is the depth $[m]$, $K$ is the unsaturated hydraulic conductivity function $[cm$ $    day^{-1}]$, and $R$ $[cm^3 cm^{-3}day^{-1}]$ is a sink term when root water uptake is considered. This equation has three unknowns, namely $h$, $K$, and $\theta$, and thus requires two constitutive relations for closure: the water retention function $\theta(h)$ and the unsaturated conductivity function $K(\theta)$. A commonly used description of these hydraulic properties is the Mualem-van Genuchten (MVG) model \cite{1980_vanGenuchten, 1976_Mualem}: 
\begin{equation}
    \theta(h) = \begin{cases}
        \theta_r + \frac{\theta_s - \theta_r}{\big( 1 + |\alpha h|^n  \big)^m}  & \quad h < h_s  \\
        \theta_s  &  \quad  h > h_s \\
    \end{cases}  
\end{equation}

\begin{equation}
    K =K_{sat}\Theta^{l} \big[ 1-\big( 1-\Theta^{\frac{1}{m}}    \big)^m \big]^2,   
\end{equation}
with the effective saturation 
\begin{equation}
 \Theta = \frac{\theta - \theta_r}{\theta_s - \theta_r},
\end{equation}

where $\theta_r$ and $\theta_s$ refer to the residual and saturated water content $[-]$ respectively; $\alpha$ $[cm^{-1}]$ and $n$ $[-]$ are shape parameters, $K_{sat}$ refers to the saturated hydraulic conductivity $[cm\cdot day^{-1}]$, $m=1-\frac{1}{n}$ is an empirical coefficient, and $l$ $[-]$ is the pore connectivity parameter by \citeA{1976_Mualem}. We use a modified form of the MVG model with an air-entry value of $h_s = $-2 cm \cite{1988_Vogel}.

In one of the considered test cases (definition see Section \ref{sec:Cases}), we use alternatively the dual porosity model by \citeA{1994_Durner} that describes the retention properties as a linear combinations of the van Genuchten  curves for different pore spaces:

\begin{equation}
 \Theta =  \sum^k_{i=1} w_i \Big[ \frac{1}{1+( \alpha_i|h|)^{n_i}}  \Big]^{m_i} ,
\end{equation}

where $w_i$ are weighting factors for the subclass of the soil with $w_i \in (0,1)$ and $ \sum w_i=1$. One of the advantages of the dual porosity model is the similar interpretation of the parameters as in the van Genuchten model \cite{1994_Durner}.

\subsection{Study Area and Field Data} 
\label{sec:FieldData}

The field data used in this research is from the Spydia experiment at site in the northern Lake Taupo catchment, New Zealand. Tensiometric pressure head at five depths was measured every 15 minutes with tensiometer probes (UMS T4e, Germany, accuracy 0.5 kPa) at five different depths between 0.4 and 5.1 m \cite{2008_Eddy}, but in this research we only use the field data of the three depths: 0.4, 1.0, 2.6 m. The daily measurements of the top layer (0.4 m, the closest layer to the surface among the three layers) show the most interesting dynamics and are therefore used in this research. Meteorological data from the 500-meters-distant Waihora station is applied for potential evaporation with the Penman-Monteith equation \cite{1981_Monteith}. 
Detailed information about the field experiment is given in \citeA{2008_Eddy, 2009_Eddy,2011_Eddy} and is therefore not repeated here.  
\citeA{2008_Eddy_2} investigated 7 different soil hydraulic models and found that none  of the models fitted to the field data for all time periods, i.e., that all models showed structural error for at least some time period. The study focused on improved predictive performance by Bayesian model average or Bayesian model combination (BMA/C) techniques rather than an the identification of structural errors. In the spirit of model diagnostics and process understanding, it is therefore interesting to investigate the model deficiencies in more detail.   

\subsection{Numerical Implementation}
\label{sec:Implementation}

In our simulation, we use the model ensemble by \citeA{2008_Eddy}, which consists of $N_{MC} = 900,000 $ realizations to represent uncertainty in the five MVG parameters $\theta_s$, $K_{sat}$, $\alpha$, $n$, and $l$ per soil layer, adding up to 15 uncertain parameters for three soil layers. $\theta_r$ is fixed to zero, because it is not sensitive in the considered calibration setup \cite{2008_Eddy_2}. The prior ranges for the uncertain parameters are given in Table \ref{tab:para_prior}, with all three soil layers sharing the same parameter bounds. As stated in Section \ref{sec:FieldData}, we focus on the simulations of pressure head in the top soil layer for model evaluation and model error detection.

\begin{table}[ht]
\caption{Bounds of the Mualem-van Genuchten (MVG) model parameters used for uniform prior sampling.}
\centering
\begin{tabular}{l ccccc}
\hline
 &   $\theta_s$ $[m^3m^{-3}]$ & $\alpha$ $[m^{-1}]$  & $n$ $[-]$ & $K_{sat}$ $[m s^{-1}]$ & $l$ $[-]$ \\
\hline
 Lower bound  &   0.3 & 1  & 1.1  & $10^{-7}$ & 0.1 \\ 
 Upper bound  &   0.7 & 20 & 9.0  & $10^{-3}$ & 1 \\
\hline
\end{tabular} \label{tab:para_prior}
\end{table}

For a given dataset $\bm{D}_{T_j(\tau)}$ of a specific time-window size $\tau$, we determine tBME (Eq. \ref{eqn:BME2}) by brute-force Monte Carlo integration (Eq. \ref{eqn:BME}). For this step, we need to define the likelihood function $p(\bm{D}_{T_j(\tau)}|\bm{\theta}_i,M)$ (Eq. \ref{eqn:likelihood}). We here assume a measurement error standard deviation of $0.09$ cm for the soil water pressure head data. This evaluation is exactly like a brute-force Monte Carlo implementation for traditional Bayesian model evaluation, only that the calibration dataset is deliberately chosen to be very short. 

To assess the convergence of our numerical BME scheme, we determine the effective sample size (ESS) \cite{2004_Liu}.  The ESS counts how many parameter realizations significantly contribute to the BME estimate. We also perform a visual analysis of the BME convergence behavior. Both analyses together provide a good indicator of how stable the obtained numerical result is. Although we have monitored BME convergence in all investigated cases, we wish to emphasize that, for real-world applications, a (too) low ESS is not annoying for two reasons: (1) Low ESS arise if only very few realizations show a non-zero likelihood, i.e., if most of the ensemble is performing really bad. This is the situation that we are in fact most interested in: it reveals severe model error. In this case, we will obtain a low BME value; and we do not care whether it is low, very low, or extremely low - the numerical BME estimate does not have to be that accurate. (2) We do not use the obtained BME value for any further procedure (such as model-averaged prediction), but we are satisfied with BME serving as more of a qualitative model error detector. We are therefore not relying on an accurate estimate of BME for further processing. 

To now obtain a complete tBME curve for a chosen time-window size $\tau$, we start by evaluating the first tBME value ($tBME_j$ with $j=\tau$) for the period of simulation days 1 to $\tau$. Then we slide this window by one day, determine $tBME_j$ for $j=\tau+1$, and so on, until we hit the end of the full calibration dataset (cf. Section \ref{sec:tBME}). In this way, we construct a tBME curve for a given dataset, e.g. for the observed time-series of pressure heads. 

We repeat this routine for a number of $30,000$ randomly selected output realizations used as synthetic data, in order to construct the tBME sampling distribution as reference (cf. Section \ref{sec:BME_reference}). Convergence of the sampling distribution was checked visually and by monitoring ESS. Note that ESS does become relevant in this synthetic part of the procedure, because we want the reference distribution to be informative and reliable. We therefore ensure that, during these synthetic runs, ESS does not drop below the value of $200$. This is usually not a limiting factor, because the synthetic analysis by definition assures that the model is true, so BME performance does not drop that dramatically, and neither does ESS. 

Finally, we repeat this whole procedure (sliding time-window applied to real and synthetic data) for distinct increasing values of window size $\tau$, i.e. $\tau=5,10,15,20$ days. 

For each time window that we use for calibration (and to determine tBME), we obtain a posterior parameter sample as a ``side product''. We simply weight our prior parameter ensemble with the corresponding vector of likelihoods (dimension 1 x $N_{MC}$) and approximate the posterior PDF by kernel density estimation on the weighted sample. As an analogue to the tBME curve, we obtain a series of posterior parameter distributions that we can inspect for further investigation of the potential source of model error.

All our analyses are implemented and performed in MATLAB$^\copyright$ R2019a on a standard desktop computer.
 

\subsection{Synthetic Test Cases}
\label{sec:Cases}
Before applying our proposed method to the real dataset, we wish to validate the method under fully-controlled conditions. We use exactly the same model setup as in the real case, but construct the dataset of ``observations'' to be used for the tBME analysis manually. To test different types of model error that could occur in reality, we define three synthetic scenarios where we knowingly introduce model error. The goal is to confirm that our method will successfully identify these predefined error periods; and thereby we wish to demonstrate how to interpret the model error indicator. Then, we finally use the real data to test our approach under realistic conditions.

To generate the synthetic datasets for our analysis, we extract a single realization of model output (soil water pressure head) from the ensemble (base case with no model error, see definition below). In the next step, we perturb this realization with different types of error to generate synthetic datasets that deviate from the model behavior and, hence, model error shall be detected. The perturbations aim to mimic the scenario when a model fails to resolve an only temporary occurring process in the natural system.

While, philosophically, model error is of course in the model predictions and not in the observed dataset, we here pragmatically introduce ``errors'' into the synthetic datasets. This approach is more straight-forward to implement, easier to interpret, and allows us to modify and arbitrarily increase the complexity of the introduced error without having to set up a new model ensemble for each test case. Whether we introduce the error in the dataset or in the ensemble of predictions does not have an impact on the functioning of our proposed approach.

\subsubsection{Base Case: No model error}
\label{sec:Base_Case}

To demonstrate how our method reacts to the ideal case of no model error, we pick one realization of the model ensemble as synthetic data for the tBME analysis. This realization lies within the high-density range of the predictive prior to represent ``typical model behavior'' (Fig. \ref{fig:Plot3_obs35}b). The model parameters of this case are listed in Table \ref{tab:para_ref}. The layers 1, 2 and 3 refer to the depth of  0.4, 1.0, 2.6 m, respectively. This dataset is considered as a reference case for comparison for the following synthetic cases 1 and synthetic case 2. For the synthetic case 3, we pick a realization which lies within the low-density range of the predictive prior to represent a ``low-probability synthetic truth''. Also, this parameter set (Table \ref{tab:para_2}) produces pressure head data with similar dynamics to the field data.

\begin{table}[ht]
\caption{MVG parameter values chosen to represent error-free data (base case)}
\centering
\begin{tabular}{l cccccc}
\hline
 & $\theta_s$ & $\alpha$ & $n$ & $K_{sat}$  & $l$ \\
\hline
  Layer 1 & 0.5455 & 12.9879 & 4.1150  & 0.0009 & 0.1945 \\ 
  Layer 2 & 0.4783 & 13.3246 & 3.6499  & 0.0006 & 0.5233 \\
  Layer 3 & 0.5278 & 5.2775  & 2.8336   & 0.0001 & 0.8315 \\
\hline
\end{tabular} \label{tab:para_ref}
\end{table}

\begin{table}[h]
\caption{MVG parameter values applied to perturb the base-case dataset during error periods in the case of superimposed errors (case 3).}
\centering
\begin{tabular}{l cccccc}
\hline
 & $\theta_s$ & $\alpha$ & $n$ & $K_{sat}$  & $l$   \\
\hline 
Layer 1 & 0.4146 & 6.1609 & 2.5970 & 0.0001 & 0.4049  \\
Layer 2 & 0.3860 & 2.2569 & 5.5489 & 0.0004 & 0.3108  \\
Layer 3 & 0.5719 & 2.9203 & 4.5101 & 0.0008 & 0.9006    \\
\hline 
\end{tabular} \label{tab:para_2}
\end{table}

\subsubsection{Synthetic Case 1: Error in Model Structure}

The synthetic dataset of the base case is now perturbed by structural error during specific time periods. We assume that ``reality'' behaves like a dual-porosity model during these periods (with strong precipitation events kick-starting this behavior, and approximately like a single-porosity model otherwise. Technically, we run HYDRUS-1D with the error-free setup described for the base case, and switch to the dual porosity model setup during three selected error periods: days 31 to 40, 81 to 90, and 161 to 170 (cf. Fig. \ref{fig:Plot3_obs44}a).  

The following points are considered for the selection of these error periods: 1. The error periods do not start before $t=20$, which is the largest window size used in our method demonstration. 2. The error period should not be too long so that the time window does not stay inside too long. 3. The error periods cover simulation periods of different spread in the predictive prior. 4. The largest resulting residual per day (i.e., each bar in Fig \ref{fig:Plot3_obs36}b) is not larger than 0.2 m.

For constructing the perturbed dataset, the parameters $K_{sat}$, $\theta_s$ and $l$ remain the same as in the base case; parameter values for the first pore subsystem are chosen as in the base case, and parameter values for the second pore subsystem are given in Table \ref{tab:para_syn1}. Here the larger values of $\alpha$ and $n$ than in the base case imply that a macropore system is considered as the second soil subsystem. Note that we assign the same dual-porosity parameter values to layers 1 and 2, while layer 3 is assumed to not show a dual-porosity behavior at all.

The ensemble that we test remains the same throughout all test cases: it consists of single-porosity runs (MVG model) with prior parameter ranges given in Section \ref{sec:Implementation}. Hence, we see significant pressure head residuals during error periods and beyond (Fig. \ref{fig:Plot3_obs44}b), because the tensiometric pressure responses in a dual-porosity soil structure to heavy precipitation in ``reality'' is quite different from what our tested single-porosity model predicts.

\begin{table}
\caption{Dual-porosity MVG parameter values applied to perturb the base-case dataset during error periods in the case of model structural error (case 1); parameters for the first pore subsystem are identical to those given in Table \ref{tab:para_ref} (base case), with $w_1=1-w_2$.}
\centering
\begin{tabular}{l ccccccccc}
\hline
 & $w_2$ & $\alpha_2$ & $n_2$ \\
\hline 
Layers 1 and 2 
&  0.85 & 31 & 15 \\
 \hline
\end{tabular} \label{tab:para_syn1}
\end{table}

\subsubsection{Synthetic Case 2: Error in Model Forcing}

As a second scenario, we consider error in the model forcing. We mimic local effects that are quite common in reality: The nearby meteorological station has recorded precipitation, but the event is such localized (e.g. convective shower), that it actually did not rain at the experimental site. As a consequence, soil moisture measurements show no response (or rather, pressure head becomes even more negative due to the drying soil, see Fig. \ref{fig:Plot3_obs36}), while our model predicts increased soil moisture due to a heavy precipitation event (pressure head becomes less negative, see Fig. \ref{fig:Plot3_obs36}). Technically, the synthetic dataset is generated by running HYDRUS-1D in the setup of the error-free base case (Table \ref{tab:para_ref}), but the input of precipitation is removed within the error periods (days 36 to 40, 89 to 90, and 166 to 170). This causes significant residuals during error periods and beyond due to the ``storage memory'' of the soil hydraulic model, see Fig. \ref{fig:Plot3_obs36}b. Note that, here, the time of the error period (forcing error of, e.g., just one day) is much shorter than the residual period (time span during which the impact on simulations is obvious from the residuals, approximately 10 days).

\subsubsection{Synthetic Case 3: Superimposed Errors in Model Structure and Forcing}

With the third test case, we aim to study the impact of realistic, superimposed errors. We run HYDRUS-1D in single-porosity mode with the parameter set in Table \ref{tab:para_2}. During the error periods defined in Case 1, we apply the dual-porosity model by switching on $w2$, $\alpha_2$, and $n_2$ as shown in Table \ref{tab:para_2Error}. Forcing errors are then superimposed by removing precipitation input on simulation days 44 to 45, 81, 82, 84, 161, 163, 164, and 165. These superimposed error periods also cause superimposed residual periods.

This leads to a combined impact on residuals between the synthetic dataset and the model ensemble predictions on different time scales, see Fig. \ref{fig:Plot3_obs56}b. Both residuals tend to partially cancel out, such that net residuals are temporarily reduced: Switching to dual-porosity mode means considering macro-pore flow, which compensates the missing precipitation since water moves quicker to greater depths just like it does when it rains.

\begin{table}[h]
\caption{MVG parameter values applied to perturb the base-case dataset during error periods in the case of superimposed errors (case 3), where $w2$, $\alpha_2$, and $n_2$ are switched on only during error periods defined in Case 1.}
\centering
\begin{tabular}{l ccccccccc}
\hline
 & $\theta_s$ & $\alpha$ & $n$ & $K_{sat}$  & $l$ & $w_2$ & $\alpha_2$ & $n_2$ \\
\hline 
Layer 1 & 0.4146 & 6.1609 & 2.5970 & 0.0001 & 0.4049 &  0.2 & 31 & 15\\
Layer 2 & 0.3860 & 2.2569 & 5.5489 & 0.0004 & 0.3108 &  0.2 & 31 & 15 \\
Layer 3 & 0.5719 & 2.9203 & 4.5101 & 0.0008 & 0.9006  & 0  & 31  & 15 \\
\hline 
\end{tabular} \label{tab:para_2Error}
\end{table}

\section{Results and Discussion}
\label{sec:Results}

We first present the results of the tBME analysis applied to the error-free base case, then use our findings for the interpretation of the results of synthetic test cases 1 to 3 (Section \ref{sec:Results_test_cases}), and finally show the results of applying our proposed approach to the observed field data in Section \ref{sec:Results_field_data}. 

\subsection{Results of tBME Analysis for Synthetic Test Cases}
\label{sec:Results_test_cases}

\subsubsection{Base Case: tBME Given a Correct Model}
\label{sec:Results_0}

\begin{figure}[ht]
\centering
\includegraphics[scale=0.9]{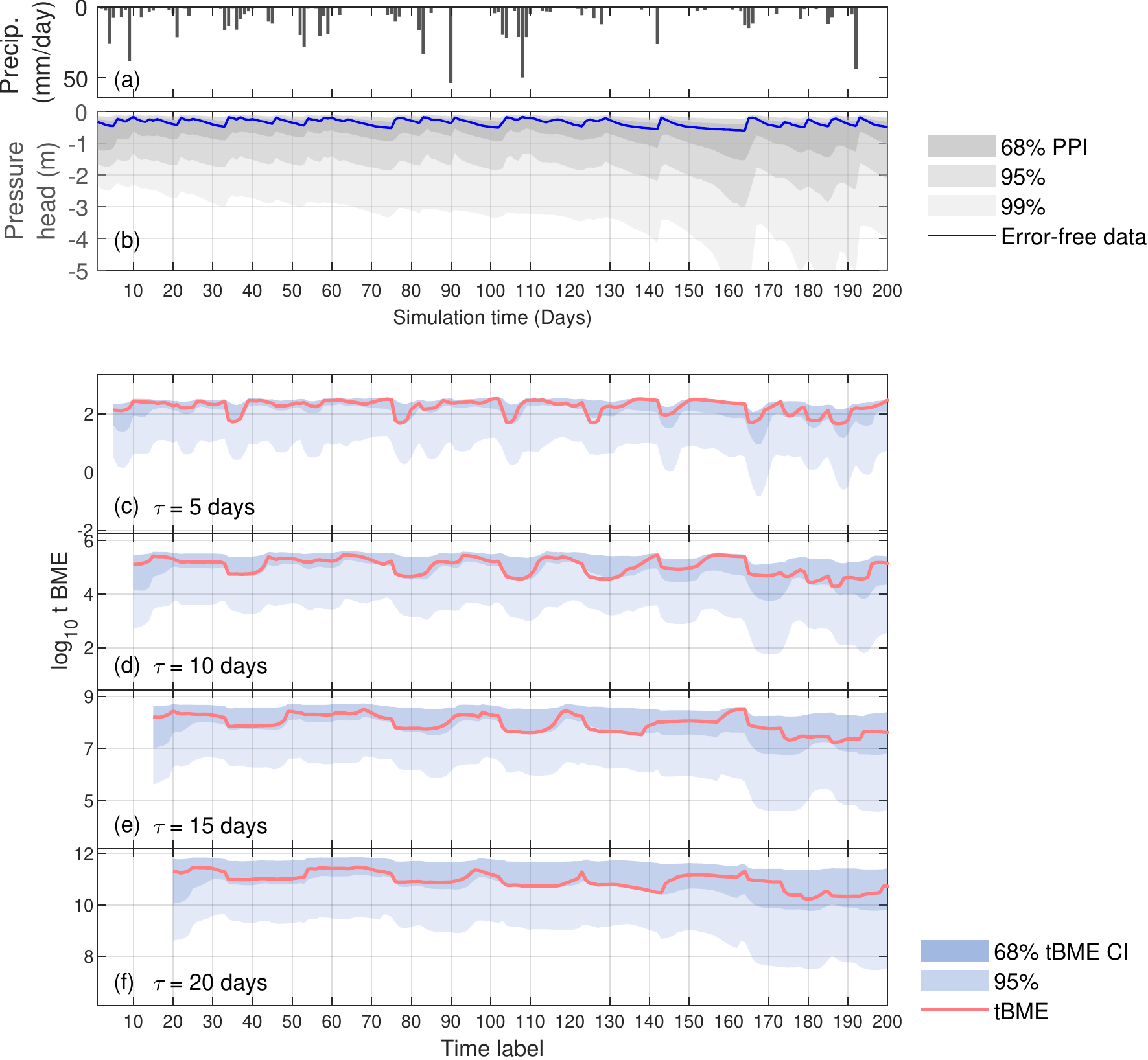}
\caption{Results for tBME analysis if the model is correct (error-free). 68 \% PPI denotes prior prediction interval;  68/95 \% tBME CI denotes 68/95 \% confidence interval }

\label{fig:Plot3_obs35}
\end{figure}

The base case illustrates what the tBME curve should look like if the model is correct. Fig. \ref{fig:Plot3_obs35}a records the precipitation (model input), and Fig. \ref{fig:Plot3_obs35}b presents the model's prior predictive distribution (with the gray shaded areas indicating the 68\%, 95\%, and 99\% prediction intervals) and the dataset (blue) used for the evaluation of tBME values. Fig. \ref{fig:Plot3_obs35}c to Fig. \ref{fig:Plot3_obs35}f show the tBME curves (red curves) resulting for four different time-window sizes $\tau$. Their corresponding tBME sampling distributions are shown as blue shaded areas indicating the 68 \% and 95 \% confidence intervals. Note that we label these as ``confidence intervals'', because the resampling procedure invokes a frequentist perspective, although our analysis is otherwise embedded in a Bayesian framework. The x-axis of the tBME curve and the tBME sampling distribution is shifted by $\tau$ days. For example, the first tBME value of Fig. \ref{fig:Plot3_obs35}c is plotted at $t = \tau = 5$, which is evaluated with the observed data from the first day to the fifth day (cf. Fig. \ref{fig:movingTau}). 

\subsubsection*{Model-Internal Variability of tBME}

As shown by the red curves in Figures \ref{fig:Plot3_obs35}c to \ref{fig:Plot3_obs35}f, all tBME curves fall inside the range of their tBME sampling distributions during all time periods, which confirms that no significant model error occurs. This meets our expectation, since this dataset is generated by the model. 

In Fig. \ref{fig:Plot3_obs35}c, the tBME curve of $\tau = 5$ contains various drops. This is due to the model's parameter uncertainty: even though the base case dataset is a model realization sampled from the high-density range of the predictive prior, other realizations in the ensemble do not necessarily follow the exact dynamic trend of this individual realization. This effect can be observed when the model is calibrated with a small window size. With increasing time-window size, those small deviations become less influential, because the bias from short-term effects is diluted over a larger window size. This can be observed in the tBME curve for $\tau = 20$, which is smoother than the one for $\tau = 5$. The important conclusion is: no matter how wiggly the tBME curve, if it stays within the tBME sampling range, there is no substantial (statistical) reason to believe in model error - it can just be due to parameter (or other sources of) uncertainty. 

\subsubsection*{Dynamics of tBME sampling distribution}

It is worth noting that the effects of parameter uncertainty is increasing over time in our case study (Fig. \ref{fig:Plot3_obs44}a), leading to wider tBME sampling distributions at later times (Fig. \ref{fig:Plot3_obs44}c to \ref{fig:Plot3_obs44}f ). This demonstrates that the criterion to trigger an error signal (falling below a certain quantile of the sampling distribution) is dynamic in time.

\subsubsection*{tBME Threshold for Error Detection}

The challenge is hence to identify a meaningful lower threshold for tBME to label a drop as ``model error occurrence''. As noted in Section \ref{sec:BME_reference}, there cannot be a general guidance for all cases, but rather we recommend to perform this base-case test before interpreting the results of real tBME curves as done here: by choosing a realization from the high-density region of the predictive prior with a small window size, modellers can check to which quantile of the sampling distribution the tBME curve drops. This level is the highest one that should be used for model error detection (to avoid false-positive error detection); rather, we would expect so see much more pronounced drops with real data if severe model error occurs. Looking at increasing window sizes $\tau$ will help gaining confidence in the choice of the threshold. This is discussed in more detail in the context of test case 1 in the following Section.

\subsubsection{Case 1: tBME Given Model Structural Error}

\begin{figure}[ht]
\centering
\includegraphics[scale=0.9]{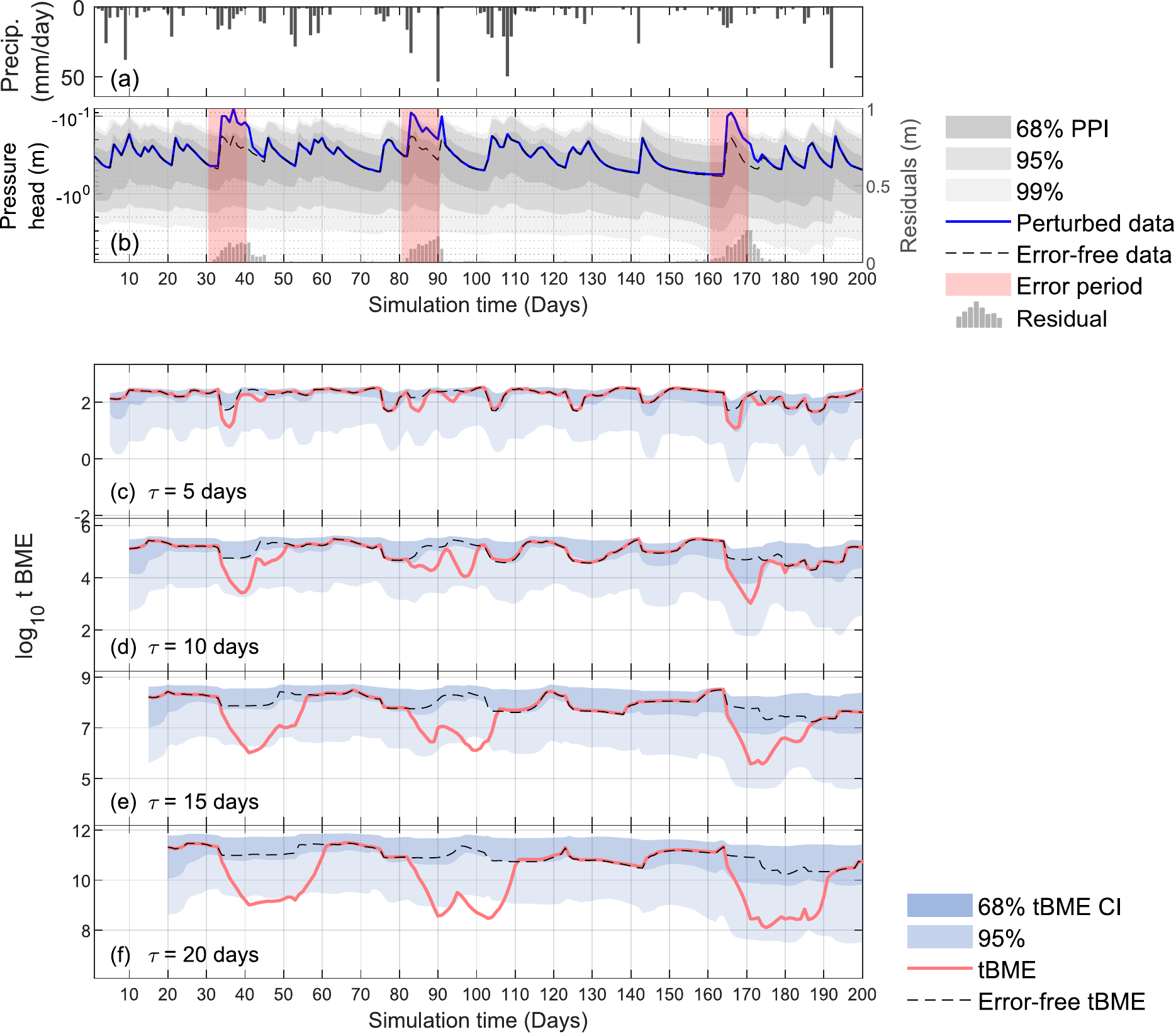}
\caption{tBME curves given model structural error (synthetic case 1: reality switches to dual-porosity behavior, model stays in single-porosity setup).  
}
\label{fig:Plot3_obs44}
\end{figure}

Our first synthetic test case demonstrates the time-windowed BME analysis to detect model structural errors. Fig. \ref{fig:Plot3_obs44} presents the results in the same order as for the base case (Fig. \ref{fig:Plot3_obs35}). 

The red-shaded boxes in Fig. \ref{fig:Plot3_obs44}b represent the \emph{error periods} during which the ``truth'' is simulated with the dual- instead of the single-porosity model (see Section \ref{sec:Cases}). These time periods now feature an ``observed'' process that is not contained in the model ensemble. 

The corresponding perturbed dataset is shown by the solid blue line. The base-case dataset is shown as dotted line for comparison. The difference between those two datasets (i.e., the residuals between the model realization corresponding to the base case dataset and the dataset now used for case 1) are plotted as gray bars. These residuals provide an impression of how wrong the model ensemble is compared to the given dataset (with variations among model realizations due to parameter uncertainty). Recall that we use the term \emph{residual period} to clearly distinguish this time span of error-induced residuals from the actual error period. 

It becomes apparent that the first two error periods only really have an effect when precipitation occurs at $t=32$ days and $t=82$ days, while during the third error period, residuals are accumulating and are intensified during moderate precipitation from $t=165$ days. Note that the residuals persist longer than the actual time span of error (residual period longer than error period) due to memory effects caused by water storage.

\subsubsection*{Error Detection} 
Figures \ref{fig:Plot3_obs44}c - \ref{fig:Plot3_obs44}f present the tBME curves for the four chosen window sizes. The tBME curves for $\tau =5$ and $\tau =10$ show various small drops, which are still within the bounds of the corresponding tBME sampling distributions. This means, the model calibrated on periods of five or ten days still fulfills the statistical condition of the model to be quasi-true. With increasing window size, the signal of error occurrence becomes more pronounced. At $t$ around 40 and 100 days, the tBME curves of $\tau =15$ and $\tau =20$ both leave the high-density interval and even the complete range of the tBME sampling distribution, which indicates significant error occurrence. Ideally, a clear turning point can be observed when the time window enters an residual period, as shown by the immediate decline at $t=33$ and $t=83$ days in the tBME curve for $\tau = 20$ (Fig. \ref{fig:Plot3_obs44}f). 

\begin{figure}
    \centering 
    \includegraphics[scale=0.8]{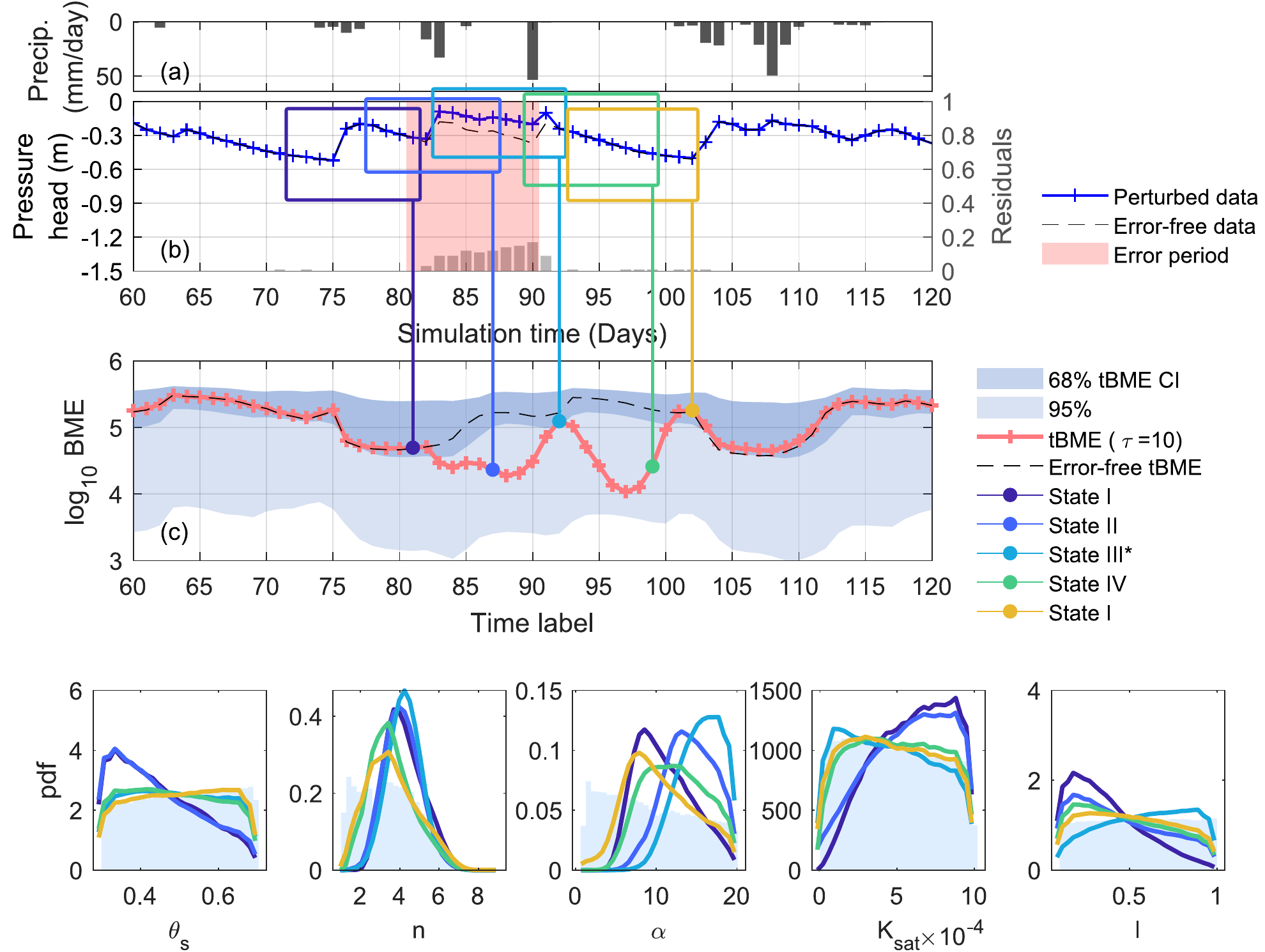}
    \caption{Zoomed-in view of tBME curve and dynamic posterior parameter distributions for window size $\tau$ = 10 (synthetic case 1).}
    \label{fig:zoom_Case1_10}
\end{figure}

By inspecting the tBME curve in more detail (Fig. \ref{fig:zoom_Case1_10}, for the example of $\tau = 10$), we identify the four major states of the tBME curve introduced in Section \ref{sec:tBME_interp}:
\begin{itemize}
    \item State I: The time window is outside of an residual period and the tBME curve stays within the high-density interval of the tBME sampling distribution.   
    \item State II: The time window starts to contain the residual period and the tBME curve declines.
    \item State III$^*$: For the time during which the time window completely captures the residual period, tBME temporarily returns back to the high-density interval of the sampling distribution. This shows that the parameters successfully compensate model error. 
    \item State IV: The time window starts to move out of the residual period (residual-free) and the tBME curve rises to the high-density interval of the sampling distribution.
\end{itemize}

\begin{figure}
    \centering
    \includegraphics[scale=0.8]{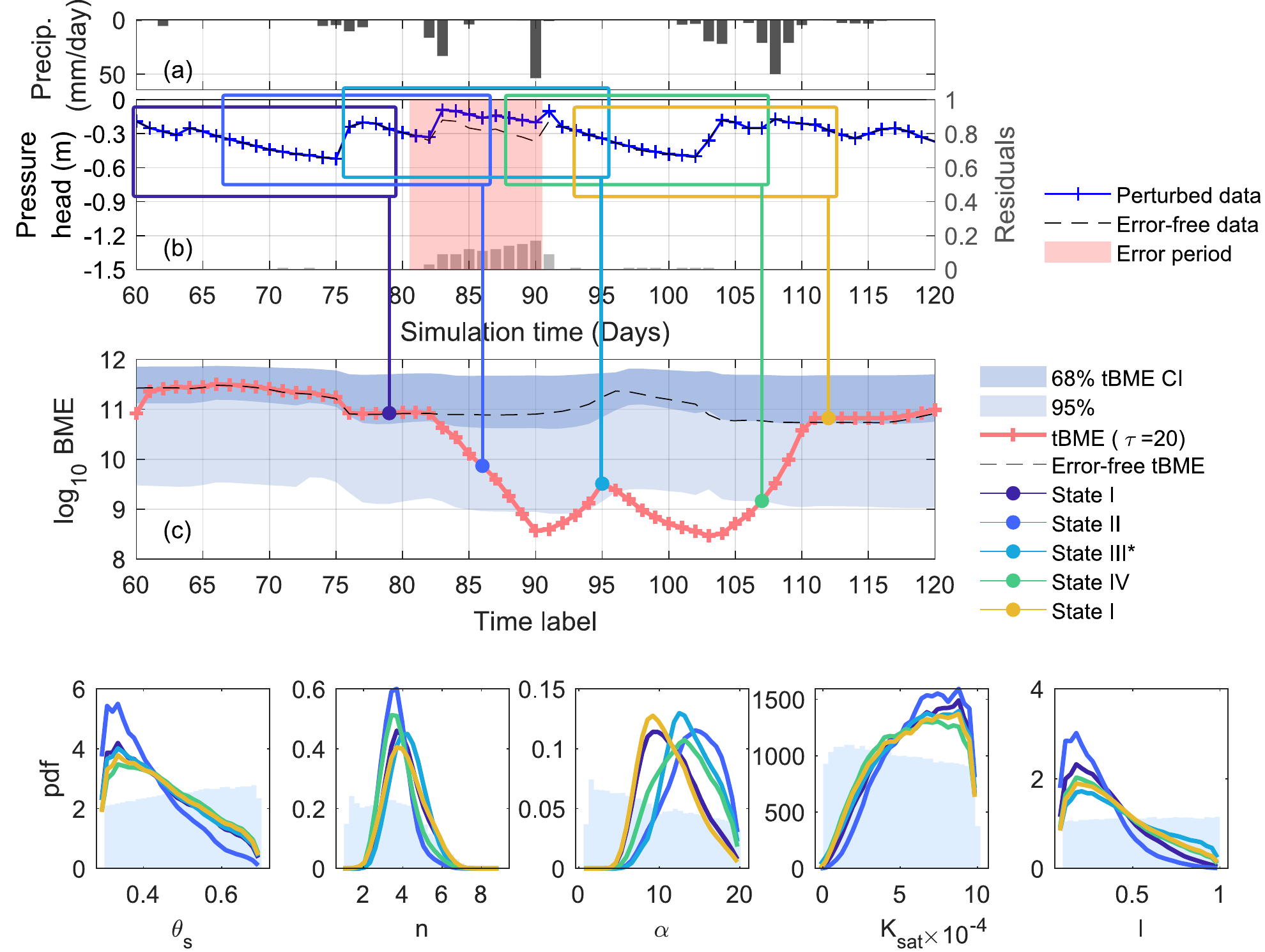} 
    \caption{Zoomed-in view of tBME curve and dynamic posterior parameter distributions for window size $\tau$ = 20 (synthetic case 1).}
    \label{fig:zoom_Case1_20}
\end{figure}

For easier recognition, we have marked and color-coded the respective time windows and corresponding tBME values in Figure\ref{fig:zoom_Case1_10}b and \ref{fig:zoom_Case1_10}c. The same four states can be identified for window sizes larger than the perceived residual period length (e.g., for $\tau=20$, Fig. \ref{fig:zoom_Case1_20}), with the only difference that state III$^*$ (compensation) is less pronounced. This is because the time window used for tBME calculation will never only contain the residual period, but will always partially contain an residual-free period. Hence, although compromise parameter sets are preferred, they will not achieve a high-density tBME value. With $\tau=20\approx 2L_e$, we would have expected a flat plateau of low tBME values according to our theoretical considerations in Section \ref{sec:tBME_interp}; Instead, we see a slight intermediate peak of the tBME curve in Figure \ref{fig:zoom_Case1_20}b (day 95). A possible explanation could be the wetting event at $t=90$ days that seems to alleviate the model's difficulty to fit the data; a similar effect can be observed for the next wetting event after $t=100$ days. Since the residuals are due to a lack of macro-pore flow behavior in the model (which would make water achieve greater depths faster), a wetting event helps to put the model into conditions that are more similar to the synthetic data. This can also be seen from the residuals plotted as gray bars: their accumulation over time is abruptly stopped with the precipitation event at $t=90$ days. If we look at the dip of the tBME curve, it starts decreasing at $t=82$ and returns at $t=111$. The total signal length is hence 29 days (111-82), which corresponds to the theoretically expected length of $\tau+L_e$. 
 
\subsubsection*{Sensitivity of tBME to Window Size}

In this test case, the misfit between the model ensemble and the data is around 0.1 m on average and spans a period of 10 to 15 days. The comparison of the four tBME curves reveals that a time-window size of 50 to 100 \% residual period length is here not yet conclusive, because such a short period is not sufficient to disentangle the error signal from measurement noise (which is chosen to be on a similar order of magnitude as the residuals here) and model-internal variability due to parameter uncertainty. Also, under real-world conditions, the tBME curve for a narrow window is expected to be sensitive to short-term deviations, as daily deviations take up a larger fraction of the total residuals. For instance, a short-term variation in the system, e.g., misfit during one day, can have one fifth contribution in a time window of five days. Therefore, major signals might not yet be distinguishable from noise in the tBME diagnosis with small window sizes. 

The BME indicator strength increases with the calibration time-window length, because the task of fitting all contained data points well with any single parameter combination becomes increasingly hard (especially, since we assume uncorrelated errors in our likelihood, cf. Eq. \ref{eqn:likelihood}). Hence, wider time windows smooth out the noise, if the error duration is longer than the typical scale of noise (or unresolved, short-term processes of less importance to the modeller), and enlarges true error signals. This is what we observe here: when the time window is larger than the actual residual period (e.g., $\tau = 20$ days), the problematic periods can be clearly detected. 

Overall, we take away from this first test case that a variation of $\tau$ can be very insightful and strengthens our interpretation of results.

\subsubsection*{Dynamics of Posterior Parameter Distributions}
Alongside the tBME curve, we obtain a series of posterior parameter samples for each time-window size $\tau$. We can inspect how each parameter's posterior PDF changes with increasing or decreasing tBME. Figures \ref{fig:zoom_Case1_10} and \ref{fig:zoom_Case1_20} show the posterior PDFs of all five uncertain model parameters corresponding to the five illustrated time windows for window sizes $\tau = 10$ and $\tau=20$ days, respectively. We observe the clearest trends in parameter value shifts for short window sizes $\tau$, because parameters can be fine-tuned to very short calibration datasets, and generally do not have to satisfy varying conditions simultaneously. 

For $\tau=10$ in Fig. \ref{fig:zoom_Case1_10}, we conclude that the shape parameters $\alpha$ and $l$ are most sensitive to the residual periods: During phases when dual-porosity behavior would be needed to adequately mimic the data, $\alpha$ and $l$ tend to take on higher values than during error-free (single-porosity) periods. We nicely see the error compensation attempt in those two parameters with their posterior PDFs shifting from state I to state III$^*$ (tBME returns to the high-density region due to parameter compensation) and back. 

\begin{figure}
\centering
\includegraphics[scale=0.8]{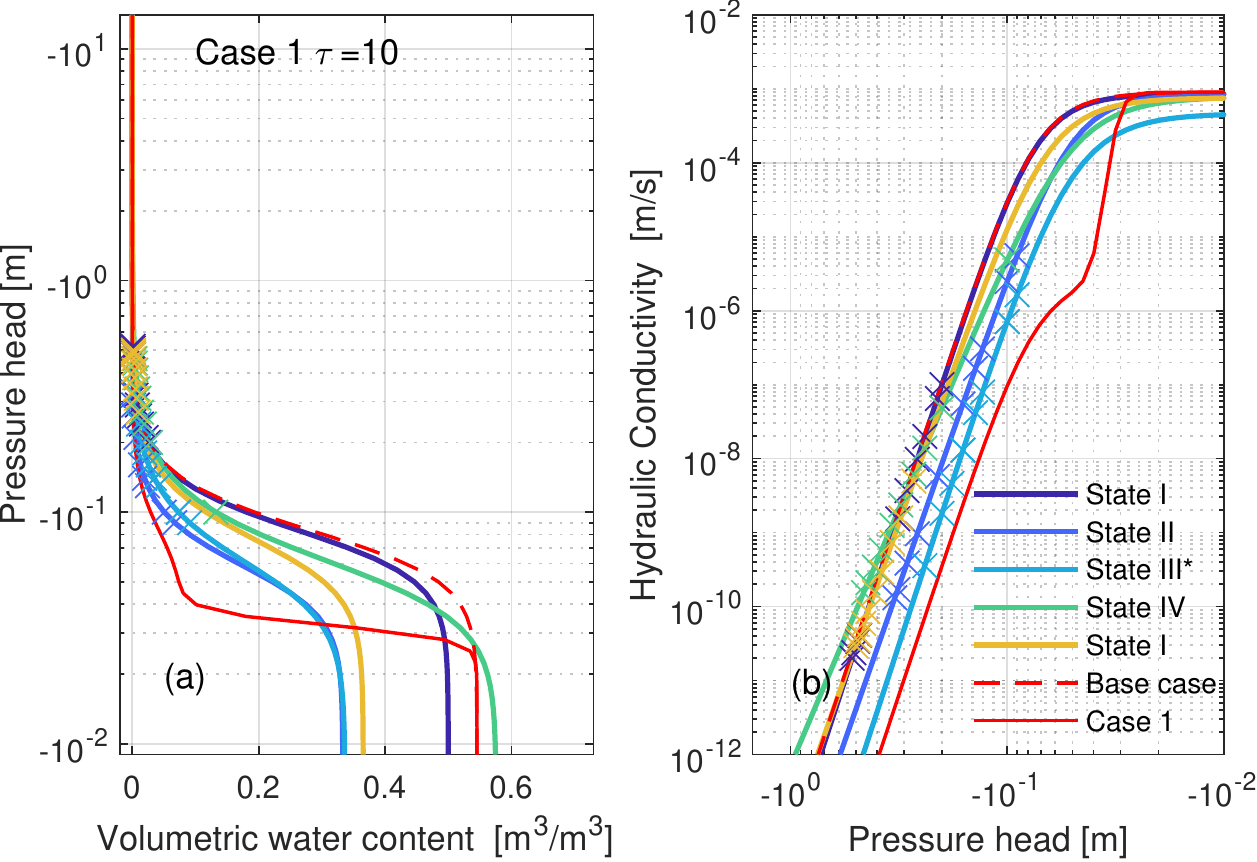}
\caption{Maximum-likelihood water retention curves (left) and unsaturated hydraulic conductivity functions (right) of the top soil layer for the five states color-coded in Fig. \ref{fig:zoom_Case1_10}b (synthetic case 1, window size $\tau=10$ days). Dashed red lines show true curves of the base case; solid red lines show the curves corresponding to the perturbed dataset. x-symbols indicate the } 
\label{fig:WRC_ML_C1_t10}
\end{figure}

These two parameters have the effect of scaling the water retention function. Higher values correspond to lower water contents for a specified pressure head or higher pressure for a specified water content, respectively. To confirm this conclusion, we look at maximum-likelihood water retention curves obtained in the five states discussed above (Fig. \ref{fig:WRC_ML_C1_t10}a). We also show the true water retention curves for the error-free base case and test case 1 for comparison. As expected, we see a shift of the water retention curve from state I being close to the base case toward case 1, induced by higher $\alpha$ and $l$-values that better represent the water retention function of the dual-porosity model parameterization. Hence, although the synthetically-introduced model error here is related to the model concept, the model manages to compensate the error by adjusting those parameters that determine the shape of the water retention curve.

Note that the information content of the calibration data for identification of the true water retention curve is limited, as can be seen from Fig. \ref{fig:WRC_ML_C1_t10}a: the pressure head observations in each state (data in the five different time windows) are marked as crosses on the corresponding simulated curve. These data mostly cover the dry range, where the water retention curve becomes a vertical line. None of the pressure head observations in that range is informative, because differences are very subtle and hence the true water retention curve cannot be identified. This is a particular result of the highly porous volcanic soils present in our case study region \cite{ 2008_Eddy_2}. In these soils (similar to sand), the water content drops dramatically over just a narrow range in matric potential. 


Hydraulic conductivity as a function of pressure head (shown in Fig. \ref{fig:WRC_ML_C1_t10}b) is dominated by the parameter compensation effect that causes the curves to shift from state I closest to the base case toward the curve of case 1 for states II and III$^*$. The unsaturated conductivity of case 1 is much lower than the base case, leading to a slower movement of infiltration fronts and larger pressure heads. 

As expected, parameter compensation effects become less dominant if the window size is increased beyond the residual period length (illustrated for $\tau=20$ in Figs. \ref{fig:zoom_Case1_20} and \ref{fig:WRC_ML_C1_t20} in \ref{sec:app}). Parameters now have to find an optimal compromise solution whenever (parts of) the error period are contained in the calibration time window, and this reduces the shifts in PDFs and water retention curves or unsaturated hydraulic conductivity functions between time window positions. Yet, differences between the individual parameters can still be found, which can help gain further understanding of how these parameters act in the model and how they possibly compensate for model error.

\subsubsection{Case 2: tBME Curve Given Forcing Error}

\begin{figure}[ht]
\centering
\includegraphics[scale=0.9]{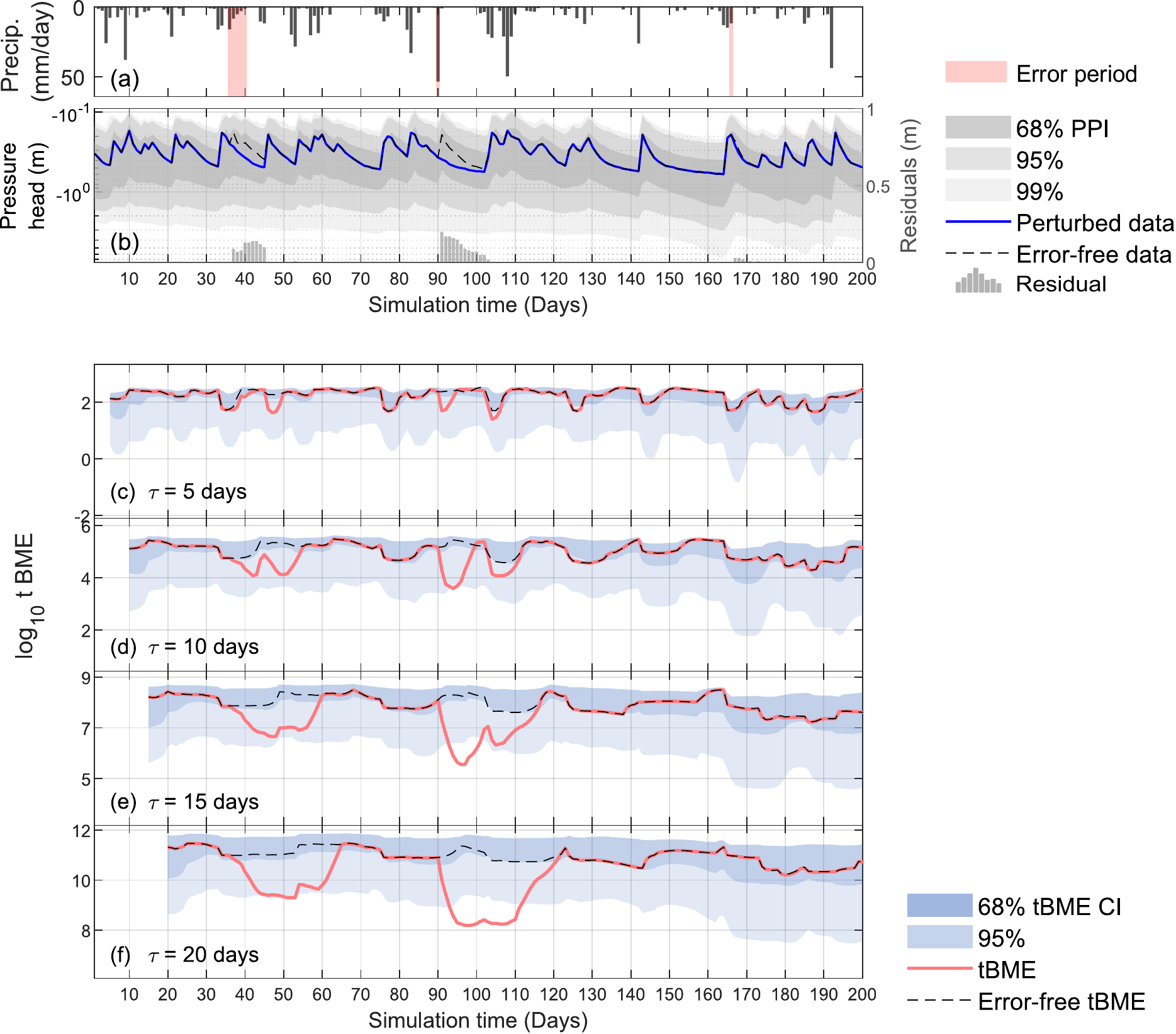}
\caption{tBME curves given forcing error (synthetic case 2: precipitation marked in red has not happened at the field site, but is used as input for our model).   
}
\label{fig:Plot3_obs36}
\end{figure}

The second case illustrates the analysis of tBME curves in the presence of forcing errors. Fig. \ref{fig:Plot3_obs36} shows the forcing (precipitation), the synthetic perturbed dataset, the error-free base case dataset for comparison, and the resulting tBME curves together with their sampling distributions for the four window sizes $\tau$. Recall that, to construct the dataset for this test case, we have run HYDRUS-1D with the base case setup but deleted the precipitation during the pre-defined error periods, which are marked with red-shaded boxes in Fig. \ref{fig:Plot3_obs36}a. 

\subsubsection*{Error Detection}
It becomes apparent that only the first two error periods lead to a significant accumulation of residuals between perturbed data and the model's base case realization. Both periods are clearly detected by the tBME indicator at window sizes larger than five days, with tBME leaving its high-density region and, for the error period with largest residuals, even the complete range of the sampling distribution.

\begin{figure}[ht]
\centering
\includegraphics[scale=0.8]{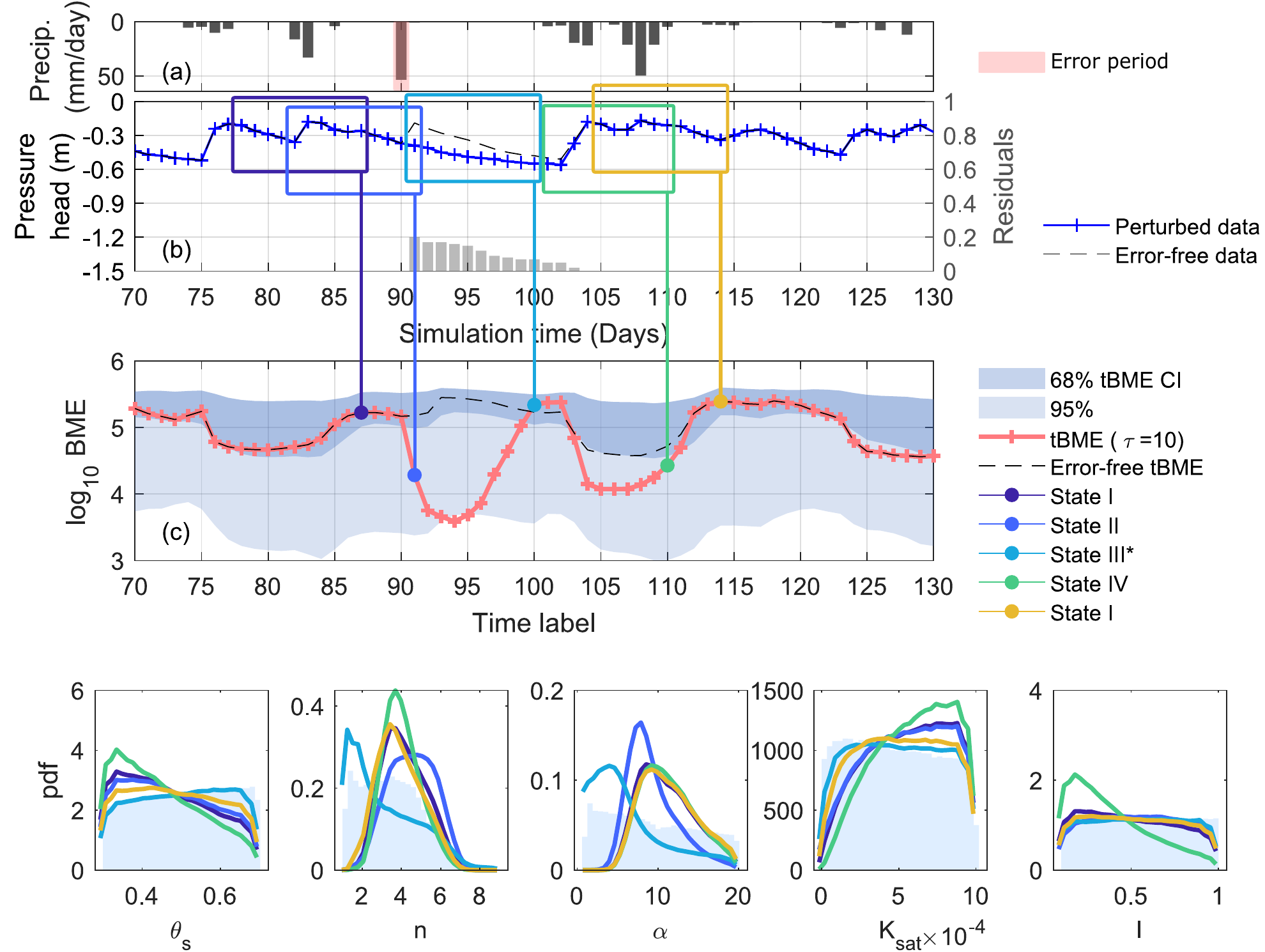}
\caption{Zoomed-in view of tBME curve and dynamic posterior parameter distributions for window size $\tau$ = 10 (Synthetic case 2).}
\label{fig:zoom_Case2_10}
\end{figure}

The theoretically expected behavior of the tBME curve can be seen beautifully at the example of $\tau=10<L_e$ in Fig. \ref{fig:zoom_Case2_10}c. The fall and rise of tBME takes approximately $\tau/2=5$ days, and the intermediate plateau of high tBME values persists for exactly $L_e-\tau=3$ days. The second dive is somewhat distorted, with a fast fall and a slow rise. This effect might be due to the precipitation events that hit the system right in this ``recovery'' period. 

\subsubsection*{Sensitivity of tBME to Window Size}
As discussed in the context of case 1 (model-structural error detection), moving towards longer time windows increases true error signals and flattens out noise and fluctuations of the tBME curve due to parameter variability within the model. Further results refer to Appendix. 

Based on the two investigated test cases so far, we conclude that the selection of an appropriate window size $\tau$ depends on many factors, such as magnitude of model residuals, level of measurement noise, duration of residual occurrence, and internal variability of the model ensemble.  
Although many of these factors might be unknown in practice, it helps if the initial $\tau$ selection is based on (expert) knowledge of the expected error and its duration, which hydrologists or generally modellers typically have some ideas about. If expert knowledge is not available, we recommend analyzing tBME curves with a prior selection of window sizes to assist model error detection and and to provide insights about temporal scales of model error occurrence.

\subsubsection*{Dynamics of Posterior Parameter Distributions}
We find that also in the case of forcing error, a compensation effect is achieved by shifting parameter distributions. This can be seen from the climb of the tBME curve for $\tau=10$ days around $t=100$ days, and the second drop right after (Fig. \ref{fig:Plot3_obs36}d). Around this point in time, the time window almost exclusively captures the residual period, and parameter values are reweighted in Bayesian updating to achieve an optimal fit of the perturbed data. A rise of the tBME curve during an residual period can only be explained by shifting parameter values, and this explanation is confirmed by inspecting the dynamic posterior parameter distributions for selected windows of size $\tau=10$ days in Fig. \ref{fig:zoom_Case2_10}.

The MVG shape parameters $\alpha$ and $n$ seem to be responsible for error compensation: both take on distinct low values only during the residual period. Inspecting the maximum-likelihood water retention curves obtained for the five states, we find a clear shift toward higher water content at a given pressure head during residual periods. This is the only possible way for the model to fit the observed pressure head data despite the surplus of precipitation in the soil. It is achieved by the interplay of low $\alpha$, low $n$ and high $\theta_s$. Similarly, hydraulic conductivity is shifted to higher values at a given pressure head. This leads to faster drainage of the soil profile which there in turn will result in lower $theta$ and higher $h$.This is clearly a compensation effect.

\begin{figure}
\centering
\includegraphics[scale=0.8]{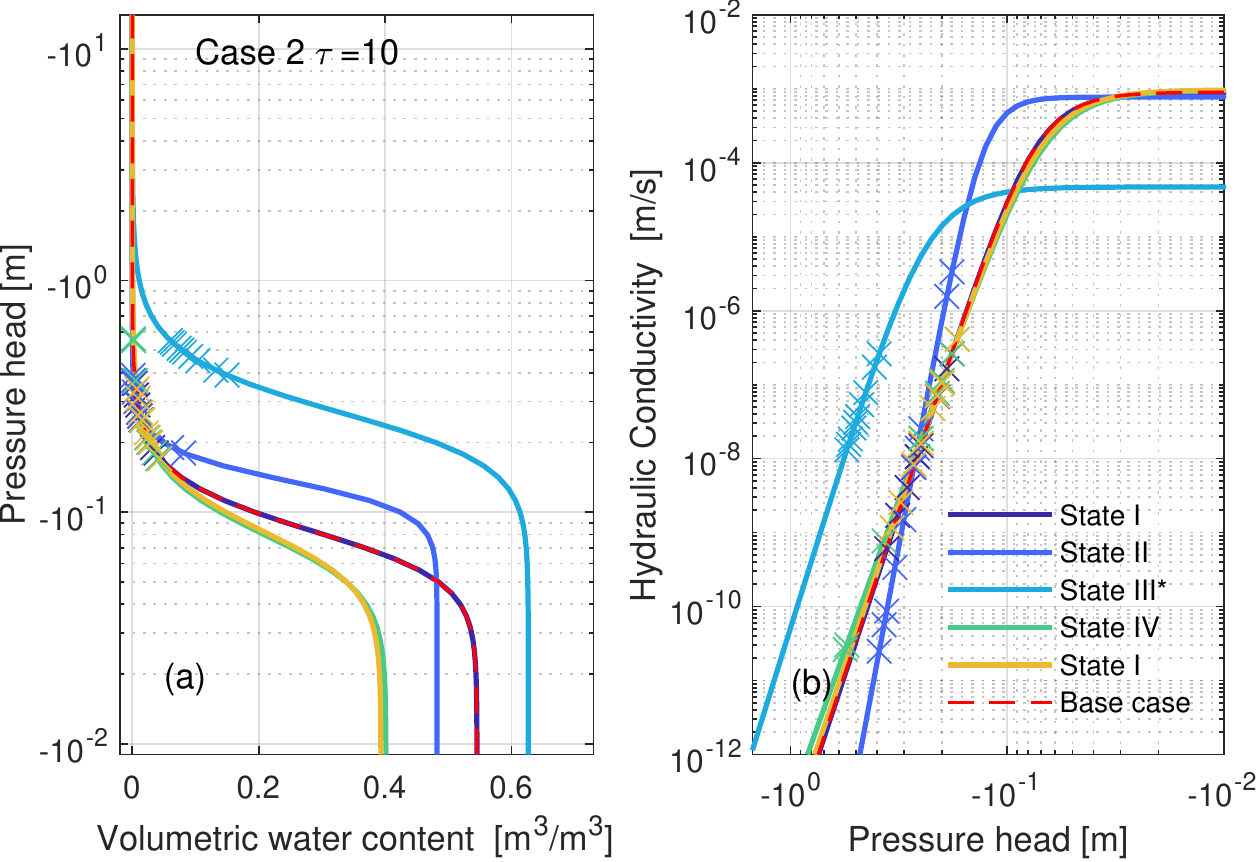}
\caption{Maximum-likelihood water retention curves (left) and unsaturated hydraulic conductivity functions (right) of the top soil layer for the five states color-coded in Fig. \ref{fig:zoom_Case2_10}b (synthetic case 2, window size $\tau=10$ days). Dashed red lines show true curves of the base case; there are no ``true'' curves for the perturbed dataset because forcing error does not impact the soil hydraulic characteristics.} 
\label{fig:WRC_ML_C2_t10}
\end{figure}

\begin{figure}
\centering
\includegraphics[scale=0.8]{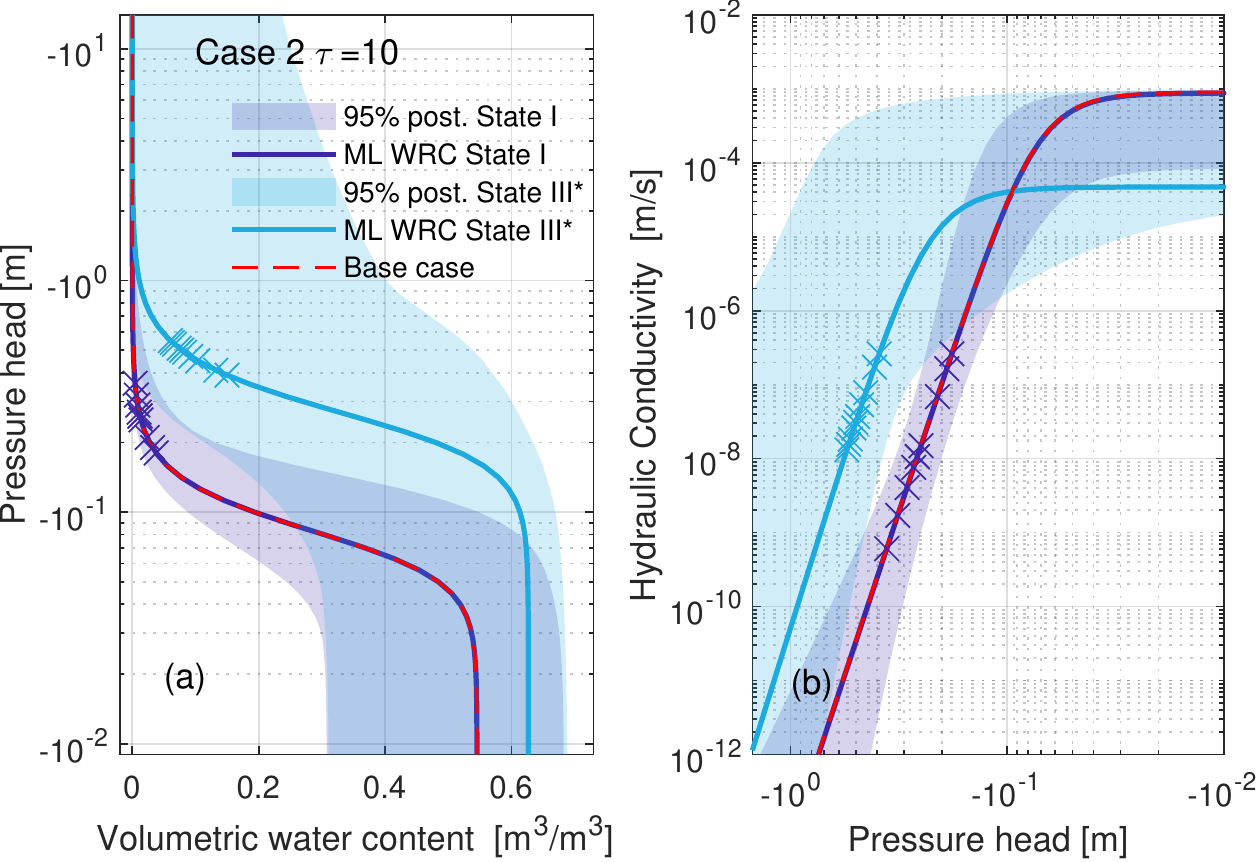}
\caption{Posterior credible intervals of water retention curves (left) and unsaturated hydraulic conductivity curves (right) of the top soil layer for states I and III* (synthetic case 2, window size $\tau=10$ days).}  
\label{fig:WRC_C2_t10}
\end{figure}

To find out whether these successfully compensating parameter sets are clearly identifiable from calibration within the short window of 10 days, we additionally show the $95 \%$ posterior credible interval of the water retention curves and the unsaturated conductivity functions of states I and III$^*$ in Fig. \ref{fig:WRC_C2_t10}. The intervals of both states are extremely large (much wider than in case 1, cf. Fig.  \ref{fig:WRC_C1_t20} in \ref{sec:app}) and allow for a wide range soil hydraulic functions. This warns us that there is no dominant pattern of error compensation - which is plausible since model error is caused by model forcing in this test case, not by modifications of the model parameterization.

Sliding out of the residual period, it seems that $K_{sat}$ and $\theta_s$ are now taking care of the excess water in the model (since it had to cope with precipitation input that is missing in the true system). $K_{sat}$ is clearly increased during this phase (and rather insensitive else), and $\theta_s$ is reduced (cf. posterior PDFs in Fig. \ref{fig:zoom_Case2_10}). The model seems to struggle when going back into the residual-free period, because what happens in the perturbed data is that the current saturation is taken as initial value to continue calculations with the original parameter setup. But the model is in a different state because it tried to compensate for excess water and hence carries some ``history'' with it. On top, another heavy precipitation event is hitting the catchment during the recovery phase after $t=100$ days. One might expect that this event should equalize the tensiometer traces; but since the model is not doing very well during infiltration events anyway, the original dip in the reference BME curve is enhanced. All parameters are shifting in a way distinct from all other phases (green time window and PDFs in Fig. \ref{fig:zoom_Case2_10}). Both effects combined might explain why pre- and after residual-period parameter values are not the same for $\theta_s$ and $K_{sat}$, and why water retention curves look different between state I \emph{before} the residual period and state I \emph{after}. But, importantly, $K(h)$ has gone back to the same state I as before.

We conclude that inspecting the dynamic posterior parameter PDFs, posterior water retention curves and posterior unsaturated hydraulic conductivity curves including their remaining uncertainty can help put the detected model error into context. The sensitivity of individual parameters to the sliding calibration time window can provide information about the type and source of error.

\subsubsection{Case 3: tBME Curve Given Superimposed Errors in Model Structure and Forcing}

In reality, model errors occur at an unexpected frequency and at various temporal scales, so multiple residual periods are expected to overlap each other. The analysis of multiple superimposed errors is demonstrated with the setup of case 3 described in Section \ref{sec:Cases}.  

\begin{figure}[ht]
\centering
\includegraphics[scale=0.9]{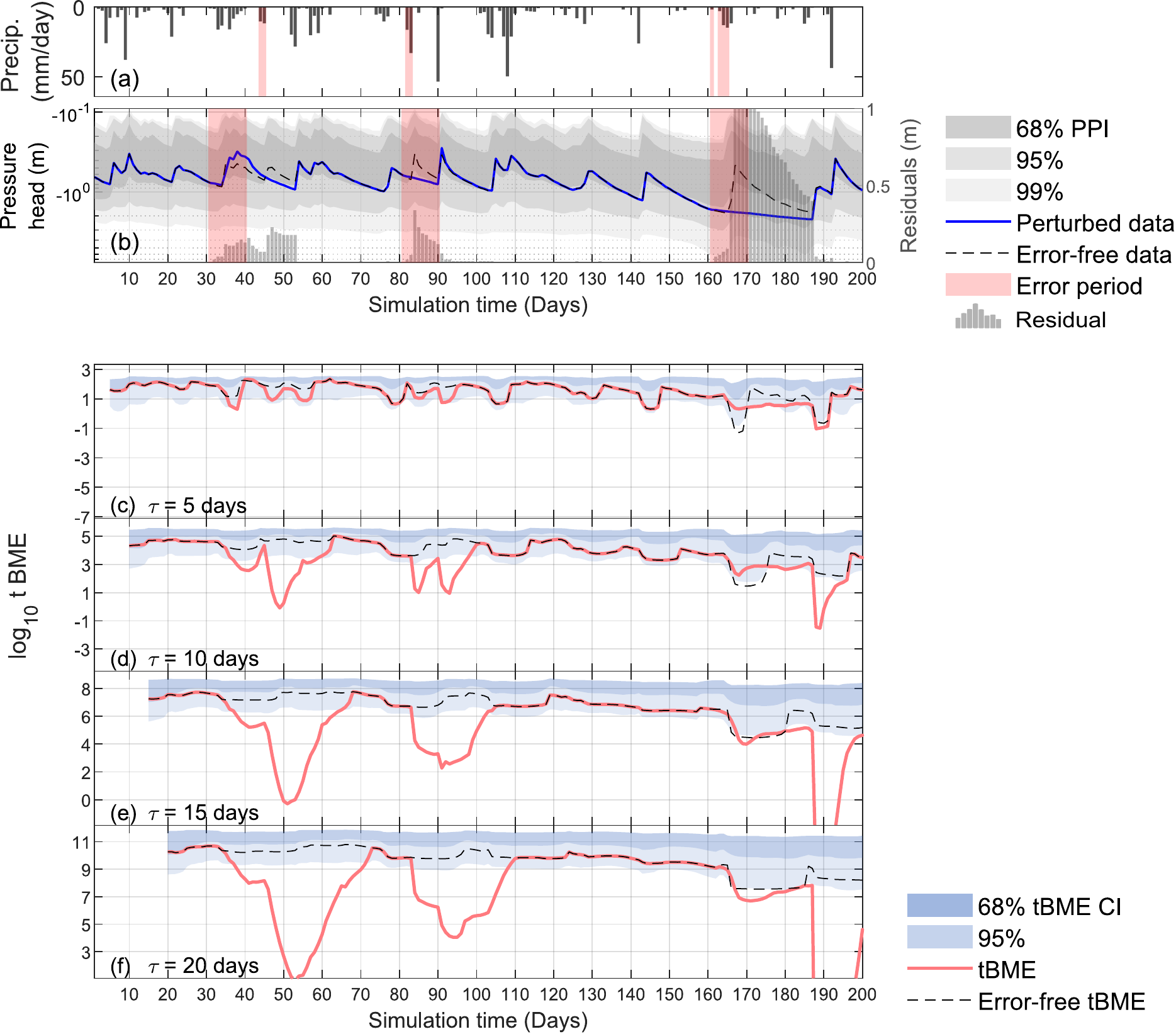}
\caption{Results of tBME analysis if model structural and forcing errors are superimposed (synthetic case 3).  
} 
\label{fig:Plot3_obs56}
\end{figure}

Fig. \ref{fig:Plot3_obs56}a records the forcing (precipitation); Fig. \ref{fig:Plot3_obs56}b shows the synthetic perturbed dataset and the error-free base case dataset for comparison. The red-shaded boxes in Fig. \ref{fig:Plot3_obs56}b mark the periods when the model suffers from structural error (reality is in dual-porosity mode), and the red-shaded boxes in Fig. \ref{fig:Plot3_obs56}a show the periods when the model suffers from erroneous forcing. The model misfits caused by these two errors partially cancel out. The net residuals are presented by the gray bars in Fig. \ref{fig:Plot3_obs56}b. The resulting tBME curves are plotted together with their sampling distributions for the four window sizes $\tau$ in Fig. \ref{fig:Plot3_obs56}c to Fig. \ref{fig:Plot3_obs56}f.

\subsubsection*{Error Detection through Varied Window Size}

Consistent with the previous two cases, we do not perceive clear signals from the shortest window size $\tau=5$ days. From $\tau=10$ days on, we see strong signals during all three residual periods; and these signals are massively intensified when increasing the window size toward $\tau=20$ days.

Recall that the dotted lines represent the tBME curves of the model realization that was perturbed to generate the synthetic data set for this test case 3 (labelled with ``error-free tBME'' in Fig. \ref{fig:Plot3_obs56}c to Fig. \ref{fig:Plot3_obs56}f). 

Note that this dataset is chosen to be a different one than the base-case dataset underlying cases 1 and 2; with the motivation to now move as closely to realistic dynamics as possible. The realization used here does not fall within the high-density region of the prior predictive distribution (cf. e.g. days 160 to 170 in Fig.\ref{fig:Plot3_obs56}b), so short segments of the error-free tBME curves fall outside the shown intervals of the BME sampling distribution (e.g. days 160 to 170 in Fig.s \ref{fig:Plot3_obs56}c and \ref{fig:Plot3_obs56}d). Note that, per construction, it must fall inside the sampling distribution, just at a very extreme tail. However, we observe that this signal is \emph{not} enlarged when increasing the time-window size. This finding again supports our faith in the method: true model errors will be detected by varied window sizes through drastically intensified signals; model behavior within the statistically plausible range will not produce a strong signal.   

The superposition of errors can be seen when comparing the tBME curves for $\tau=10$ days with the tBME curves for larger window sizes during days 30 to 70: a time window of ten days exactly separates the two residual periods (the residual caused by structural error first, the residual caused by precipitation input error second); at larger time windows, these two signals triggered by the two residuals merge. This points us effectively to the time scales of both errors, which are about ten days of large residuals each.

\begin{figure}
\centering
\includegraphics[scale=0.8]{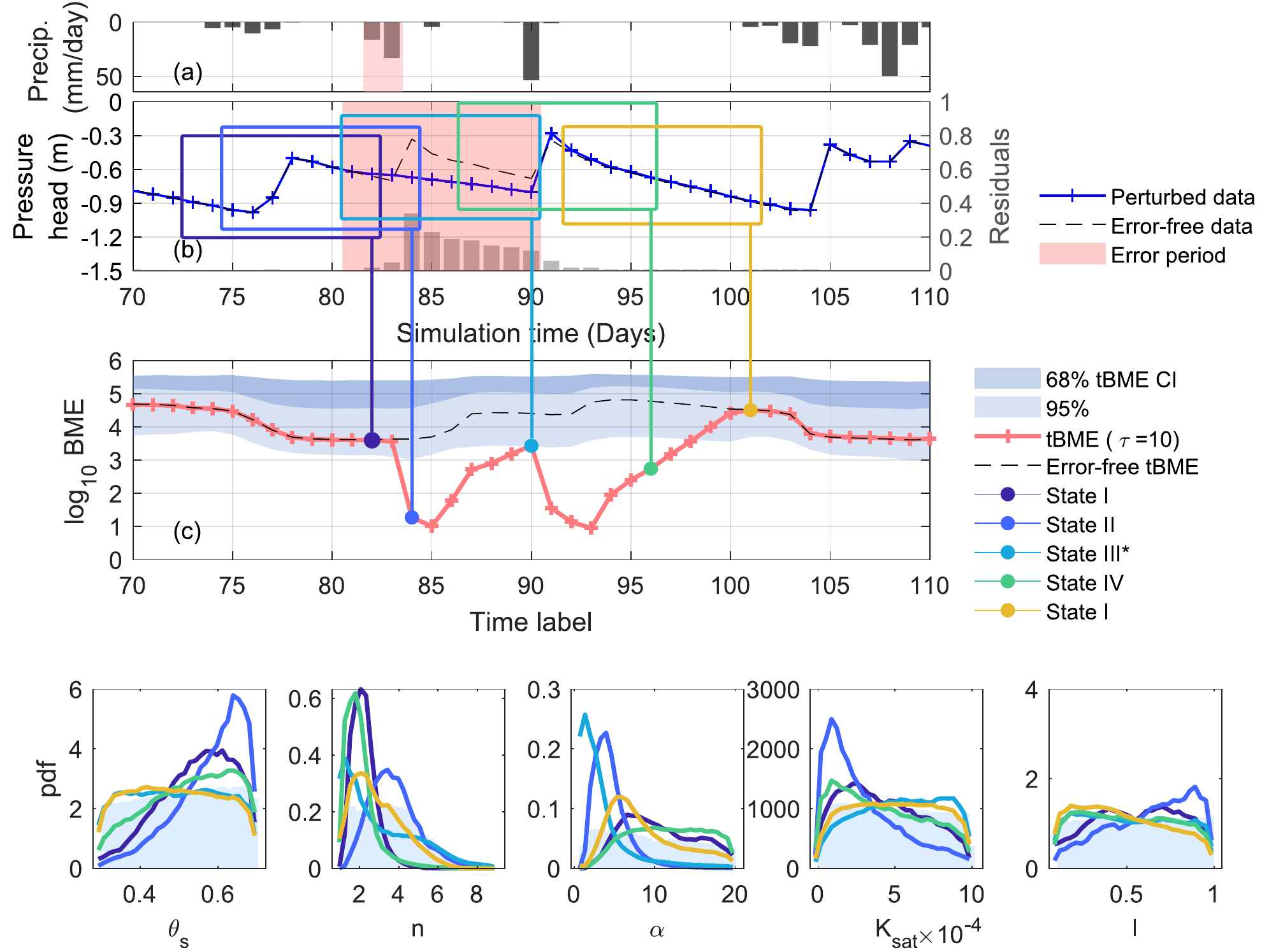}
\caption{Zoomed-in view of tBME curve and dynamic posterior parameter distributions for window size $\tau$ = 10 (synthetic case 3).   
} 
\label{fig:zoom_Case3_10}
\end{figure}

When one residual period overlaps the other (days 80 to 90), we see a double peak for $\tau=10$ days, and a merged signal for larger window sizes. Inspecting the double peak in more detail in Fig. \ref{fig:zoom_Case3_10} reveals that the rise of the tBME curve after its first dive into the residual period comes too early - theoretically, we would expect it at $\tau/2$ days after the onset of the signal because $\tau \approx L_e$ (cf. Section \ref{sec:tBME_interp}); i.e., at $t=88$ days. Here, instead, we see a rise already at $t=86$ days. This is due to the fact that the superimposed errors cancel out to some degree, and hence, tBME has a chance to rise again even earlier. As expected, it arrives at its temporary peak at $L_e$ days after the onset of the signal (because then, the residual period is completely captured by the window), here at $t=90$ days. Sliding out of the residual period again leads to a decline of tBME, which should arrive at its intermediate minimum after $\tau/2$ days, but again is too early here. This is why the rise to the end of the error signal appears very long.

The steep decline and slow return of the tBME curve is indeed the very characteristic of superimposed, counteracting errors: the partial cancelling of residuals affects the originally expected tBME behavior for individual errors in that way. The effect becomes even more clear if we carefully think about which timing to expect here for an ideal case of a single error signal: since $\tau$ is in fact slightly larger than $L_e$, the two falls of the tBME curve should last $\tau/2$ days, and the two rises should last $L_e-\tau/2$ days; i.e. the fall would theoretically take longer (five days) than the rise (two days) to a plateau of three days length. What we see is the opposite behavior, and this is due to the counteracting effect of the error sources.

\begin{figure}
\centering
\includegraphics[scale=0.8]{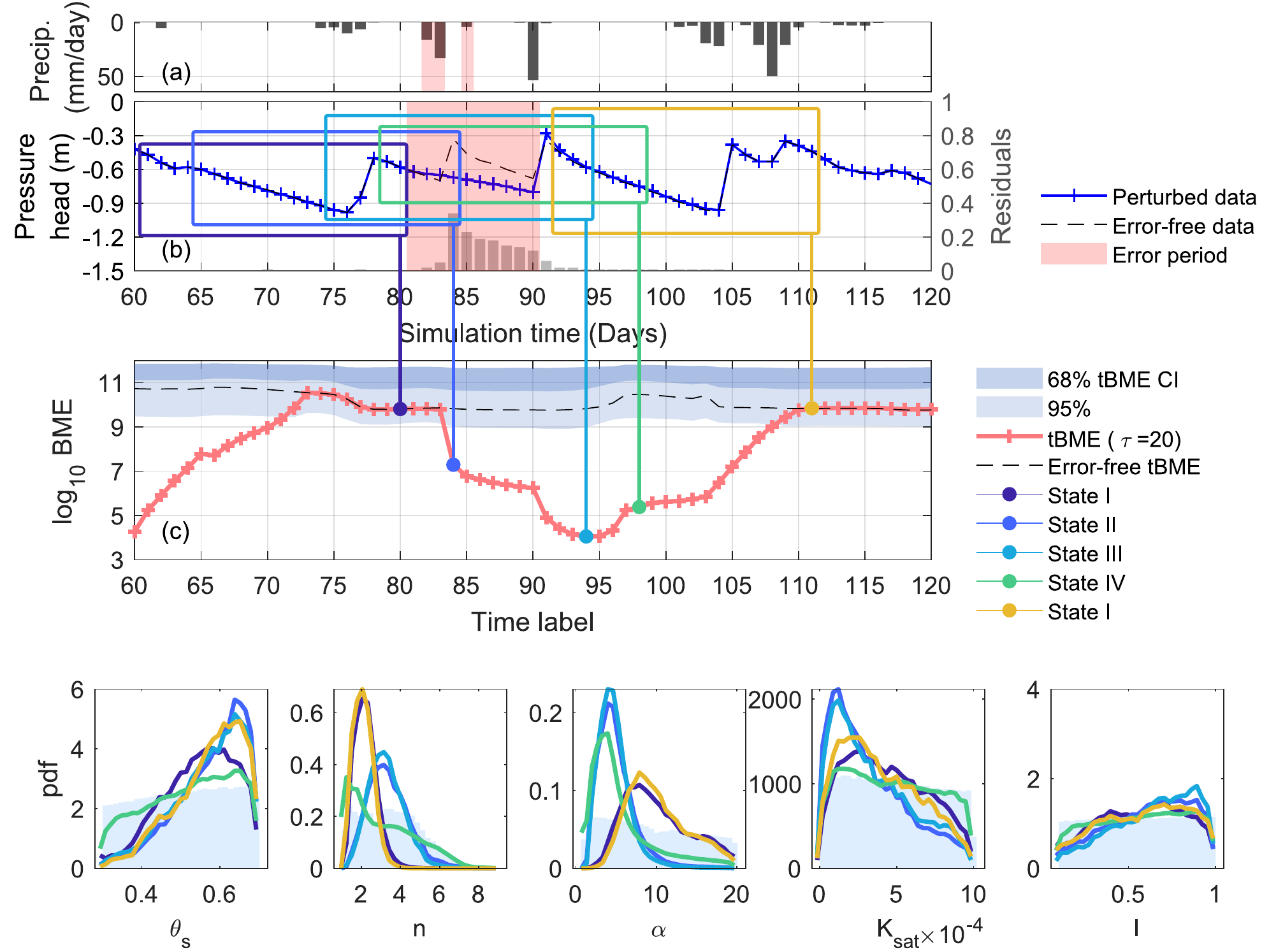}
\caption{Zoomed-in view of tBME curve and dynamic posterior parameter distributions for window size $\tau$ = 20 (synthetic case 3).  
} 
\label{fig:zoom_Case3_20}
\end{figure}

Also, the merged signals for larger $\tau$ in both residual periods (days 30 to 70 and 80 to 90) look distinct and can be distinguished from the case of parameter compensation as seen in cases 1 and 2. In the case of parameter compensation during a single-error period, we observed a stable plateau in the tBME curve when increasing the window size. Here, we see distinct steep drops from such a plateau. Similar to our interpretation above for $\tau=10$ days, we find that the fall to the minimum plateau, expected to occur for $t=91$ to 103, seems to be dampened ($\tau=20$ days, Fig. \ref{fig:zoom_Case3_20}). The plateau is interrupted by an additional dive, before the tBME starts to climb back up as expected from $t=104$ days on. 

We found that determining the residual period length as total signal length $L_s$ minus $\tau$ yielded seven days, which only identifies the period of large superimposed residuals. Hence, for directly superimposed residuals, inferring the individual time scales will be hard (at least for the less pronounced residual), but varying the time-window size together with low-noise data could help identify also this one.

\subsubsection*{Dynamics of Posterior Parameter Distributions}


When inspecting the dynamic posterior parameter distributions for this test case 3 at a window size of 10 days (Fig. \ref{fig:zoom_Case3_10}), we see attempts to parameter compensation that seem disrupted: judging from the tBME value during state III, a parameter shift can accomplish a return to the edge of the high-density region of the tBME sampling distribution; however, this shift is not consistent throughout the residual period. For all parameters but $\alpha$, the first reaction of the parameter distributions during state II produces much more pronounced shifts away from state I (residual-free period) than during state III. This again confirms a ``mitigating'' effect of error superposition in this case. The role of $\alpha$ has indeed not changed - it takes on very small values during error compensation as in test case 2 (erroneous forcing), but the interplay with $n$ seems different here. 

\begin{figure}
\centering
\includegraphics[scale=0.8]{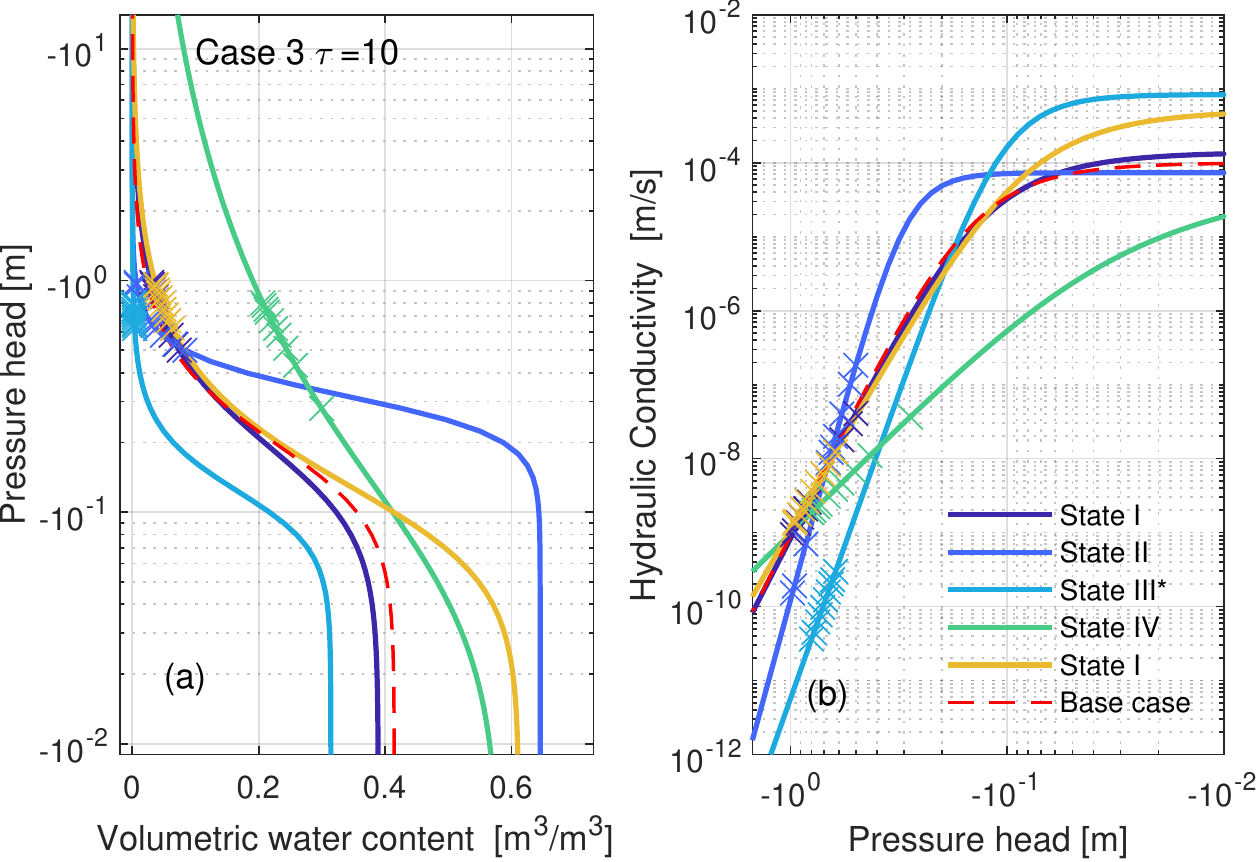}
\caption{Maximum-likelihood water retention curves (left) and unsaturated hydraulic conductivity functions (right) of the top soil layer for the five states color-coded in Fig. \ref{fig:zoom_Case3_10}b (synthetic case 3, window size $\tau=10$ days). Dashed red lines show true curves of the base case; again, there are no ``true'' curves for the perturbed dataset.} 
\label{fig:WRC_ML_C3_t10}
\end{figure}

We further inspect the maximum-likelihood water retention curves and unsaturated hydraulic conductivity functions for the five states in Fig. \ref{fig:WRC_ML_C3_t10}. Also here we see mixed compensation signals from the two previous cases. 

The fact that $\alpha$ clearly shifts to counteract the forcing error, but not the structural error ($\alpha$ moved to higher values in test case 1), combined with the much larger shift of the water retention curve towards higher water contents as in case 2, suggests that the forcing error is playing the dominant role here. Such interpretations thus help identifying the source and role of model errors. For interpreting real-world analysis results (as to be done in Section \ref{sec:Results_field_data}), we can thus recommend to create synthetic scenarios with hypothesized errors to investigate the corresponding outcomes of tBME, posterior parameter distributions and posterior soil hydraulic functions for comparison. 

For $\tau=20$ days, we see more consistent parameter compensation attempts in both marginal parameter PDFs (Fig. \ref{fig:zoom_Case3_20}) and soil hydraulic functions (Fig. \ref{fig:WRC_ML_C3_t20}). ``Averaging'' of the impact of both errors seems to produce clearer compromise solutions in parameter values. $\alpha$ is decreased during the residual period, $n$ is increased during the residual period. Again, this represents a mitigation of the forcing error, not so much of the structural error. $\theta_s$ and $K_{sat}$ show trends of increased saturated water content and decreased conductivity during (and after) the residual period, which is similar to our findings from case 1 (structural error), and not case 2 (forcing error). 

\begin{figure}
\centering
\includegraphics[scale=0.8]{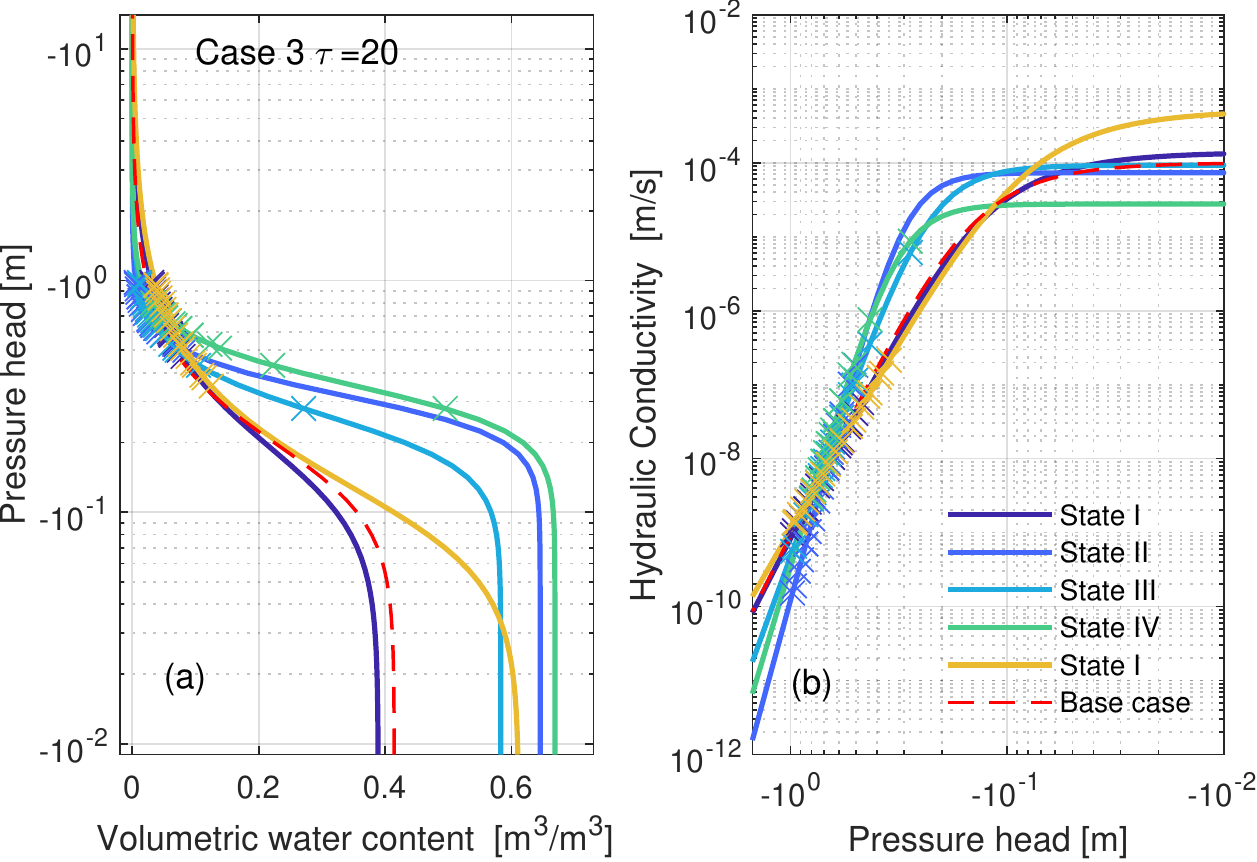}
\caption{Maximum-likelihood water retention curves (left) and unsaturated hydraulic conductivity functions (right) of the top soil layer for the five states color-coded in Fig. \ref{fig:zoom_Case2_10}b (synthetic case 3, window size $\tau=20$ days). Dashed red lines show true curves of the base case; there are no ``true'' curves for the perturbed dataset because forcing error does not impact the soil hydraulic characteristics.}  
\label{fig:WRC_ML_C3_t20}
\end{figure}

Note that the error timing in case 3 is different and does not allow for a detailed 1-to-1 comparison: while the structural error period in this third test case is the same as in case 1, the forcing error has been chosen to overlap, so it is earlier in case 3 than in case 1. This leads to a severe precipitation event hitting the catchment during its error recovery phase in case 3; in case 2, this event was subject to forcing error inside the error period. Hence, both scenarios naturally request for different strategies to bring the model back into a state that matches the original error-free data set well.

\subsubsection{Summary of Findings from Synthetic Test Cases}

We found that tBME curves can be quite wiggly due to parameter uncertainty or other factors that increase the dynamic and variability in ensemble predictions. Also, the tBME sampling distribution can show dynamics over time. In conclusion, no matter how wiggly the tBME curve obtained for real data, if it stays within the tBME sampling range, there is no substantial (statistically significant) reason to believe in model error.

We do not give any general recommendations on a significance threshold for ``low tBME'' as an indicator for error occurrence; rather, we encourage the modeler to test increasing window sizes and identify those periods that consistently produce clearer signals. For more confidence, the modeler could perform synthetic tests as demonstrated here to learn about the model's tBME reaction to a certain level of error under idealized conditions.

Results have shown that different types of errors and even superimposed errors can be detected with our proposed method. The interpretation of the tBME behavior is guided by the theoretical considerations in Section \ref{sec:tBME_interp}. We can, e.g., easily identify the residual period length from visual inspection of ``tBME valleys''. We found that mostly the tBME curves in our test cases followed the theoretically expetected behavior closely; however, changing conditions (such as in forcing) make the tBME curve leave its theoretical trajectory. Also, superimposed errors that tend to cancel out have a specific impact on the tBME curve: they produce a steep decline and a slow return. Hence, by comparing the actually obtained tBME curve with the theoretical sketches and by paying attention to other factors that could change the model ensemble's behavior before and after an error period, we can extract valuable information about the type and duration of model error.

Analyzing of dynamic shifts in time-windowed posterior parameter distributions allows to gain an understanding of how parameters possibly compensate for model error. These insights may, in turn, provide avenues for model refinement. In the specific case of soil hydraulic modeling, interpreting posterior water retention curves and unsaturated hydraulic conductivity functions proves highly valuable. In the considered case study setup, the range of ``synthetically observed'' data values covered only a very small (dry) portion of the full saturation range and hence, data are only little informative for identifying most plausible soil hydraulic parameter values. Although this is a well-known and important phenomenon, it is often forgotten during the analysis of hydrological model results.

Potential users of our proposed method might ask themselves whether just looking at residuals would be sufficient to obtain a model error signal, avoiding the hassle to go through Bayesian updating altogether? Unfortunately, inspection of residuals only would not do the trick: remember that what we showed as the gray bars in Figs. \ref{fig:Plot3_obs44} to \ref{fig:zoom_Case3_20} are the residuals between the error-free data set and its perturbed version - this is unknown in practice. We would have to look at average residuals of ensemble predictions vs. data instead, or at best-fit-residuals, and both can be misleading if parameter compensation occurs and/or superimposed errors tend to cancel out. Thus, it is beneficial to make use of the Bayesian framework: it allows error detection via tBME and deeper interpretation with respect to error compensation based on dynamic parameter posteriors.

\subsection{Results of tBME Analysis for Real Observation Data}
\label{sec:Results_field_data}

We now use the actually observed Spydia data to construct the tBME curves and wish to identify model error in our tested ensemble. The dataset has not been perturbed with extra errors. Note that the overall time period of the calibration dataset and the simulation is the same as in all synthetic cases; hence, we can use our findings from the base case and the test cases 1 to 3 to help us interpret the results from this real case - since the source of error is obviously unknown now. This case is a demonstration of the potential usefulness of our method under realistic conditions, while the synthetic cases have served as a validation of our approach.

\begin{figure}
\centering
\includegraphics[scale=0.9]{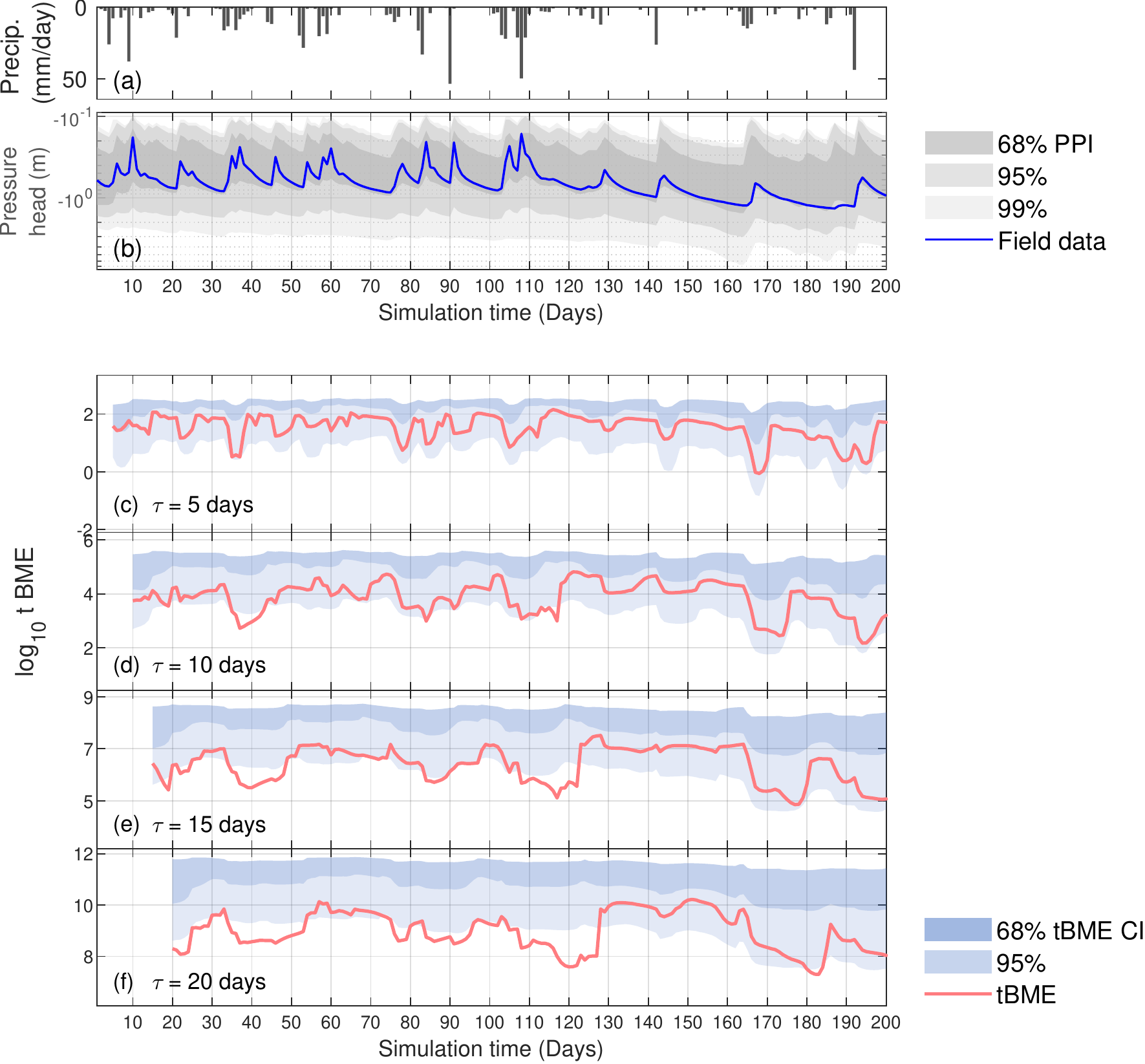}
\caption{Results of tBME analysis for real field data.}
\label{fig:obs_true}
\end{figure}

Fig. \ref{fig:obs_true}a visualizes the model forcing (recorded precipitation); Fig. \ref{fig:obs_true}b shows the real dataset and the prior predictive range of our model. Error periods and residuals are obviously unknown now. The obtained tBME curves are plotted together with their sampling distributions for the four window sizes $\tau$ in Figures \ref{fig:obs_true}c- \ref{fig:obs_true}f.

\subsubsection*{Error Detection through Varied Window Size}

The drops in the tBME curve of $\tau=5$ indicate multiple potential error occurrences, which could, however, be clouded by measurement noise as noted in the synthetic test cases. Compared with the observations, we notice that drops in the tBME curve always occur when a recession phase abruptly changes to a wetting event, i.e. when the soil water pressure jumps to higher moisture contents. This implies that the model cannot describe highly dynamic responses well, and consequently tBME goes down.

With the wiggly nature of the tBME curve under real-world conditions, the sampling distribution proves to be very helpful for interpretation. That is also why we designed synthetic case 3 such that it mimics real-world dynamics, to make ourselves familiar with interpretation of such curves.

To distinguish between statistically significant errors and noise, we search for enlarged signals along the vertical axis of Fig. \ref{fig:obs_true} (tBME curves of increasing $\tau$). From the tBME curve for $\tau=20$, we observe amplified signals that fall out of the tBME sampling distribution during days 34 to 53, 75 to 96, and 105 to 128, which are considered as error signals. The interval between days 130 and 160 does not show noticable model errors, since the tBME curve does not shift downward significantly with increasing window size. 

The tBME curves of larger window sizes do not indicate well-separated single peaks, which suggests that either the time scale of model error is on the order of ten days or less; or model error events are overlapping and hence leading to a smeared out signal. Both hypotheses are probably true if we assume that model error is related to the time of wetting (one to three days): with increasing $\tau$, often the next event is already included in the window. 

\begin{figure}
\centering
\includegraphics[scale=0.8]{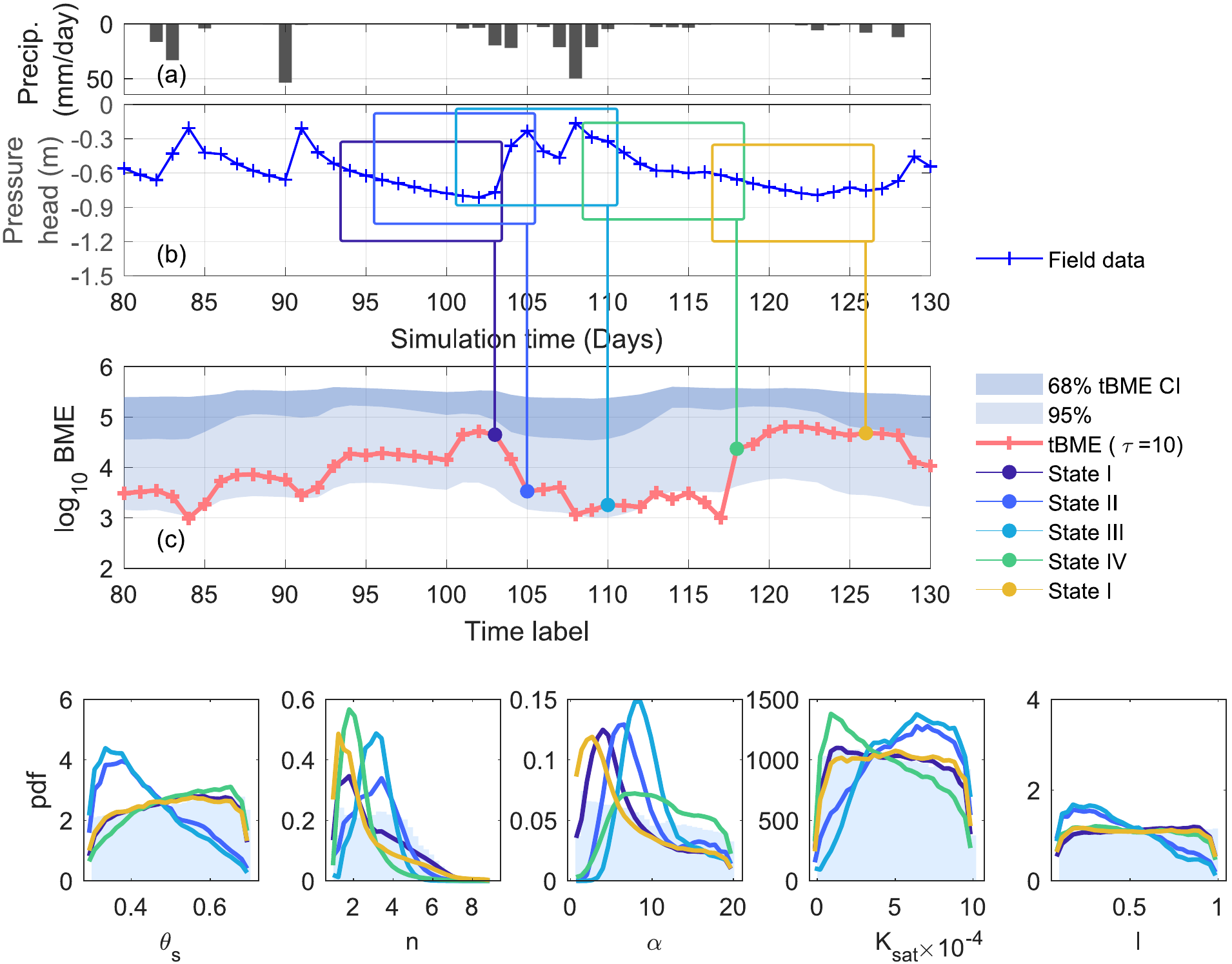}
\caption{Zoomed-in view of tBME curve and dynamic posterior parameter distributions for window size $\tau$ = 10 (real observation data).  }  
\label{fig:zoom_Case4_10}
\end{figure}

In Fig. \ref{fig:zoom_Case4_10}, we can inspect the dive of the tBME curve between days 103 and 118 in more detail. The model clearly leaves the high-density tBME trajectory for a period of $L_s=14$ days. Recalling the theoretical properties of the tBME indicator, the underlying error period (time period with significant residuals) should be of length $L_s-\tau=4$ days. Further, the pattern of the tBME curve looks more similar to superimposed errors (cf. synthetic case 3) than to individual errors (cf. cases 1 and 2). Again, superimposed, short-term errors seem plausible here if we assume that they are related to soil wetting, given that several days of precipitation with varied intensity occur during this period.  

For simulation days 160 to 170, we perceive a signal that has a very clear, familiar shape, with a widening valley for increasing time-window size. This signal formally does not leave the reference interval obtained by sampling tBME from the model; however, the sampling range is here very wide due to the  increased model-interval variability. We therefore consider this to be a sign of model error - during this period, the model structure itself may not be disqualified as ``quasi-true'', but most of the parameter ensemble is. Certainly, the model is struggling here, and reasons should be investigated.

\subsubsection*{Dynamics of Posterior Parameter Distributions}

Inspecting the dynamic posterior parameter distributions for $\tau=10$ days in Fig. \ref{fig:zoom_Case4_10}, we see clear attempts to parameter compensation: the MVG shape parameters $\alpha$ and $n$ both are increased during the low-tBME period, the pore connectivity $l$ is decreased, and low values of saturated water content $\theta_s$ are preferred paired with higher conductivity $K_{sat}$. This reaction is similar to the pattern of parameter compensation that we have seen in test case 1 (structural error, missing dual-porosity component in the model). 

Yet, judging from the tBME curve, we conclude that these compensation attempts through shifting parameter values are not highly successful. They seem to dampen the fall of the tBME curve to some degree, but they cannot bring its value up to the higher-density region of the model. This underlines that the model most likely suffers from structural error related to the wetting phases (e.g., missing macropore flow, surface runoff, hydrophobicity) that needs to be treated by modifying the model structure; time-dependent parameters could here only help to a limited extent. 

\begin{figure}
\centering
\includegraphics[scale=0.8]{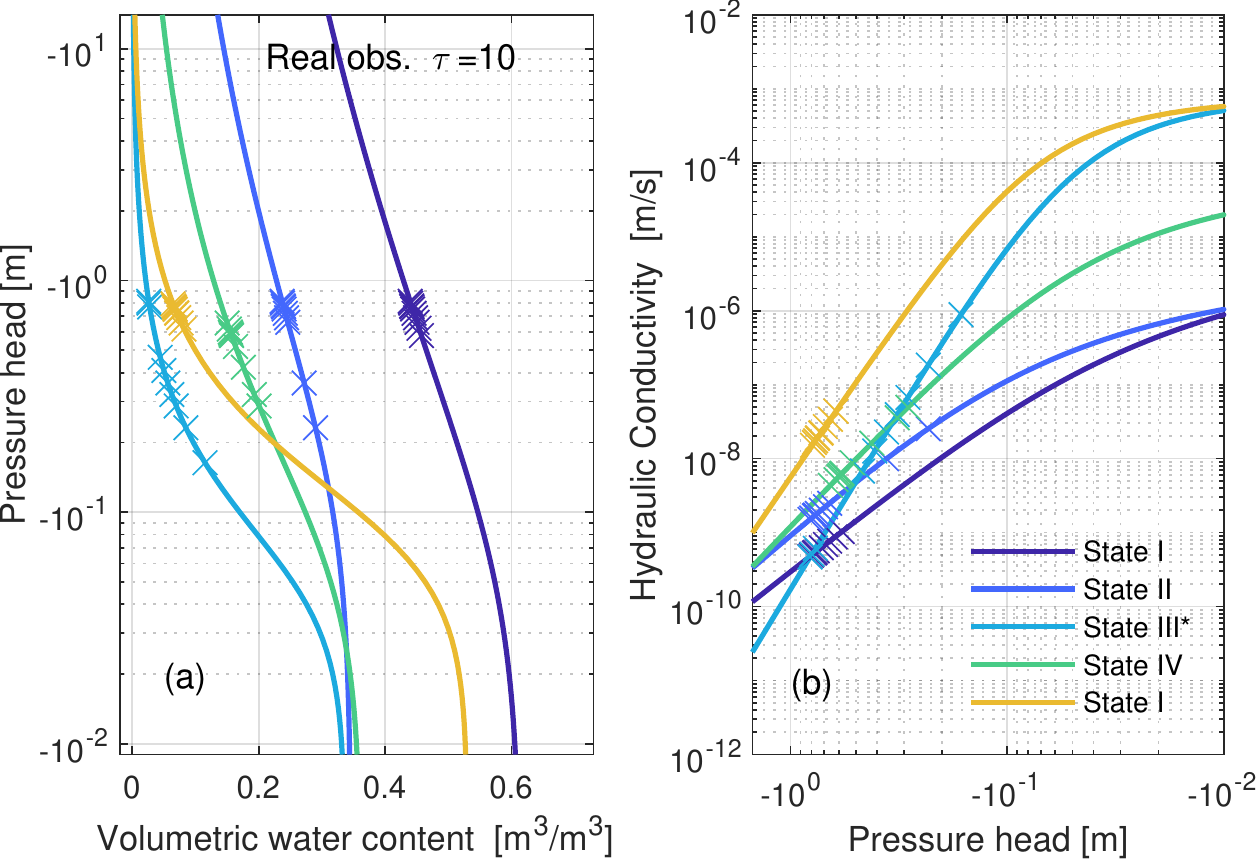}
\caption{Maximum-likelihood water retention curves (left) and unsaturated hydraulic conductivity functions (right) of the top soil layer for the five states color-coded in Fig. \ref{fig:zoom_Case4_10}b (real observation data, window size $\tau=10$ days).} 
\label{fig:WRC_ML_C4_t10}
\end{figure}

\begin{figure}
\centering
\includegraphics[scale=0.8]{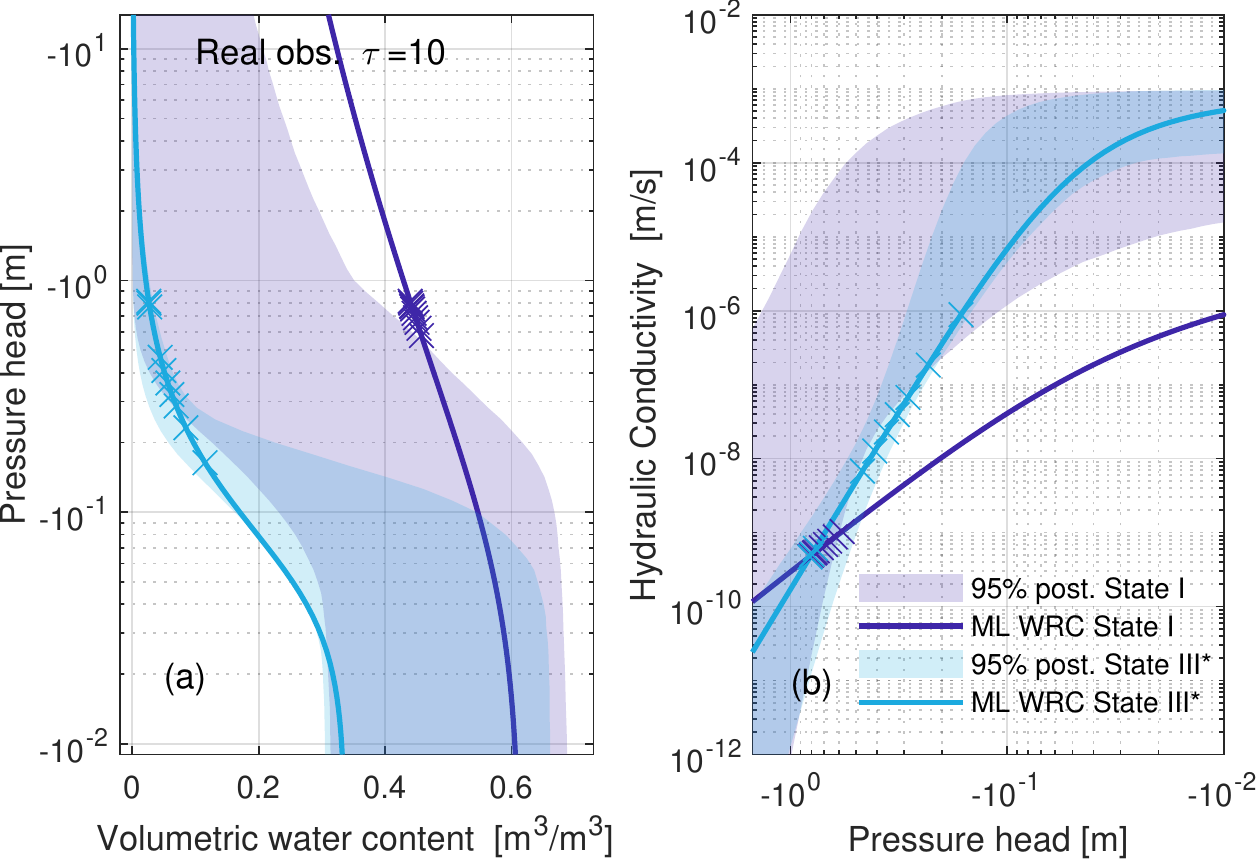}
\caption{Posterior credible intervals of water retention curves (left) and unsaturated hydraulic conductivity curves (right) of the top soil layer for states I and III* (real observation data, window size $\tau=10$ days).}  
\label{fig:WRC_C4_t10}
\end{figure}

When inspecting the dynamic water retention curves and unsaturated hydraulic conductivity curves in Fig. \ref{fig:WRC_ML_C4_t10}, we find that their shape in most states does not resemble the typical shape that our model ensemble favors. This supports the hypothesis that a structural error dominates which is not related to parameter misspecification. The remaining uncertainty after calibration on this short time window is large (see Fig. \ref{fig:WRC_C4_t10}), but still the maximum-likelihood curve of state I falls outside of the $95 \%$ credible interval. This emphasizes the struggle of the model to fit the data well, even during periods with a high-level tBME value. 
 
\begin{figure}
\centering
\includegraphics[scale=0.8]{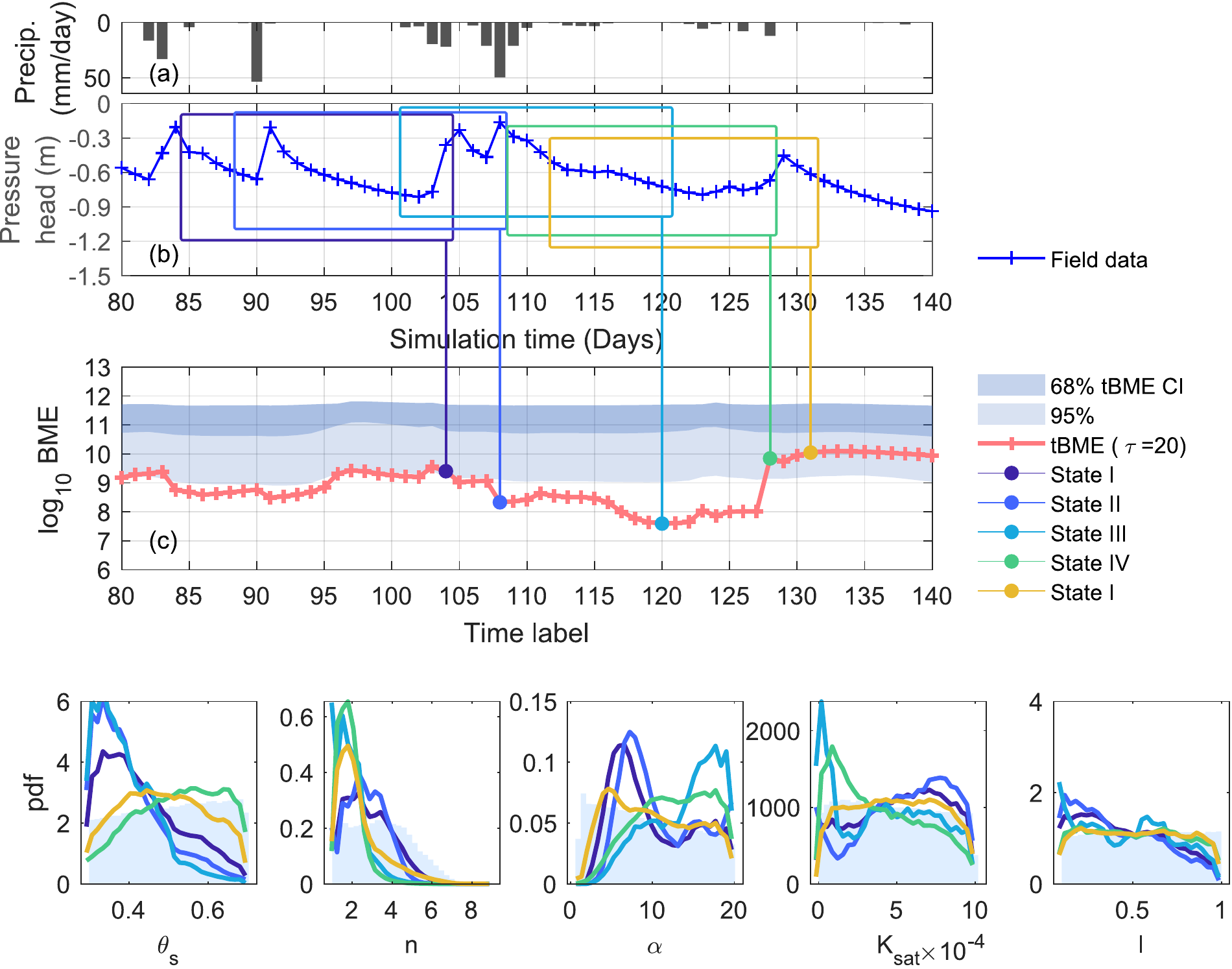}
\caption{Zoomed-in view of tBME curve and dynamic posterior parameter distributions for window size $\tau$ = 20 days (real observation data).  }
\label{fig:zoom_Case4_20}
\end{figure}

The discussed findings are confirmed when inspecting the posterior parameter distributions for $\tau=20$ days in Fig. \ref{fig:zoom_Case4_20}. For this window size, we see a total signal length of about 24 days, which yields the same suggested residual period length of four days. Also the steep return to the higher-density region of our model is at the same relative position and suggests that the model is coping much better with the period after $t=109$ days. This exactly corresponds to the transition between heavy precipitation and drier conditions, and supports our hypothesis that the model is much better at representing soil water redistribution (soil drying) than at simulating infiltration events (soil wetting).

Within these longer time windows, $\alpha$ and $K_{sat}$ seem to try and compensate heavily during the recession phase of the precipitation event at $t=108$ days. The other parameter distributions look similar to those for $\tau=10$ days. This suggests that averaging over longer periods of time does not require opposite directions of parameter compensation, and hence indicates potentially superimposed errors of the same type. This again supports our hypothesis of structural errors of our model related to infiltration processes. To improve this model, we could allow the macropore flow fraction $w2$ as another uncertain parameter, and see whether it could effectively compensate most of the detected model-structural error. Given the current error patterns revealed by our analysis, we would expect that $w2$ would be larger than zero during the residual period, and with the effective soil subsystem the model misfit would be improved. However, is beyond the scope of the current study.

\section{Summary and Conclusions}
\label{sec:Conclusion}
Model structural errors are omnipresent in hydrological modelling. Identifying such errors is vital to deepen our system understanding, to adequately judge the skill of hydrological models, and to improve their forecast reliability in subsequent research. We propose a data-driven method to detect time-dependent structural errors. Our method builds on the statistically rigorous Bayesian framework and does not require any prior assumptions about the error type or pattern. The novelty of our method is to apply a sequence of Bayesian hypothesis tests by sliding a calibration time window through the given dataset. By testing the assumption that the model has generated the data along the time axis, we can identify the time points when the model fails to describe the system sufficiently well. The core idea of our approach is hence to perform a time-windowed Bayesian analysis that identifies short periods in time where the model statistically disqualifies itself as (pseudo-)true.

We track model performance, measured as Bayesian model evidence (BME), over time to detect the onset of structural error, as well as potential error compensation mechanisms during the error period. As a threshold for reliable error detection, we introduce reference intervals for time-windowed BME (tBME) by sampling the model's predictive distribution as synthetic datasets. If a tBME value falls into this reference range, the model's behavior in this time window is interpreted to be quasi-true; if the tBME value falls outside of that range, statistically-significant model error is detected. If tBME drops from the high-density region of the sampling distribution to a very low quantile, this can also be suspected a sign of model error; statistically, this does not reject the hypothesis that the model structure has generated the data, but it rejects a large part of the model's prior parameter space for that time window.

Our proposed approach is designed to detect error occurrence on various temporal scales, which is especially useful in hydrological modelling. As a side product, we obtain a time sequence of posterior parameter distributions (or of best-fit calibration parameters) that help investigate the reasons for model error and provide guidance for model improvement. Inspecting the dynamic parameter distributions (and, in the specific soil hydraulic context: posterior soil hydraulic functions) allows to put the detected model error into context, and the sensitivity of individual parameters to the sliding calibration time window can provide information about the type and source of error.

We have first provided guidance on how to interpret resulting tBME curves theoretically: their expected behavior is determined by the interplay of the chosen time-window length $\tau$, the residual period length (duration of significant residuals between model and data) $L_e$, and the residual recurrence interval $R$. The proposed method works best under the following conditions: 

\begin{enumerate}
    \item The data set should cover the full residual period to be detected. 
    \item The signal-to-noise ratio in residuals needs to be reasonably high (and ideally error and noise show on different time scales). 
    \item If potential parameter compensation effects shall be detected, the sliding time-window needs to be shorter than the residual period. 
    \item If multiple errors shall be identified individually, the recurrence interval of residuals should be larger than the time-window size (although superimposed errors can also be detected, probably with varying degree of clarity in practical applications). 
\end{enumerate}
  
Since the tBME curve purely reflects the residuals, its interpretation regarding potential error sources requires expert knowledge due to potential delays of the system response (e.g., time lags due to storage effects). 

Then, we have ``calibrated'' the reader on how to interpret tBME curves under known error scenarios by designing three synthetic test cases that vary in the error source and error time scales, including a scenario of superimposed errors. Since superposition of two errors of the same sign (both overestimating or underestimating the target value) naturally increases the net residuals (and simplifies the task for our proposed model error indicator), we here deliberately picked error types that could potentially cancel out.

The test cases are inspired by a real case study on soil hydraulic modeling of the Spydia experiment site in the northern Lake Taupo catchment, New Zealand. We have used daily measurements of tensiometric pressure head of the top soil layer to detect (1) hypothetical structural error (single-porosity model vs. dual-porosity system behavior), (2) forcing error (model input not representative of actual precipitation on-site), and (3) a superposition of both error types. Finally, we have presented insights from applying our proposed method to the actual field observations. 

Results of our synthetic test cases demonstrate that we can detect the start and end of residual periods for all investigated error sources and temporal scales. Even for superimposed errors that tend to cancel out, we are able to reveal the ``fingerprint'' of each error with our method. With the sampling distribution of tBME as a reference, we can clearly differentiate between true error signals and model-internal variations that lead to fluctuations in the tBME curve. Varying the time-window size $\tau$ has been found to be very insightful, as it intensifies true error signals and filters out noise. 

The theoretically expected patterns and the timing of the tBME curve could be discovered from all investigated test cases. Of course, our theoretical considerations hold for idealized, constant residuals during the error period; in reality, we have arbitrary patterns of residuals that affect the shape of the tBME curve. In hydrology in specific, we tend to have rising and falling residuals due to the memory in the system and the model. This means that the timing of, e.g., the lowest point of tBME might be different in real cases. However, we found that the expected patterns are surprisingly clear to identify even from real data. Hence, we can use the simple calculation $L_e=L_s-\tau$ to determine the temporal scale of the detected error, i.e., the length of the residual period (producing significant residuals between model predictions and observed data) is equal to the total tBME signal length minus the chosen length of the moving time window.

In application of the real dataset, our analysis pointed to the model's weaknesses with respect to soil moisture recession and revealed very short temporal scales of error. Superimposed errors of longer scale were not identified; overall, the hypothesis that this model has generated the data could not be rejected during most periods. This, in turn, means that the model is doing respectably well most of the time under the assumed scenario of measurement error and parameter uncertainty. For real-world predictions, we might want to reduce both sources of uncertainty to obtain more precise predictions; that, however, could possibly lead to statistical rejection of the model during larger time periods, because then the hypothesis test through the tBME analysis becomes more strict. Such investigations are left for future studies. 

In principle, our concept of tBME could be generalized to a time-windowed calibration analysis using any other model performance metric based on the residuals between model predictions and data. However, we prefer using BME as a model error indicator because it is consistent with the surrounding Bayesian framework, it takes into account measurement noise, and it is cheap to obtain upon calculation of likelihoods that are needed for Bayesian calibration (i.e., to obtain posterior parameter and predictive distributions).

Overall, we have shown the theoretical rigor of the proposed method, and its usefulness for detecting model errors in various scenarios. We are confident that the concept of the tBME analysis will be valuable in a wide range of disciplines that rely on predictions by dynamic models with potential model errors.

\appendix

\newpage
\section{Further results}
\label{sec:app}

\subsection{Synthetic Case 1, $\tau=10$ Days}

\begin{figure}[ht]
    \centering 
    \includegraphics[scale=0.8]{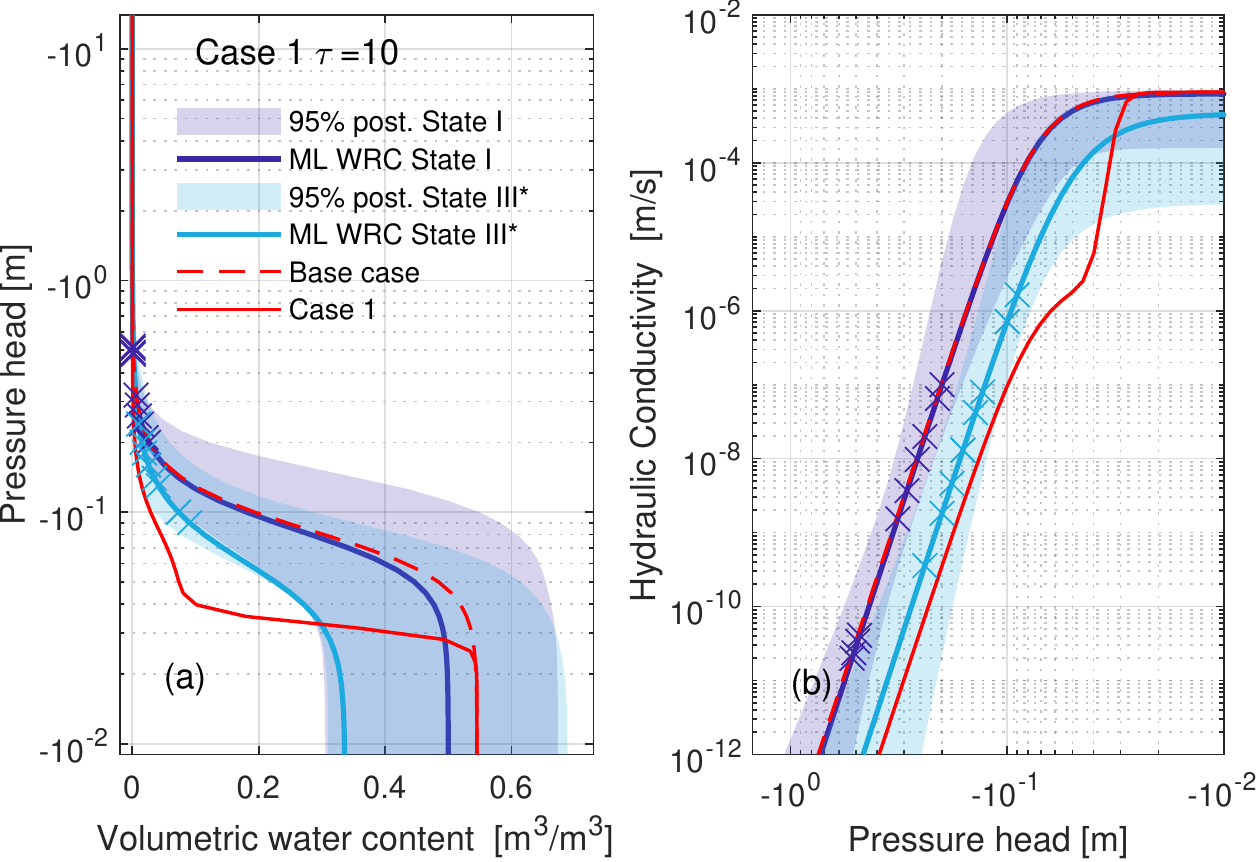}
    \caption{Comparison of the uncertainty range of WRC at state I and state III* in in Fig. \ref{fig:zoom_Case1_10} (Synthetic case 1 $\tau$ = 10). }
    \label{fig:WRC_C1_t10}
\end{figure}

Fig. \ref{fig:WRC_C1_t10} presents the 95\% posterior credible intervals of the water retention curves and the unsaturated hydraulic conductitvity functions at state I and state III* for a window size of $\tau =10$ days. Comparing the uncertainty intervals of these two states in Fig \ref{fig:WRC_C1_t10}a, we observe slightly decreasing uncertainty when the time window moves from the error-free state (state I) to a parameter compensation state (state III*). This is caused by the observation points leaving the high-density area of the predictive prior.   

\pagebreak
\subsection{Synthetic Case 1, $\tau=20$ Days}

\begin{figure}[ht]
\centering
\includegraphics[scale=0.8]{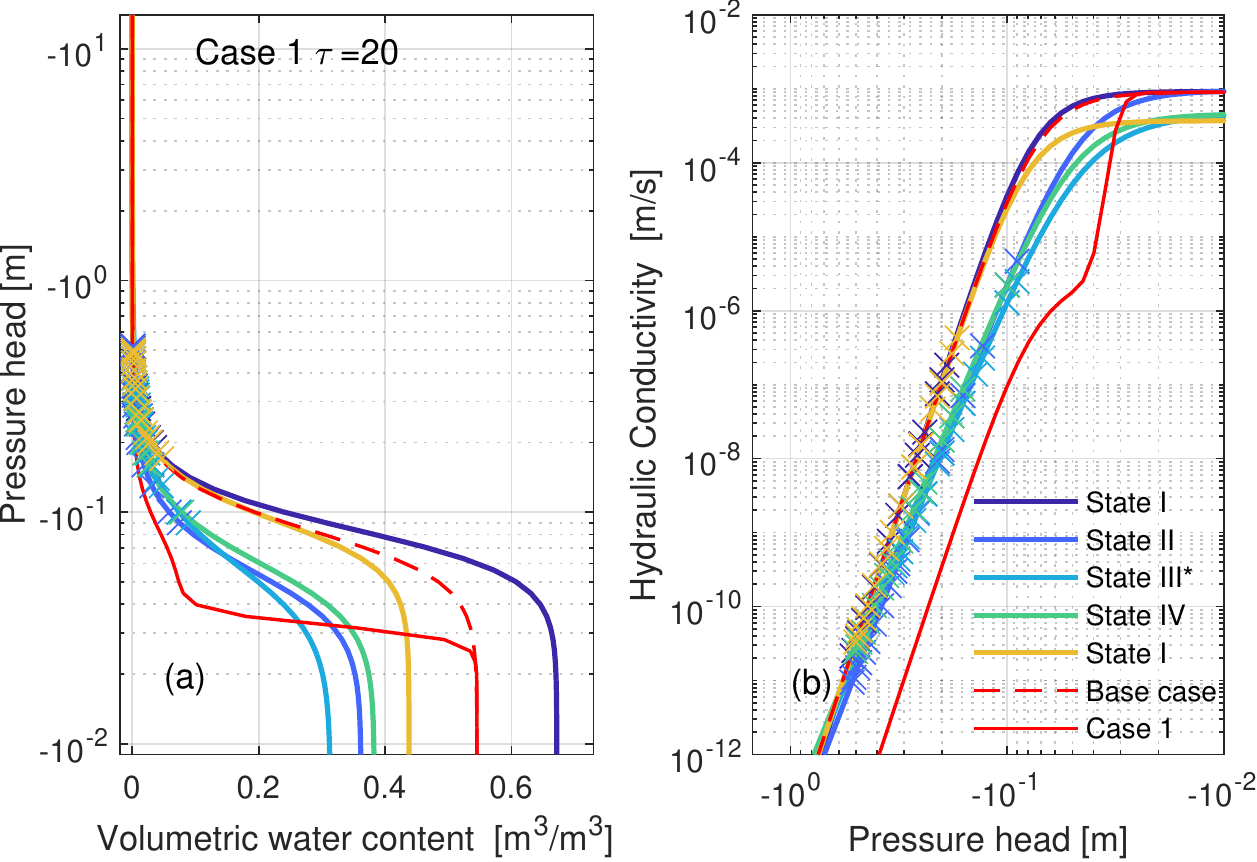}
\caption{Maximum-likelihood water retention curves (left) and unsaturated hydraulic conductivity functions (right) of the top soil layer for the five states color-coded in Fig. \ref{fig:zoom_Case1_10}b (synthetic case 1, window size $\tau=20$ days). Dashed red lines show true curves of the base case; solid red lines show the curves corresponding to the perturbed dataset.} 
\label{fig:WRC_ML_C1_t20}
\end{figure}

\begin{figure}[ht]
\centering
\includegraphics[scale=0.8]{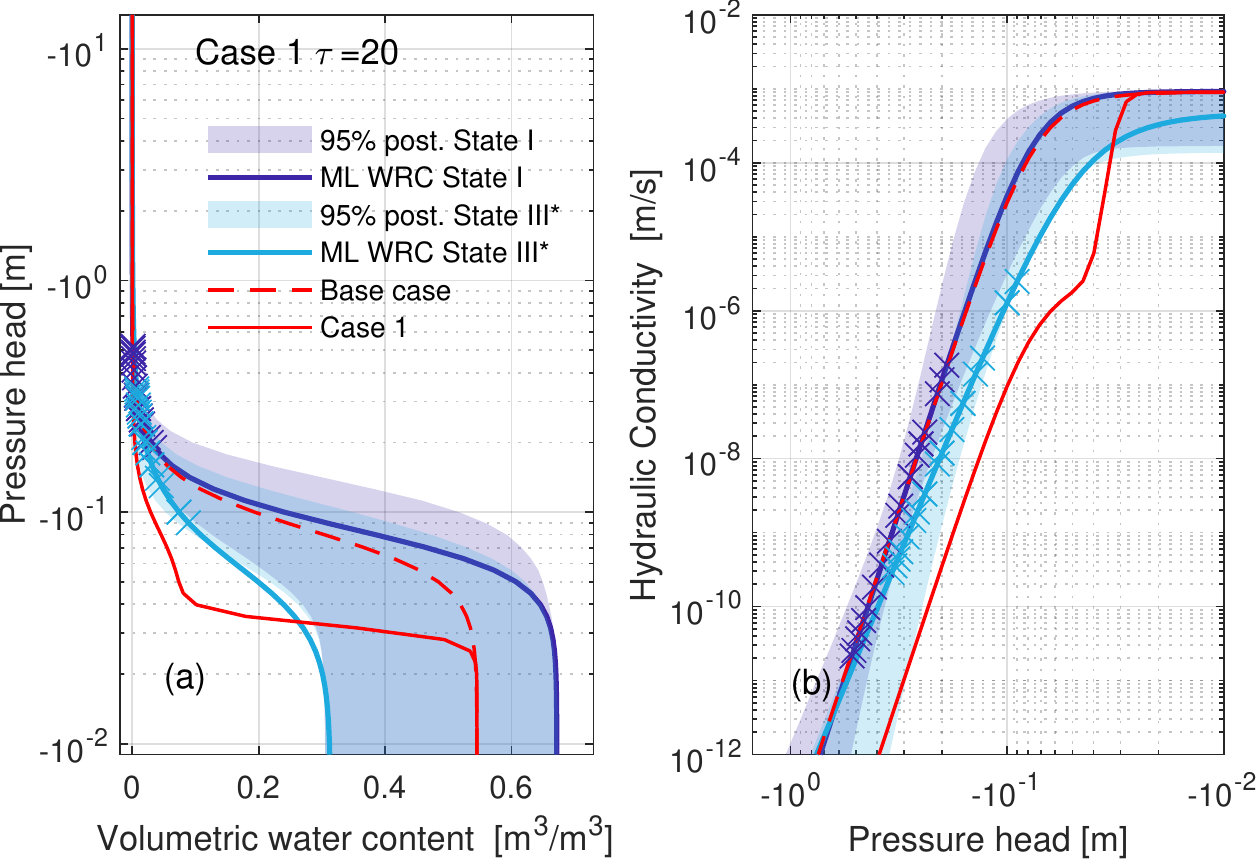}
\caption{Posterior credible intervals of water retention curves (left) and unsaturated hydraulic conductivity curves (right) of the top soil layer for states I and III* (synthetic case 1, window size $\tau=20$ days).}  
\label{fig:WRC_C1_t20}
\end{figure}

\pagebreak
\subsection{synthetic case 2, $\tau=20$}

\begin{figure}[ht]
\centering
\includegraphics[scale=0.8]{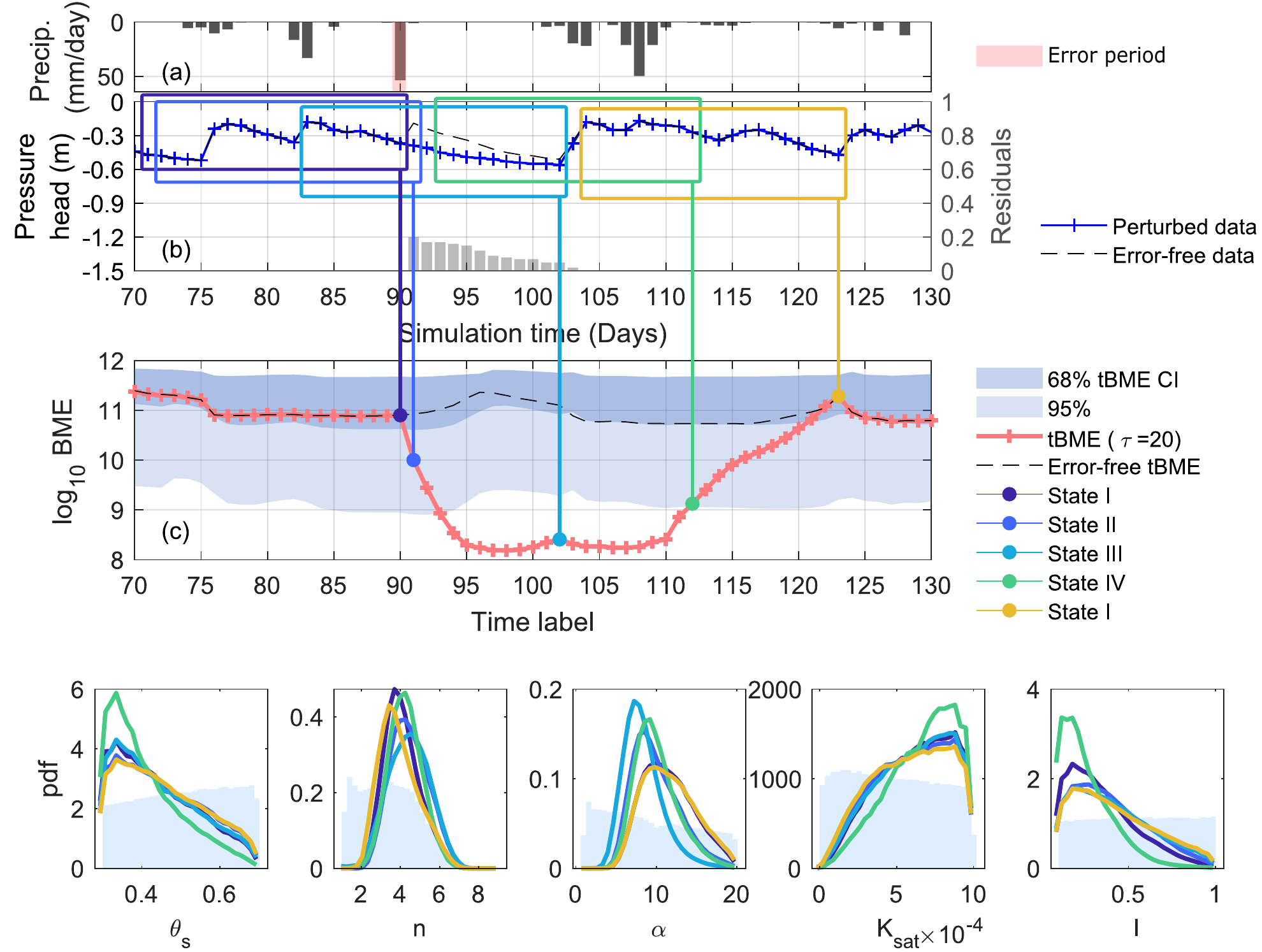}
\caption{Zoomed-in view of tBME curve and dynamic posterior parameter distributions for window size $\tau$ = 20 days (synthetic case 2). }
\label{fig:zoom_Case2_20}
\end{figure}

We also note the familiar pattern of two drops in the tBME curve (e.g., at $t=100$ days for $\tau = 10$ days) flattening out to be one single valley for $\tau=20$ days. This pattern indicates that the time span of model error is between those window sizes. In other words, only states II (initial decline) and IV (final rise) are clearly observed now, but state III (intermediate rise and fall) is absent. This can be investigated in more detail by comparing the zoomed-in tBME curves for $\tau=10$ and $\tau=20$ days (Figs. \ref{fig:zoom_Case2_10} and \ref{fig:zoom_Case2_20}, respectively) during days 90 to 110. The tBME curve for $\tau = 10$ shows two drops (state III$^*$ identifiable), while only one wide valley appears in the tBME curve for $\tau = 20$ (state III). This signal merge occurs when the time window is larger than the error period itself, because compensation of model error through parameter adjustment cannot happen, and so the tBME curve does not return to the high-density region of the tBME sampling distribution. 
Hence, it can be judged from Figs. \ref{fig:Plot3_obs36}f and \ref{fig:zoom_Case2_20}c that the error persists for at least ten and less than 20 days. This conclusion is correct - we can verify it from estimating the length of significant residuals (gray bars in Fig. \ref{fig:Plot3_obs36}b) produced by the second error period which is 13 days. This number is also obtained when determining $L_e=L_s-\tau$ based on theoretical considerations, with $L_s=33$.

\begin{figure}[ht]
\centering
\includegraphics[scale=0.8]{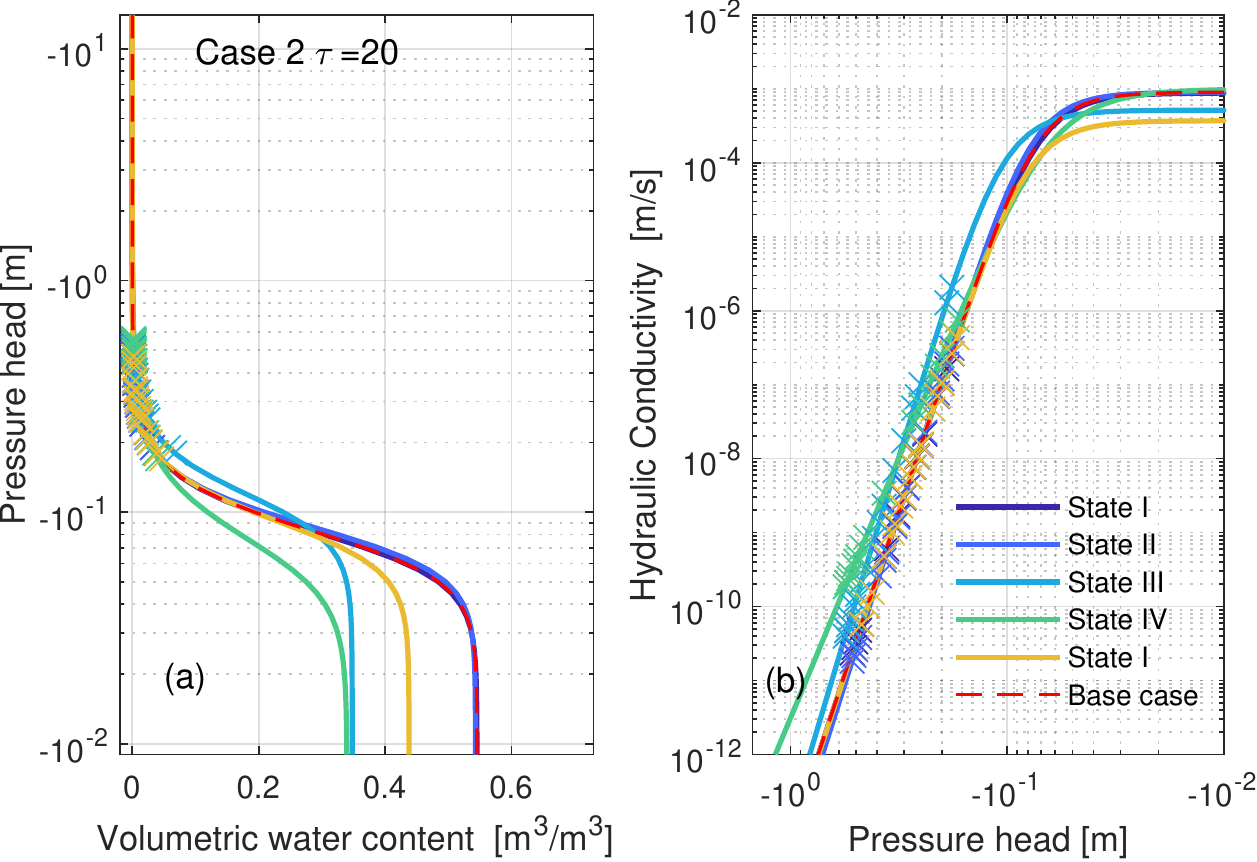}
\caption{Maximum-likelihood water retention curves (left) and unsaturated hydraulic conductivity functions (right) of the top soil layer for the five states color-coded in Fig. \ref{fig:zoom_Case2_10}b (synthetic case 2, window size $\tau=20$ days). Dashed red lines show true curves of the base case; there are no ``true'' curves for the perturbed dataset because forcing error does not impact the soil hydraulic characteristics.} 
\label{fig:WRC_ML_C2_t20}
\end{figure}

\begin{figure}[ht]
\centering
\includegraphics[scale=0.8]{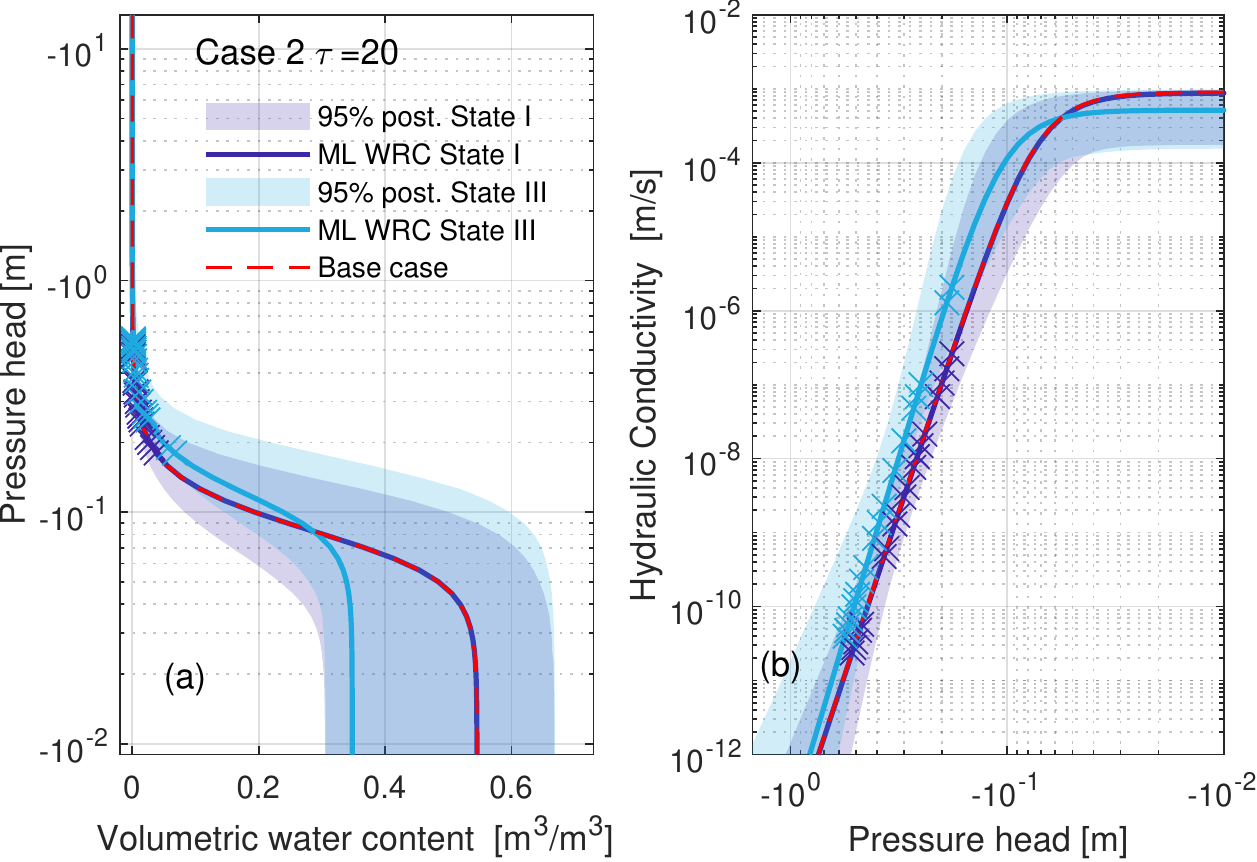}
\caption{Posterior credible intervals of water retention curves (left) and unsaturated hydraulic conductivity curves (right) of the top soil layer for states I and III* (synthetic case 2, window size $\tau=20$ days).}  
\label{fig:WRC_C2_t20}
\end{figure}

\clearpage
\subsection{synthetic case 3, $\tau=10$}

\begin{figure}[ht]
\centering
\includegraphics[scale=0.8]{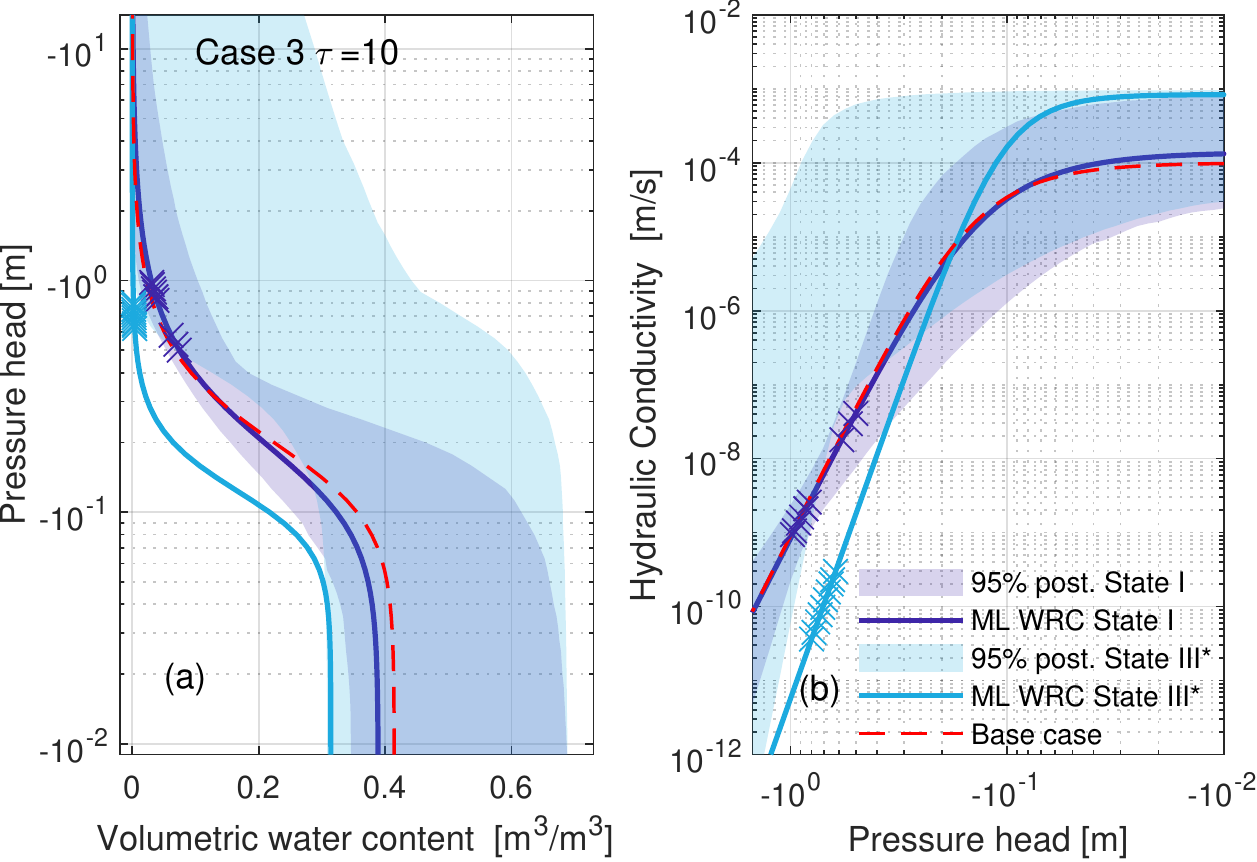}
\caption{Comparison of the uncertainty range of WRC at state I and state III* in Fig. \ref{fig:zoom_Case3_10} (Synthetic case 3 $\tau$ = 10).}  
\label{fig:WRC_C3_t10}
\end{figure}

\clearpage
\subsection{synthetic case 3, $\tau=20$}

\begin{figure}[ht]
\centering
\includegraphics[scale=0.8]{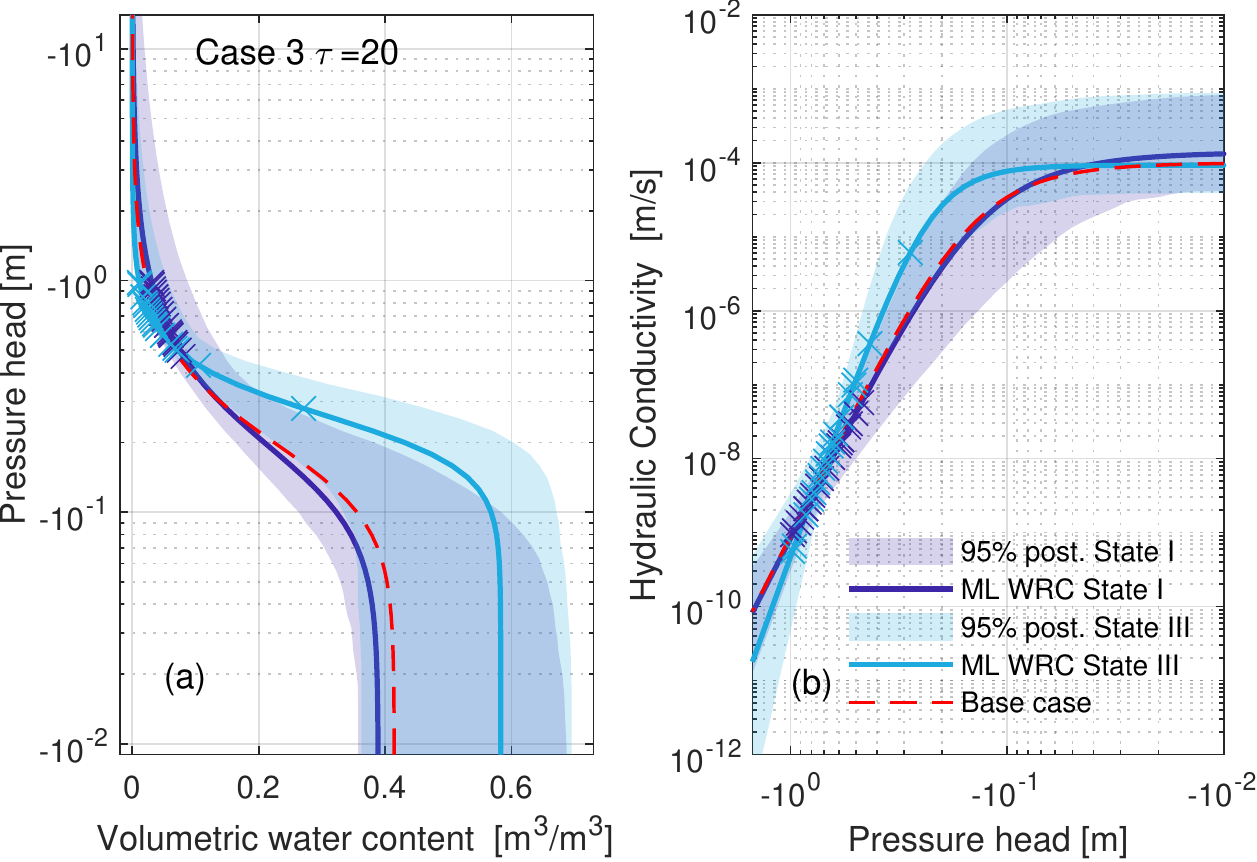}
\caption{Comparison of the uncertainty range of WRC at state I and state III in Fig. \ref{fig:zoom_Case3_20} (Synthetic case 3 $\tau$ = 20).}  
\label{fig:WRC_C3_t20}
\end{figure}

\clearpage
\subsection{Real data, $\tau=10$}
\begin{figure}[ht]
\centering
\includegraphics[scale=0.8]{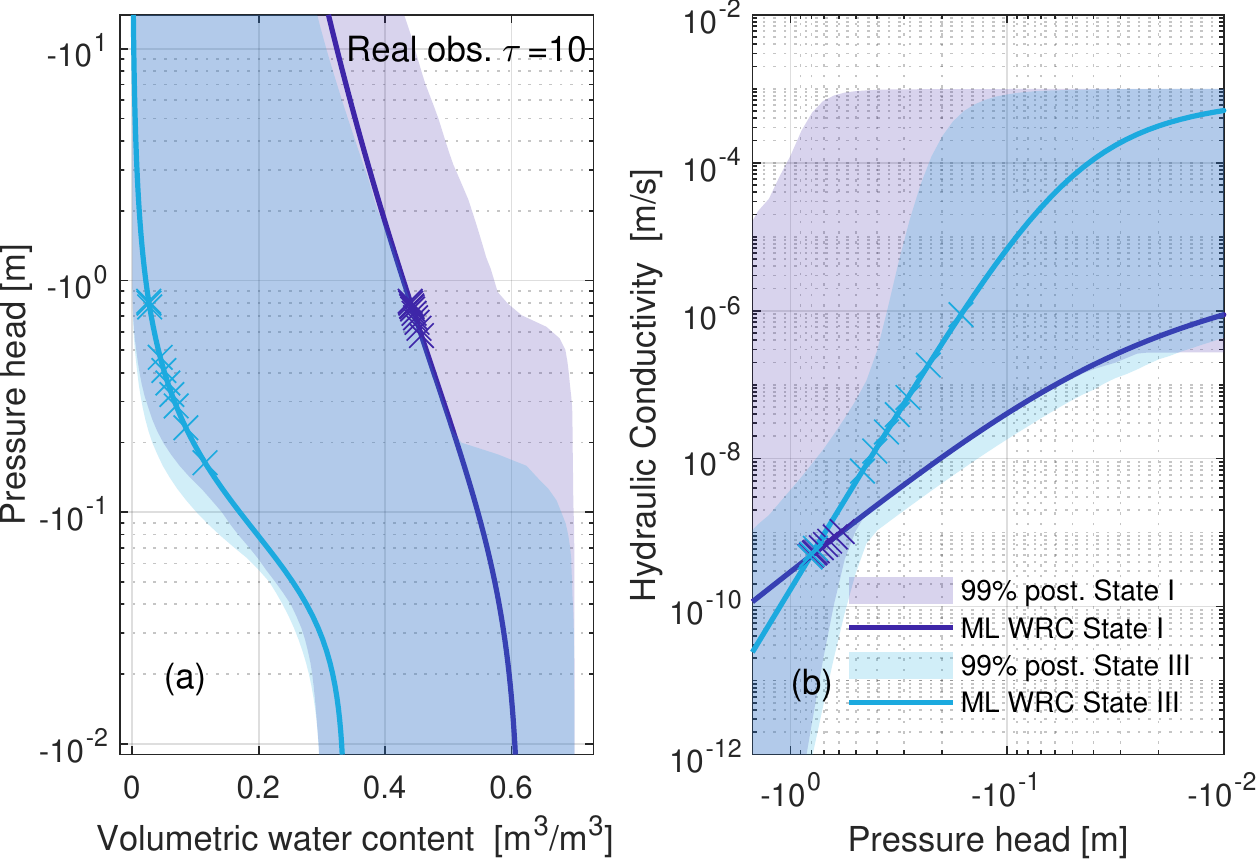}
\caption{Comparison of the 99 \% uncertainty range of WRC at state I and state III in Fig. \ref{fig:zoom_Case4_10} (Real data, $\tau$ = 10).}  
\label{fig:WRC_C4_t10_p99}
\end{figure}

Figure \ref{fig:WRC_C4_t10_p99} presents the 99 \% uncertainty range of WRC at state I and state III in Fig. \ref{fig:zoom_Case4_10}, which is the real data case analyzed with $\tau$ = 10. 

\clearpage
\subsection{Real data, $\tau=20$}
\begin{figure}[ht]
\centering
\includegraphics[scale=0.8]{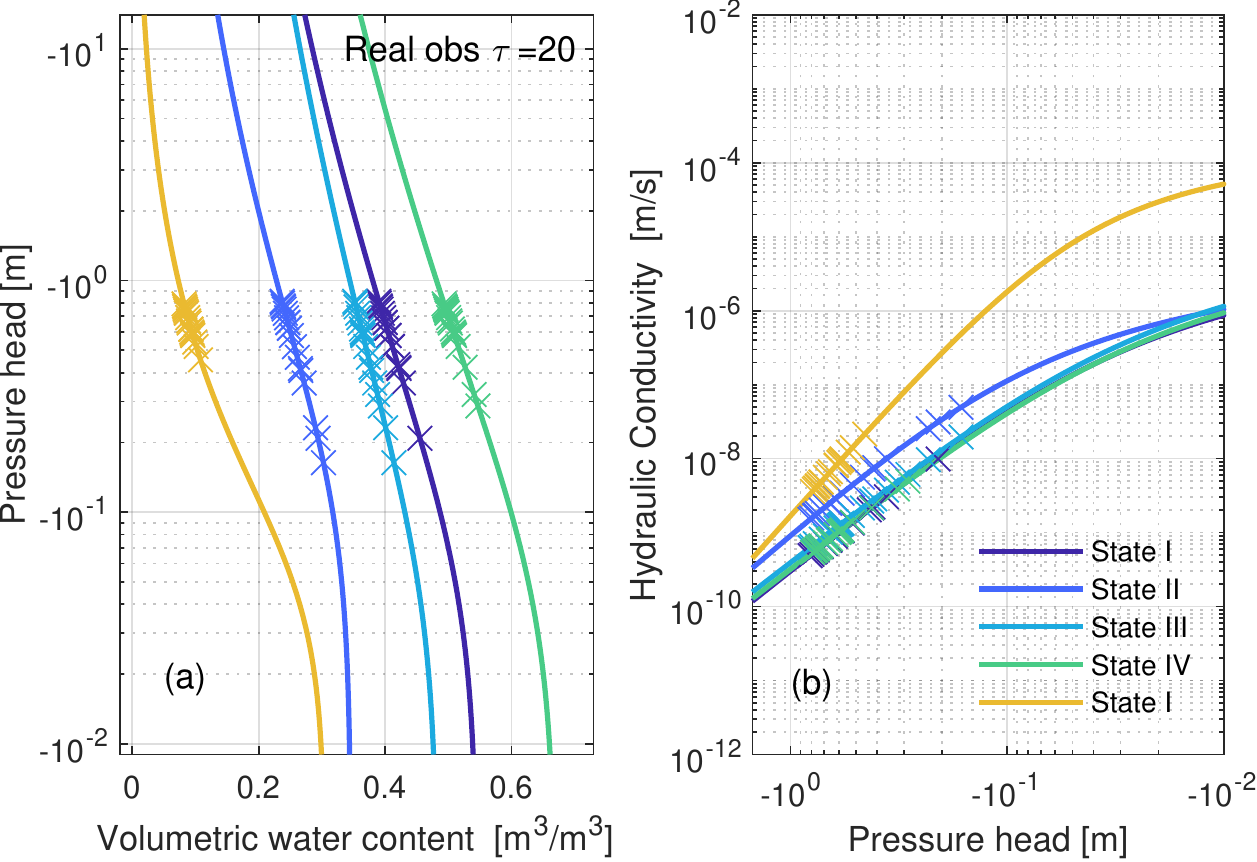}
\caption{Comparison of the WRC of MAP at each state in Fig. \ref{fig:zoom_Case4_20} (Real data, $\tau$ = 20).
}  
\label{fig:WRC_ML_C3_t20}
\end{figure}

\begin{figure}[ht]
\centering
\includegraphics[scale=0.8]{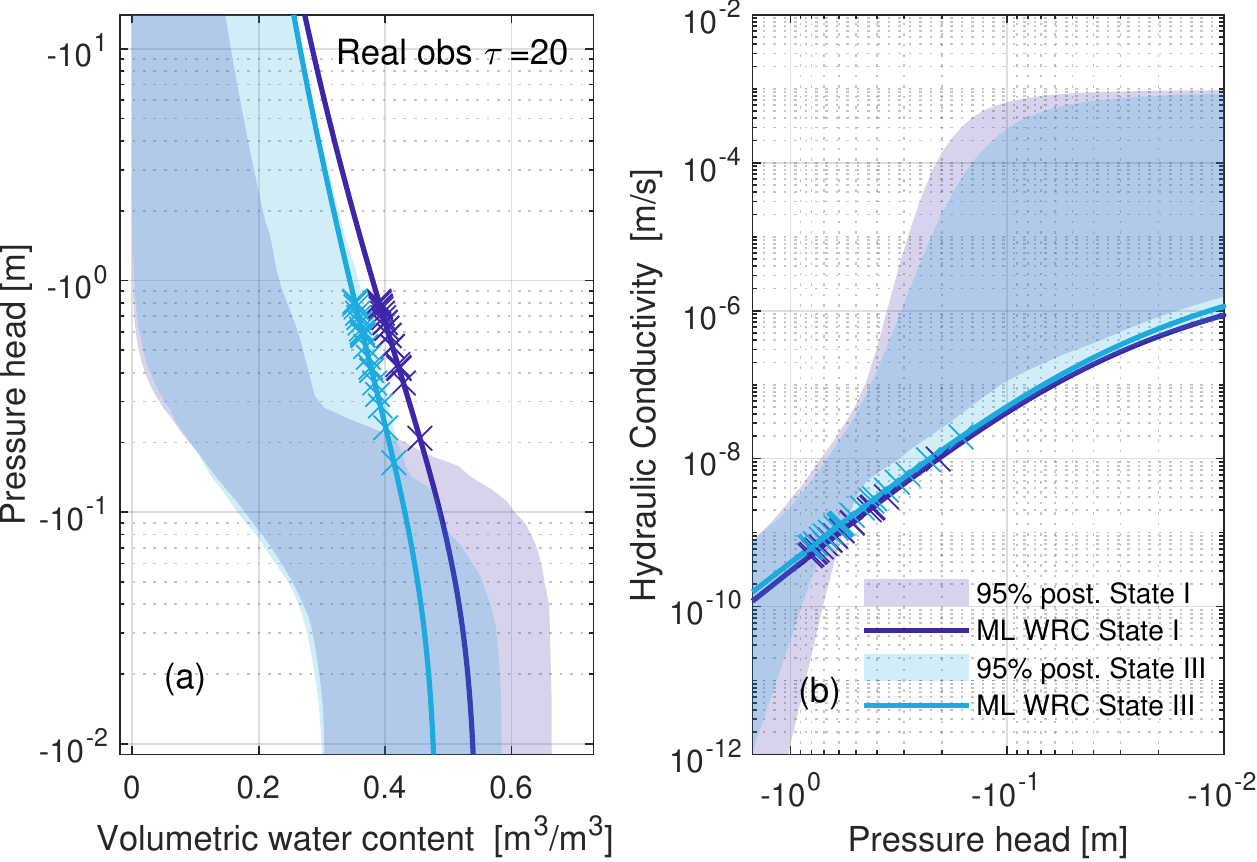}
\caption{Comparison of the uncertainty range of WRC at state I and state III in Fig. \ref{fig:zoom_Case4_20} (Real data, $\tau$ = 20).}  
\label{fig:WRC_C4_t20}
\end{figure}

\begin{figure}[ht]
\centering
\includegraphics[scale=0.8]{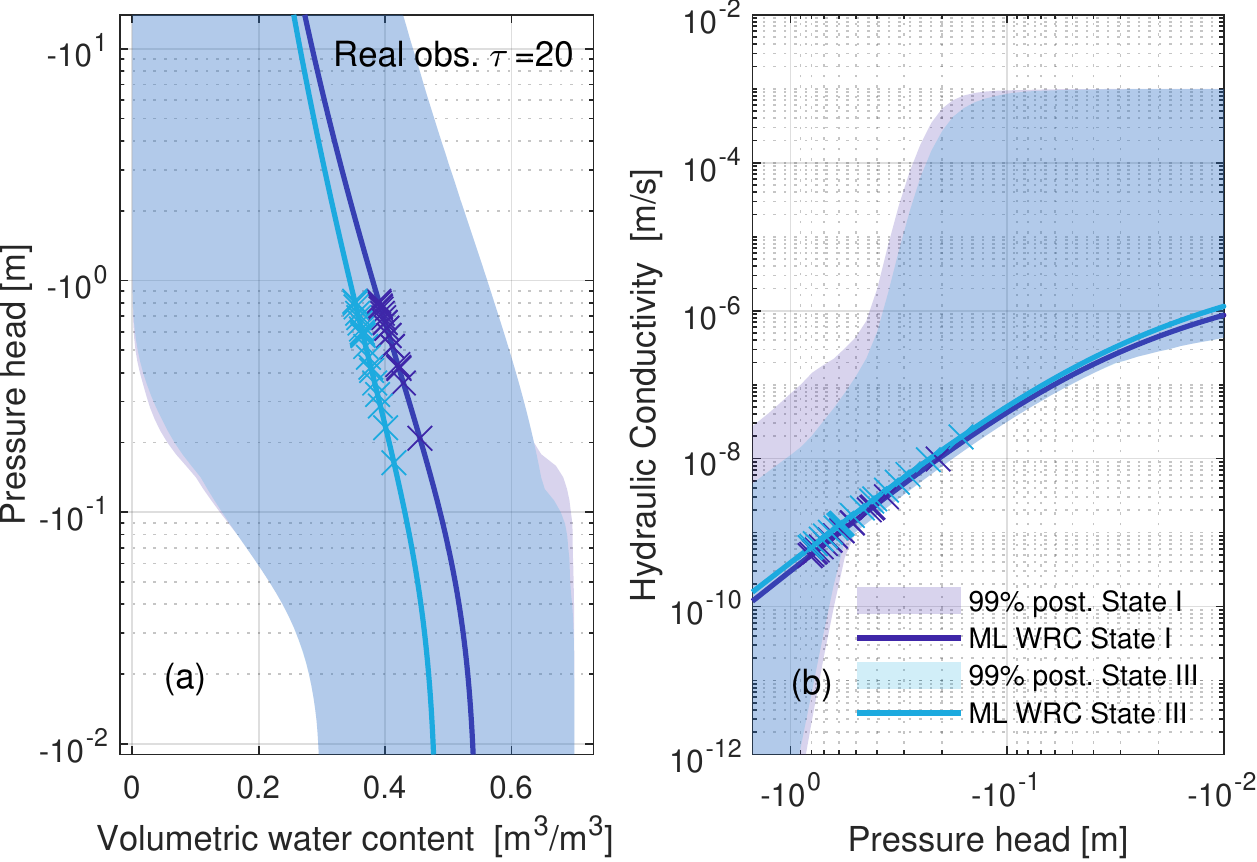}
\caption{Comparison of the 99 \% uncertainty range of WRC at state I and state III in Fig. \ref{fig:zoom_Case4_20} (Real data, $\tau$ = 20).}  
\label{fig:WRC_C4_t20_p99}
\end{figure}

\clearpage

\acknowledgments
The authors would like to thank the German Research Foundation (DFG) for financial support of the project within the Research Training Group GRK1829 ``Integrated  Hydrosystem  Modelling'' and the Cluster of Excellence EXC 2075 ``Data-integrated Simulation Science (SimTech)'' at the University of Stuttgart under Germany’s Excellence Strategy - EXC 2075 - 39074001. The Spydia field data was aquired and Hydrus models setup as part of Lincoln Agritech's  (formerly Lincoln Ventures) Groundwater Quality Protection Programme (8137-ASXS-LVL) founded by the New Zealand Foundation for Research, Science and Technology (FRST). We  like to acknowledge the LAL field team and Greg Barkle for support. The field data and model ensemble used in this study can be requested from the authors. All code files of the analysis in this research will be provided on the platform of the Data Repository of the University of Stuttgart (DaRUS).


%
%

\bibliography{tBME}

%
%
%
%
%

\end{document}